\documentclass[twocolumn,aps,prl]{revtex4-1}

	\usepackage[usenames,dvipsnames]{xcolor}
	\usepackage{amsmath}
	\usepackage{amsfonts}
	\usepackage{amssymb}
	\usepackage[colorlinks=true,citecolor=blue,linkcolor=red]{hyperref}
	\usepackage{graphicx}
	\usepackage{bbold}					
	\usepackage[makeroom]{cancel}		
	\usepackage{multirow}				
	\usepackage[normalem]{ulem}        
	\usepackage{array}
	\usepackage{pdfpages}
    \usepackage{float}
	\makeatletter
	\AtBeginDocument{\let\LS@rot\@undefined}
	\makeatother
	
	\newcolumntype{x}[1]{>{\centering\let\newline\\\arraybackslash\hspace{0pt}}p{#1}}

	\DeclareMathAlphabet{\mathbbold}{U}{bbold}{m}{n}

	\newcounter{subeqn} %
	\renewcommand{\Re}{\operatorname{Re}}
	\renewcommand{\Im}{\operatorname{Im}}

	\makeatletter
	\@addtoreset{subeqn}{equation}
	\makeatother

\definecolor{TB}{rgb}{0,0,0} 

\begin{document}

\title{Theoretical Prediction of Non-Hermitian Skin Effect in Ultracold Atom Systems}

\author{Sibo Guo$^{1,2}$}\thanks{These two authors contributed equally}
\author{Chenxiao Dong$^{1,2}$}\thanks{These two authors contributed equally}
\author{Fuchun Zhang$^{3}$}\email[Corresponding author: ]{fuchun@ucas.ac.cn}
\author{Jiangping Hu$^{1,2,3}$}\email[Corresponding author: ]{jphu@iphy.ac.cn}
\author{Zhesen Yang$^{3}$}\email[Corresponding author: ]{yangzs@ucas.ac.cn}

\affiliation{$^{1}$Beijing National Laboratory for Condensed Matter Physics,
	and Institute of Physics, Chinese Academy of Sciences, Beijing 100190, China}
\affiliation{$^{2}$School of Physical Sciences, University of Chinese Academy of Sciences, Beijing 100049, China}
\affiliation{$^{3}$Kavli Institute for Theoretical Sciences, University of Chinese Academy of Sciences, Beijing 100190, China}

\date{\today}
	
\begin{abstract}
Non-Hermitian skin effect, which refers to the phenomenon that an extensive number of eigenstates are localized at the boundary, has been widely studied in lattice models and experimentally observed in several classical systems. In this work, we predict that the existence of the non-Hermitian skin effect in the dissipative ultracold fermions with spin-orbit coupling, a continuous model that has been implemented by the Hong-Kong group in a recent experiment. This skin effect is robust against the variation of external parameters and trapping potentials. We further reveal a dynamic sticky effect in our system, which has a common physical origin with the non-Hermitian skin effect. Our work paves the way for studying novel physical responses of non-Hermitian skin effect in quantum systems.
\end{abstract}

\maketitle

{\em Introduction.}---The ultracold atom system has provided an ideal platform for studying both single-particle physics and many-body physics due to its super purity, high controllability, and intrinsic quantum nature~\cite{r47,r49,r16,r90,r30}. Some examples include the study of topological band theory~\cite{r27,r28,r29,r31,r32,r33,r35,r36,r37,r38,r41,r42,r51,r52,r127,r144,r145,r146,r147}, and strongly correlated physics~\cite{r49,r73,r74,r75,r76}. When the system is not closed, particles can escape from the system to the environment~\cite{r22,r88,r23,r89,r24,r25,r26,r12,r21,r120,r122,r126,r128,r129,r130,r132,r133,r141}, and its single-particle physics can no longer be described by a Hermitian Hamiltonian, but by an effective non-Hermitian Hamiltonian~\cite{r133,r136,r143}. For example, recently, Ref.~\cite{r12} first implemented a one-dimensional spin-orbit coupled system with highly controllable spin-dependent particle loss, which sets a new platform to study non-Hermitian physics.

Non-Hermitian skin effect (NHSE), which refers to an anomalous localization phenomenon in an open boundary system~\cite{yaoEdgeStatesTopological2018b}, has been widely studied recently~\cite{yaoEdgeStatesTopological2018b,yaoNonHermitianChernBands2018b,r123,PhysRevLett.121.026808,
yokomizoNonBlochBandTheory2019a,PhysRevLett.125.126402,PhysRevLett.124.086801,PhysRevLett.125.226402,PhysRevB.97.121401,
leeAnatomySkinModes2019b,PhysRevLett.123.170401,PhysRevLett.123.016805,PhysRevLett.124.250402,PhysRevLett.124.056802,
PhysRevResearch.1.023013,PhysRevLett.124.066602,PhysRevLett.125.186802,Li:2020aa,r118,Xiao:2020aa,Helbig:2020aa,
Weidemann311,PhysRevResearch.2.023265,r124,r125,r131,r112,r140,r142}. For a system with NHSE, the number of boundary localized eigenstates is proportional to the volume of the system. From the theoretical point of view, an important implication of the NHSE is the non-perturbative breakdown of Bloch's theorem~\cite{r64}, which brings new opportunities and challenges to the quantum theories established within the Bloch's theorem. From the experimental point of view, although the NHSE has been experimentally observed in several classical systems~\cite{Helbig:2020aa,Xiao:2020aa,Weidemann311,PhysRevResearch.2.023265,r57,r54,r94,r95,r96,r98}, its implementation in quantum systems is still absent. Therefore, the realization and detection of NHSE in quantum systems, e.g. ultracold atom systems, become the next step for the non-Hermitian community and may open a new route for further theoretical research.

\begin{figure}[b]
	\centerline{\includegraphics[scale=0.42]{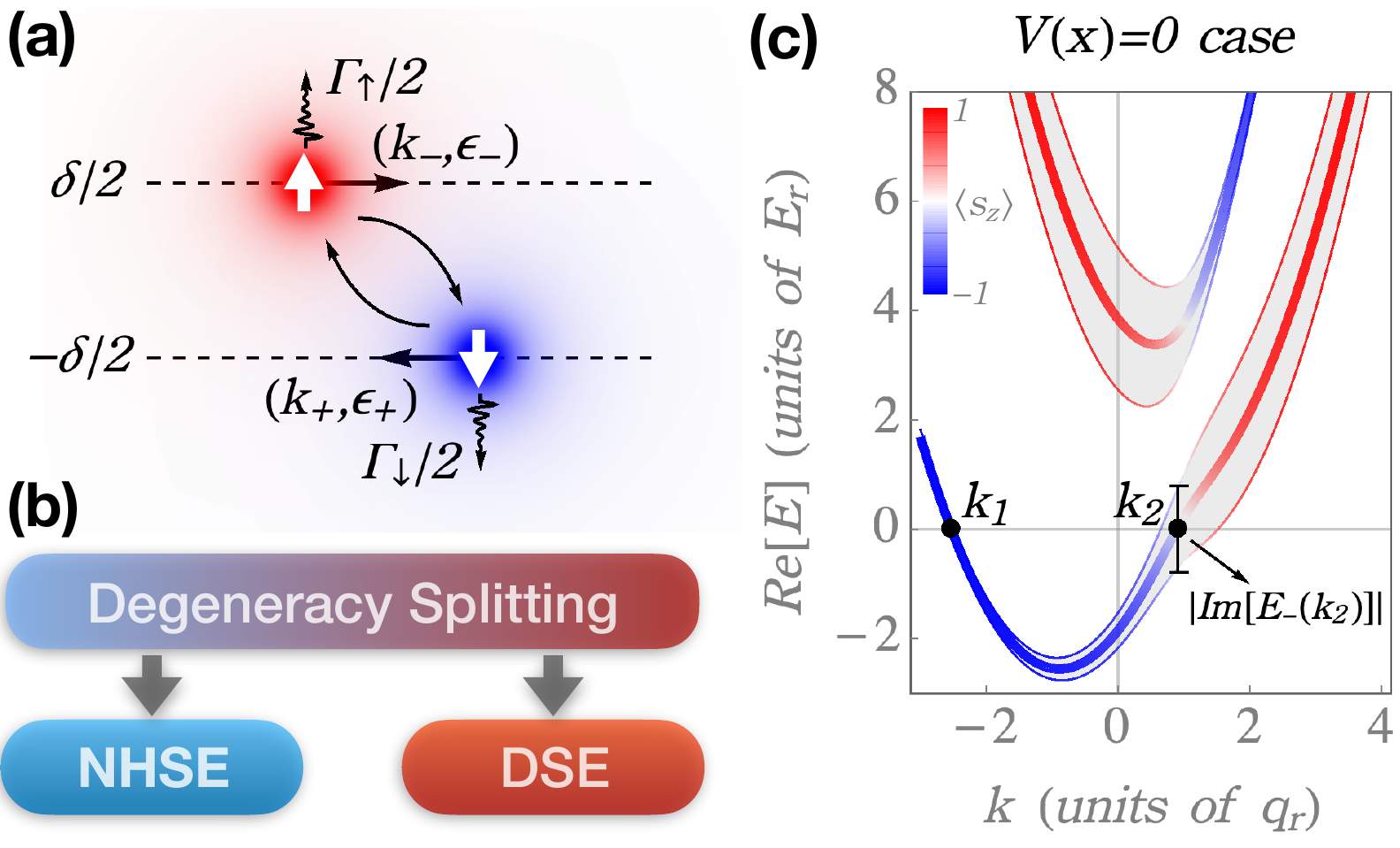}}
	\caption{(a) shows the physical system we studied, where $k_\pm=k\pm q_r$ and $\epsilon_\pm=\hbar^2k_\pm^2/2m$. (b) shows that both the NHSE and the DSE originate from the (free particle spectral) degeneracy splitting. (c) shows the presence of degeneracy splitting in our model. 
 One can find that although the real part of the energy between $k_1$ and $k_2$ states are the same, their imaginary part of the energy is different. 
 The parameters are $(\delta,\Omega_R,\Gamma_{\uparrow},\Gamma_{\downarrow})=(4,9/2,6,6/13)E_{r}$, where $E_r=\hbar^2 q_r^2/2m$.}
	\label{F1}
\end{figure}

In this paper, we predict that the one-dimensional dissipative spin-orbit coupled ultracold fermions implemented in Ref.~\cite{r12} have {\em robust} NHSE. In contrast to the previous studies in the lattice model, the NHSE discussed in this paper appears in a continuous model due to the absence of external applied optical lattice. We also generalize the non-Bloch band theory~\cite{yaoEdgeStatesTopological2018b,yokomizoNonBlochBandTheory2019a,PhysRevLett.125.226402} to the continuous case, which is dubbed as {\em generalized wave vectors} (GWV) theory. Besides, we propose that the NHSE in our model can induce a so-called momentum-dependent {\em dynamic sticky effect} (DSE), that is when a small/large momentum wave packet hits the skin-localized boundary, it will be bounded/reflected by the corresponding boundary.

{\em NHSE in ultracold atoms.}---The model implemented in Ref.~\cite{r12}, which is illustrated in Fig.~\ref{F1} (a), can be described by the following stationary Schrödinger equation,
\begin{equation}
	[H(\hat{k})+V(x)]\psi_E(x)=E\psi_E(x),
	\label{Eq3}
\end{equation}
where $\hat{k}=\hat{p}/\hbar=-i\partial_x$ is the wave vector operator, $V(x)$ is the effective trapping potential, and $H(\hat{k})$
is the free particle Hamiltonian whose form is
\begin{equation}
H(\hat{k})=
\left(\begin{array}{cc}
	\epsilon_-(\hat{k})+\delta / 2	-i\Gamma_\uparrow/2 & \Omega_{R} / 2 \\
	\Omega_{R} / 2 & \epsilon_+(\hat{k})-\delta / 2 -i\Gamma_\downarrow/2
\end{array}\right).
\label{Eq1}
\end{equation}
Here $\epsilon_\pm(\hat{k})=(\hbar\hat{k}\pm\hbar q_{r})^{2}/2m$, $m$ is the mass of the atom, $\pm \hbar q_r$ is half of the momentum transferred through the two-photon Raman coupling process with Raman coupling strength $\Omega_R$ and two-photon detuning $\delta$~\cite{r14}, and $\Gamma_{\uparrow/\downarrow}$ is proportional to the spin-dependent loss of the atom~\cite{SM3}.

\begin{figure}[t]
	\centerline{\includegraphics[scale=0.42]{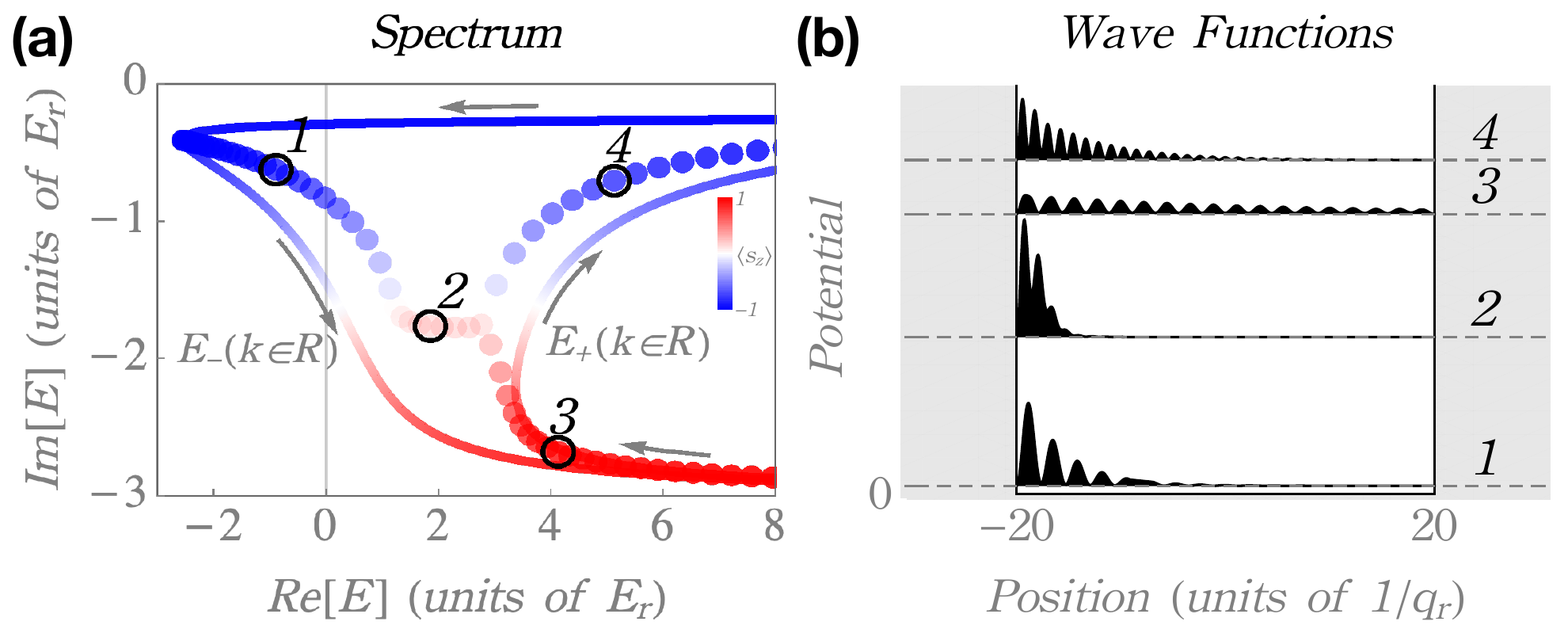}}
	\caption{(a) shows the energy spectrum of Eq.~\ref{Eq3}, where solid lines represent the spectrum under $V(x)=0$ and the dots represent the spectrum under Eq.~\ref{Eq4} with $L=40/q_r$. 
		(b) shows several eigenfunctions whose eigenvalues are labeled in (a). Both the spectral difference and the exponential localization of the eigenfunctions are the signatures of NHSE. The parameters are the same as Fig.~\ref{F1}.}
	\label{F2}
\end{figure}

When the effective trapping potential $V(x)$ is ignored, the atoms can be considered free. By diagonalizing Eq.~\ref{Eq1}, one can obtain the corresponding eigenvalues
\begin{equation}\begin{aligned}
		E_{\pm}(k)=E_0(k)
		\pm\sqrt{\left(-\frac{\hbar^2q_r}{m}k+\frac{\delta}{2}-i\frac{\Gamma_z}{2}\right)^2+\frac{\Omega_R^2}{4}},
		\label{Eq2}
\end{aligned}\end{equation}
where $k\in\mathbb{R}$,  $E_0(k)=\hbar^2(k^2+q_r^2)/2m-i\Gamma_0/2$,  $\Gamma_{0/z}=(\Gamma_\uparrow\pm\Gamma_\downarrow)/2$. As shown in Fig.~\ref{F1} (b), an important feature of Eq.~\ref{Eq2} is the presence of (free particle spectral) {\em degeneracy splitting}~\cite{SM12}, which causes the NHSE and DSE. The reason is that due to the existence of degeneracy splitting, the number of plane waves is not enough to form the eigenstate (and participate in the boundary reflection process), which indicates the emergence of NHSE (and DSE). This degeneracy splitting can be observed in Fig.~\ref{F1} (c) and Fig.~\ref{F2} (a). In Fig.~\ref{F1} (c), $\Re E_\pm(k)$ is plotted with thick solid lines; $\Re E_\pm (k)+|\Im E_\pm (k)|/2$ and $\Re E_\pm (k)-|\Im E_\pm (k)|/2$ are plotted with thin solid lines. Therefore, the vertical distance around $\Re E_\pm(k)$ represents the $|\Im E_\pm(k)|$. One can find that although $\Re E_-(k_1)=\Re E_-(k_2)=0$, their imaginary parts are different. Other parameters in the calculation are listed in the caption of the corresponding figures, and the color of the curves represent the mean value of the $z$-component spin, i.e. $\langle s_z\rangle$~\cite{SM5}. Fig.~\ref{F2} (a) further shows that the above discussed degeneracy splitting still holds in almost the whole spectrum in the plotted region. Here we note that the solid lines and arrows in Fig.~\ref{F2} (a) indicate the spectral flow in the complex energy plane when $k$ runs from $-\infty$ to $+\infty$.

When the effective trapping potential $V(x)$ is taken into account, the NHSE appears. In order to simplify the discussion, a box potential is assumed, i.e.
\begin{equation}
V(x)=\left\{\begin{array}{ll}
		0, & -L/2<x<L/2\\
		\infty, & \text { otherwise }.
	\end{array}\right.
	\label{Eq4}
\end{equation}
As shown in Fig.~\ref{F2} (a), the dots represent the numerical solutions in the drawing interval with $L=40/q_r$ for  Eq.~\ref{Eq4}. One can find that the eigenvalues in the box potential are distinct from the free particle spectrum $E_\pm(k\in\mathbb{R})$ in the complex energy plane. This is the spectral signature for the emergence of NHSE~\cite{PhysRevLett.125.126402,PhysRevLett.124.086801}. Fig.~\ref{F2} (b) further shows several typical numerical wavefunctions of the system, whose eigenvalues are labeled in Fig.~\ref{F2} (a)~\cite{SM1}. One can find that they are all exponentially localized at the left boundary and have different localization lengths. The appearance of the exponentially localized eigenstates for the quasi-continuous spectrum is the wave function signature for the emergence of NHSE~\cite{yaoEdgeStatesTopological2018b,yokomizoNonBlochBandTheory2019a}.

\begin{figure}[b]
	\centerline{\includegraphics[scale=0.42]{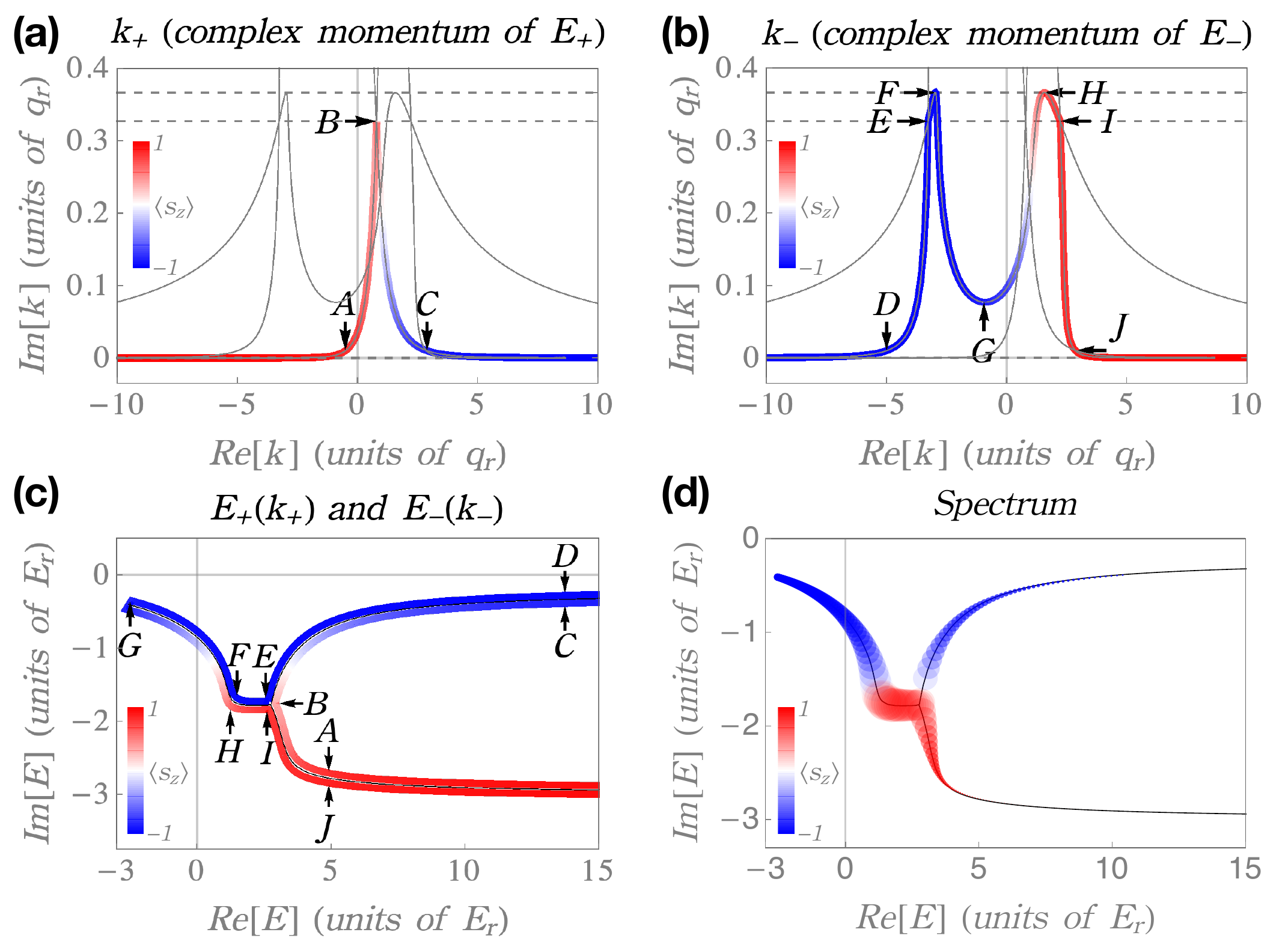}}
	\caption{(a)-(d) show the asymptotic solutions under the parameters shown in Fig.~\ref{F1}. 
In (a)-(b), the generalized wave vectors and auxiliary generalized wave vectors are shown with thick colored lines and thin gray lines, respectively. Here the dashed lines represent the largest and smallest value of the imaginary part of $k$, respectively, which will be used to characterize the strength of NHSE~\cite{SM2}. (c) represents the asymptotic spectrum, namely, $E_{+}$/$E_{-}$ maps curve A-B-C/D-E-F-G-H-I-J (represent the energy flow for $E_+(k_+)$/$E_-(k_-)$, respectively) in (a)/(b) to the curve A-B-C/D-E-F-G-H-I-J in (c). (d) shows the comparison between the asymptotic (black line) and numerical (colored dots) solutions. Here the radius of the dots in (d) is proportional to the inverse of the localization length. 
}
	\label{F3}
\end{figure}

{\em Asymptotic solutions.}---For large $L$, the asymptotic solution of Eq.~\ref{Eq3} under the box potential Eq.~\ref{Eq4} can be obtained analytically, which contains two boundary-condition-allowed complex wave vector curves $k_+$ and $k_-$, respectively. The corresponding asymptotic eigen spectrum of $E_+$ and $E_-$ can be obtained from $E_+(k_+)$ and $E_-(k_-)$, respectively~\cite{SM2}. Here we note that our method can be regarded as a continuous version of the auxiliary generalized Brillouin zone theory~\cite{PhysRevLett.125.226402}. As shown in Fig.~\ref{F3} (a) and (b), the thick solid colorful lines represent the asymptotic curves of complex wave vectors $k_+$ and $k_-$, which we call GWV, and the thin solid gray lines represent the auxiliary generalized wave vectors, which determine the analytical behaviors of $k_\pm$ and can be calculated exactly~\cite{SM2}. From Fig.~\ref{F3} (a) and (b), one can find that $\Im k_\pm$, which is related to the localization behaviors, gradually approaches zero as $|\Re k_{\pm}|$ increases~\cite{SM7}. As a result, the corresponding asymptotic eigenstate becomes extended. In contrast, the eigenstates for small $|\Re k_{\pm}|$ are localized skin modes. Fig.~\ref{F3} (c) shows the asymptotic spectrum determined by $k_\pm$, 
where the following point pairs $(F,H)$, $(E,B,I)$, $(A,J)$ and $(C,D)$ in (c) represent four different common points in the complex energy plane. It should be noted that $E_{+}(k_+)$ and $E_{-}(k_-)$ cover different regions on the complex plane, respectively~\cite{SM2}. As shown in Fig.~\ref{F3} (d), by comparing $E_\pm(k_\pm)$ with numerical solutions represented by the dots in Fig.~\ref{F3} (d), one can find that they match well.

\begin{figure}[b]
	\centerline{\includegraphics[scale=0.42]{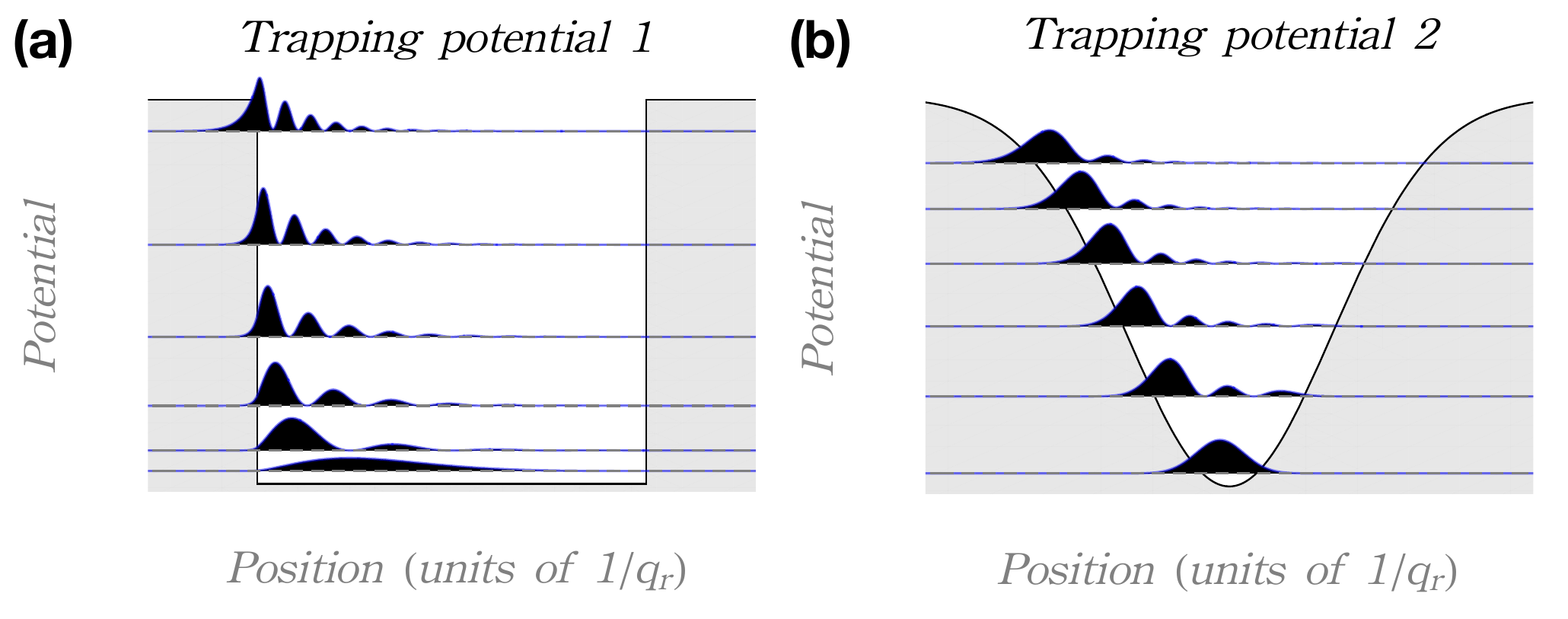}}
	\caption{(a) and (b) show the spatial distribution of several eigenfunctions under finite deep square and Gaussian well, respectively~\cite{SM2}, where the presence of the skin modes localized on the left can be clearly seen. The parameters are the same as those in Fig.~\ref{F1}.}
	\label{F3_2}
\end{figure}
	
{\em The robustness of NHSE.}---
The stability of the NHSE is related to the absence of the inversion symmetry $\mathcal{P}$ and anomalous time-reversal symmetry $\mathcal{\bar{T}}$~\cite{r67,PhysRevLett.125.186802} of $H(k)$, whose definitions are
\begin{equation}
	\mathcal{P}^{-1}H(k)\mathcal{P}=H(-k),~~~~~\mathcal{\bar{T}}^{-1}H(k)\mathcal{\bar{T}}=H(-k),
	\label{Eq5}
\end{equation}
where $k$ is a complex number, and $\mathcal{P}$ is a unitary operator, and  $\mathcal{\bar{T}}=U_{\mathcal{\bar{T}}}\mathcal{K}^t$ is an anti-unitary operator with $\mathcal{K}^t$ being the transpose operator~\cite{PhysRevX.8.031079,r67}. It can be checked that once $\Gamma_\uparrow\neq\Gamma_\downarrow$, the above two symmetries must be broken, which ensures the existence of the NHSE. Further numerical verifications are shown in the Supplemental Material.

The NHSE is also robust against trapping potentials. In Fig.~\ref{F3_2} (a) and (b), we have plotted several eigenstates localized on the left side of the boundary under different trapping potentials, which numerically indicates the robustness of NHSE~\cite{SM2}. This robustness is quite important, since if the NHSE can be destroyed by these potentials, then, the NHSE will be fragile and can hardly be observed in experiments.

We believe that the above robustness of NHSE to external potentials is a consequence of the non-Hermitian bulk-boundary correspondence. Here the bulk refers to the degeneracy splitting of the free particle spectrum, and the boundary refers to the NHSE under generic slowly varying trapping potentials $V(x)$. The reason is that in the slow varying trapping region, $V(x)$ can be approximated by a constant. As a result, the corresponding solution in this region will not all be composed of plane waves due to the presence of degeneracy splitting.

\begin{figure*}[t]
	\centerline{\includegraphics[scale=0.4]{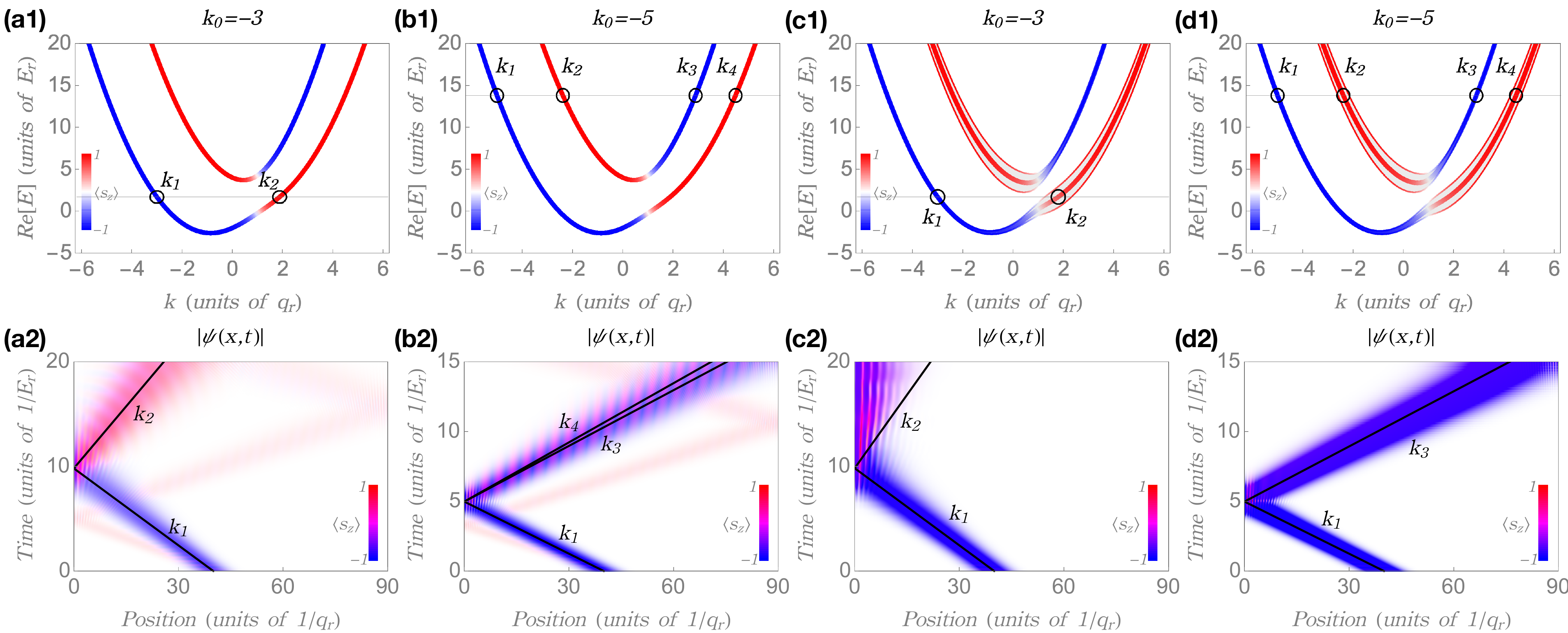}}
	\caption{(a-b) and (c-d) show the dynamical behavior of the system without NHSE, i.e. $(\delta,\Omega_R,\Gamma_\uparrow, \Gamma_\downarrow)=(4,9/2,0,0)E_{r}$ and with NHSE, i.e. $(\delta,\Omega_R,\Gamma_\uparrow, \Gamma_\downarrow)=(4,9/2,6,6/13)E_{r}$, respectively. (a1)-(d1) represent the free particle dispersion relation $\Re E_{\pm} (k\in\mathbb{R})$, in which the shadow region represents the $|\Im E_\pm(k\in\mathbb{R})|$. (a2)-(d2) show the scattering behavior of Gaussian wave packet with a different momentum, in which the solid lines represent the trajectories calculated from the group velocity. The anomalous reflection behavior in (c2) can be understood from the presence of degeneracy splitting in (c1).}
	\label{F5}
\end{figure*}

{\em Dynamic sticky effect.}---Now we reveal a new effect related to the NHSE, that is, the DSE. In general, this effect can be present in any non-Hermitian systems with NHSE, not restricted to ultracold atoms. In Fig.~\ref{F5}, (a-b) and (c-d) show the dynamics of the system without and with NHSE, respectively. Here, the initial state is $\psi_\uparrow(x,t=0)=0$, $\psi_\downarrow(x,t=0)=A\exp[-(x-x_0)^2/20+ik_0x]$, where $A$ is the normalization factor, the center position $x_0=40/q_r$, the average momentum $k_0=-3q_r$ and $-5q_r$ for Fig.~\ref{F5} (a/c) and (b/d).  In order to characterize the dynamical evolution of wave packets, we normalized the wave function at each time, and plot $|\psi(x,t)|=\sqrt{|\psi_\uparrow (x,t)|^2+|\psi_\downarrow (x,t)|^2}$ in Fig.~\ref{F5} (a2-d2). An important feature for the dynamics with NHSE is that the wave packets will be bounded/reflected at a small/large momentum $k_0$ by the left boundary as shown in Fig.~\ref{F5} (c2)/(d2). This anomalous dynamical behavior is dubbed as momentum-dependent DSE in this paper.

The reason why the DSE appears and is momentum-dependent can be well explained by the fact that the DSE has a similar physical origin to the NHSE, that is the degeneracy splitting as shown in Fig.~\ref{F1} (b). 
Now we explain it. We first focus on the incident waves. As shown in Fig.~\ref{F5} (a1-d1), the (mean) momentum of the incident wave is $k_1$, and we can define the corresponding group velocity as $v_{1}=\Re [dE_{-}(k)/dk|_{k=k_1}]$~\cite{SM2}. The solid lines labeled by $k_1$ in Fig.~\ref{F5} (a2-d2) represent the dynamics determined by $v_1$, i.e. $x(t)=x_0+v_1 t$. One can find that the dynamics of the incident waves can be well described by the group velocity calculations. Next, we focus on the reflection waves. For the case in Fig.~\ref{F5} (a1/a2), there is only one reflection channel, i.e. $k_2$, which satisfies the equal energy condition, i.e. $E_-(k_1)=E_-(k_2)$. Therefore, the reflection wave can be well described by the group velocity of $k_2$, which is demonstrated in Fig.~\ref{F5} (a2). For the case in Fig.~\ref{F5} (b1/b2), since there are two reflection channels, i.e. $E_-(k_1)=E_-(k_3)=E_-(k_4)$, the corresponding reflection wave shows an interference feature, and also, the corresponding dynamics can be well described by the group velocity calculations, namely, $k_3$ and $k_4$ in Fig.~\ref{F5} (b2). However, for one case with NHSE, i.e. Fig.~\ref{F5} (c1/c2), the presence of degeneracy splitting implies that there is no plane wave that satisfies the equation energy condition. As a result, the reflection becomes anomalous, and the DSE appears~\cite{SM9}. As a comparison, for another case with NHSE, the degeneracy splitting between $k_1$ and $k_3$ is weak, which implies that the reflection wave can be well described by the group velocity of $k_3$. The result in Fig.~\ref{F5} (d2) demonstrates this point. Here we note that, due to the degeneracy splitting between $k_1$ and $k_4$, $k_4$ does not contribute to the reflection wave, which can be observed in Fig.~\ref{F5} (d2) with the absence of interference for the reflection wave.

{\em Discussions and conclusions.}---The degeneracy splitting proposed in this work can be regarded as a {\em band criterion} for the emergence of NHSE. Before our work, a spectral criterion has been established based on the spectral winding number of the Bloch Hamiltonian~\cite{PhysRevLett.125.126402,PhysRevLett.124.086801}. Although this criterion is straightforward in some theoretical calculations, it cannot be applied to the experiments because there is no method to measure the complex energy spectrum directly. In contrast, our band criterion, i.e. the degeneracy splitting, can be measured directly, for example, by the ARPES. Therefore, this criterion will be helpful in experiments.

The GWV theory proposed in this work also provides an analytical method for solving a typical class of differential equations of the form $\mathcal{H}\left(-i\frac{\mathrm{d} }{\mathrm{d} x}\right)\Psi(x)=E\Psi(x)$~\cite{SM2} in the large $L$ limit with homogeneous boundary conditions $\frac{\mathrm{d}^{k}}{\mathrm{d}^{k}x}\Psi(x=0)=0$, $\frac{\mathrm{d}^{l}}{\mathrm{d}^{l}x}\Psi(x=L)=0$, where $\mathcal{H}(-i\mathrm{d}/\mathrm{d} x)$ can be a matrix and its elements can have arbitrary order of $-i\mathrm{d}/\mathrm{d} x$, and $k,l=0,1,2,\cdots$, which represent the order of the derivative with respect to $x$.

In summary, we predict that the ultracold fermions with dissipative spin-orbit coupling have robust NHSE~\cite{SM11}. Since both the many-body interaction can be well controlled in such a system, therefore, our proposal establishes an ideal platform to study the interplay between NHSE and many-body interactions. We further show that the NHSE is robust not only to external parameters but also to external trapping potentials. The latter can be understood from the perspective of non-Hermitian bulk-boundary correspondence. Finally, we propose a boundary effect related to the NHSE, that is the DSE~\cite{SM10}. This DSE not only provides a valuable method to identify the NHSE but also reveals that the non-Hermitian systems with NHSE violate the conventional reflection rule, which is an important physical consequence that deserves to be investigated in detail.
	
{\em Acknowledgments.}---This work is supported by the Ministry of Science and Technology of China 973 Program (Grant No. 2017YFA0303100), the National Natural Science Foundation of China (Grant No. NSFC-11888101), and the Strategic Priority Research Program of Chinese Academy of Sciences (Grant No. XDB28000000). F.Z. and Z.Y. are supported in part by the Strategic Priority Research Program of Chinese Academy of Sciences Grant No. XDB28000000 and NSF China Grant No. 11674278. Z.Y. is also supported by the fellowship of China National Postdoctoral Program for Innovative Talents BX2021300.

\bibliography{ref,aGBZ,UnivSkin}

\begin{thebibliography}{100}%
\makeatletter
\providecommand \@ifxundefined [1]{%
 \@ifx{#1\undefined}
}%
\providecommand \@ifnum [1]{%
 \ifnum #1\expandafter \@firstoftwo
 \else \expandafter \@secondoftwo
 \fi
}%
\providecommand \@ifx [1]{%
 \ifx #1\expandafter \@firstoftwo
 \else \expandafter \@secondoftwo
 \fi
}%
\providecommand \natexlab [1]{#1}%
\providecommand \enquote  [1]{``#1''}%
\providecommand \bibnamefont  [1]{#1}%
\providecommand \bibfnamefont [1]{#1}%
\providecommand \citenamefont [1]{#1}%
\providecommand \href@noop [0]{\@secondoftwo}%
\providecommand \href [0]{\begingroup \@sanitize@url \@href}%
\providecommand \@href[1]{\@@startlink{#1}\@@href}%
\providecommand \@@href[1]{\endgroup#1\@@endlink}%
\providecommand \@sanitize@url [0]{\catcode `\\12\catcode `\$12\catcode
  `\&12\catcode `\#12\catcode `\^12\catcode `\_12\catcode `\%12\relax}%
\providecommand \@@startlink[1]{}%
\providecommand \@@endlink[0]{}%
\providecommand \url  [0]{\begingroup\@sanitize@url \@url }%
\providecommand \@url [1]{\endgroup\@href {#1}{\urlprefix }}%
\providecommand \urlprefix  [0]{URL }%
\providecommand \Eprint [0]{\href }%
\providecommand \doibase [0]{http://dx.doi.org/}%
\providecommand \selectlanguage [0]{\@gobble}%
\providecommand \bibinfo  [0]{\@secondoftwo}%
\providecommand \bibfield  [0]{\@secondoftwo}%
\providecommand \translation [1]{[#1]}%
\providecommand \BibitemOpen [0]{}%
\providecommand \bibitemStop [0]{}%
\providecommand \bibitemNoStop [0]{.\EOS\space}%
\providecommand \EOS [0]{\spacefactor3000\relax}%
\providecommand \BibitemShut  [1]{\csname bibitem#1\endcsname}%
\let\auto@bib@innerbib\@empty
\bibitem [{\citenamefont {Bloch}(2005)}]{r47}%
  \BibitemOpen
  \bibfield  {author} {\bibinfo {author} {\bibfnamefont {I.}~\bibnamefont
  {Bloch}},\ }\href {https://www.nature.com/articles/nphys138\#citeas}
  {\bibfield  {journal} {\bibinfo  {journal}
  {\href{https://www.nature.com/articles/nphys138\#citeas}{Nature Physics}}\
  }\textbf {\bibinfo {volume} {1}},\ \bibinfo {pages} {23} (\bibinfo {year}
  {2005})}\BibitemShut {NoStop}%
\bibitem [{\citenamefont {Jaksch}\ and\ \citenamefont {Zoller}(2005)}]{r49}%
  \BibitemOpen
  \bibfield  {author} {\bibinfo {author} {\bibfnamefont {D.}~\bibnamefont
  {Jaksch}}\ and\ \bibinfo {author} {\bibfnamefont {P.}~\bibnamefont
  {Zoller}},\ }\href {\doibase 10.1016/j.aop.2004.09.010} {\bibfield  {journal}
  {\bibinfo  {journal}
  {\href{https://www.sciencedirect.com/science/article/pii/S0003491604001782}{Annals
  of Physics}}\ }\textbf {\bibinfo {volume} {315}},\ \bibinfo {pages} {52}
  (\bibinfo {year} {2005})}\BibitemShut {NoStop}%
\bibitem [{\citenamefont {Bloch}\ \emph {et~al.}(2008)\citenamefont {Bloch},
  \citenamefont {Dalibard},\ and\ \citenamefont {Zwerger}}]{r16}%
  \BibitemOpen
  \bibfield  {author} {\bibinfo {author} {\bibfnamefont {I.}~\bibnamefont
  {Bloch}}, \bibinfo {author} {\bibfnamefont {J.}~\bibnamefont {Dalibard}}, \
  and\ \bibinfo {author} {\bibfnamefont {W.}~\bibnamefont {Zwerger}},\ }\href
  {\doibase 10.1103/RevModPhys.80.885} {\bibfield  {journal} {\bibinfo
  {journal}
  {\href{https://journals.aps.org/rmp/abstract/10.1103/RevModPhys.80.885}{Reviews
  of Modern Physics}}\ }\textbf {\bibinfo {volume} {80}},\ \bibinfo {pages}
  {885} (\bibinfo {year} {2008})}\BibitemShut {NoStop}%
\bibitem [{\citenamefont {Giorgini}\ \emph {et~al.}(2008)\citenamefont
  {Giorgini}, \citenamefont {Pitaevskii},\ and\ \citenamefont
  {Stringari}}]{r90}%
  \BibitemOpen
  \bibfield  {author} {\bibinfo {author} {\bibfnamefont {S.}~\bibnamefont
  {Giorgini}}, \bibinfo {author} {\bibfnamefont {L.~P.}\ \bibnamefont
  {Pitaevskii}}, \ and\ \bibinfo {author} {\bibfnamefont {S.}~\bibnamefont
  {Stringari}},\ }\href {\doibase 10.1103/RevModPhys.80.1215} {\bibfield
  {journal} {\bibinfo  {journal} {Reviews of Modern Physics}\ }\textbf
  {\bibinfo {volume} {80}},\ \bibinfo {pages} {1215} (\bibinfo {year}
  {2008})}\BibitemShut {NoStop}%
\bibitem [{\citenamefont {Zhang}\ \emph {et~al.}(2019)\citenamefont {Zhang},
  \citenamefont {Zhu}, \citenamefont {Zhao}, \citenamefont {Yan},\ and\
  \citenamefont {Zhu}}]{r30}%
  \BibitemOpen
  \bibfield  {author} {\bibinfo {author} {\bibfnamefont {D.-W.}\ \bibnamefont
  {Zhang}}, \bibinfo {author} {\bibfnamefont {Y.-Q.}\ \bibnamefont {Zhu}},
  \bibinfo {author} {\bibfnamefont {Y.~X.}\ \bibnamefont {Zhao}}, \bibinfo
  {author} {\bibfnamefont {H.}~\bibnamefont {Yan}}, \ and\ \bibinfo {author}
  {\bibfnamefont {S.-L.}\ \bibnamefont {Zhu}},\ }\href {\doibase
  10.1080/00018732.2019.1594094} {\bibfield  {journal} {\bibinfo  {journal}
  {\href{https://www.tandfonline.com/doi/full/10.1080/00018732.2019.1594094}{Advances
  in Physics}}\ }\textbf {\bibinfo {volume} {67}},\ \bibinfo {pages} {253}
  (\bibinfo {year} {2019})}\BibitemShut {NoStop}%
\bibitem [{\citenamefont {Atala}\ \emph {et~al.}(2013)\citenamefont {Atala},
  \citenamefont {Aidelsburger}, \citenamefont {Barreiro}, \citenamefont
  {Abanin}, \citenamefont {Kitagawa}, \citenamefont {Demler},\ and\
  \citenamefont {Bloch}}]{r27}%
  \BibitemOpen
  \bibfield  {author} {\bibinfo {author} {\bibfnamefont {M.}~\bibnamefont
  {Atala}}, \bibinfo {author} {\bibfnamefont {M.}~\bibnamefont {Aidelsburger}},
  \bibinfo {author} {\bibfnamefont {J.~T.}\ \bibnamefont {Barreiro}}, \bibinfo
  {author} {\bibfnamefont {D.}~\bibnamefont {Abanin}}, \bibinfo {author}
  {\bibfnamefont {T.}~\bibnamefont {Kitagawa}}, \bibinfo {author}
  {\bibfnamefont {E.}~\bibnamefont {Demler}}, \ and\ \bibinfo {author}
  {\bibfnamefont {I.}~\bibnamefont {Bloch}},\ }\href {\doibase
  10.1038/nphys2790} {\bibfield  {journal} {\bibinfo  {journal}
  {\href{https://www.nature.com/articles/nphys2790}{Nature Physics}}\ }\textbf
  {\bibinfo {volume} {9}},\ \bibinfo {pages} {795} (\bibinfo {year}
  {2013})}\BibitemShut {NoStop}%
\bibitem [{\citenamefont {Jotzu}\ \emph {et~al.}(2014)\citenamefont {Jotzu},
  \citenamefont {Messer}, \citenamefont {Desbuquois}, \citenamefont {Lebrat},
  \citenamefont {Uehlinger}, \citenamefont {Greif},\ and\ \citenamefont
  {Esslinger}}]{r28}%
  \BibitemOpen
  \bibfield  {author} {\bibinfo {author} {\bibfnamefont {G.}~\bibnamefont
  {Jotzu}}, \bibinfo {author} {\bibfnamefont {M.}~\bibnamefont {Messer}},
  \bibinfo {author} {\bibfnamefont {R.}~\bibnamefont {Desbuquois}}, \bibinfo
  {author} {\bibfnamefont {M.}~\bibnamefont {Lebrat}}, \bibinfo {author}
  {\bibfnamefont {T.}~\bibnamefont {Uehlinger}}, \bibinfo {author}
  {\bibfnamefont {D.}~\bibnamefont {Greif}}, \ and\ \bibinfo {author}
  {\bibfnamefont {T.}~\bibnamefont {Esslinger}},\ }\href {\doibase
  10.1038/nature13915} {\bibfield  {journal} {\bibinfo  {journal}
  {\href{https://www.nature.com/articles/nature13915}{Nature}}\ }\textbf
  {\bibinfo {volume} {515}},\ \bibinfo {pages} {237} (\bibinfo {year}
  {2014})}\BibitemShut {NoStop}%
\bibitem [{\citenamefont {Goldman}\ \emph {et~al.}(2013)\citenamefont
  {Goldman}, \citenamefont {Dalibard}, \citenamefont {Dauphin}, \citenamefont
  {Gerbier}, \citenamefont {Lewenstein}, \citenamefont {Zoller},\ and\
  \citenamefont {Spielman}}]{r29}%
  \BibitemOpen
  \bibfield  {author} {\bibinfo {author} {\bibfnamefont {N.}~\bibnamefont
  {Goldman}}, \bibinfo {author} {\bibfnamefont {J.}~\bibnamefont {Dalibard}},
  \bibinfo {author} {\bibfnamefont {A.}~\bibnamefont {Dauphin}}, \bibinfo
  {author} {\bibfnamefont {F.}~\bibnamefont {Gerbier}}, \bibinfo {author}
  {\bibfnamefont {M.}~\bibnamefont {Lewenstein}}, \bibinfo {author}
  {\bibfnamefont {P.}~\bibnamefont {Zoller}}, \ and\ \bibinfo {author}
  {\bibfnamefont {I.~B.}\ \bibnamefont {Spielman}},\ }\href {\doibase
  10.1073/pnas.1300170110} {\bibfield  {journal} {\bibinfo  {journal}
  {\href{https://www.pnas.org/content/110/17/6736.short}{Proc. Natl. Acad. Sci.
  USA}}\ }\textbf {\bibinfo {volume} {110}},\ \bibinfo {pages} {6736} (\bibinfo
  {year} {2013})}\BibitemShut {NoStop}%
\bibitem [{\citenamefont {Sherson}\ \emph {et~al.}(2010)\citenamefont
  {Sherson}, \citenamefont {Weitenberg}, \citenamefont {Endres}, \citenamefont
  {Cheneau}, \citenamefont {Bloch},\ and\ \citenamefont {Kuhr}}]{r31}%
  \BibitemOpen
  \bibfield  {author} {\bibinfo {author} {\bibfnamefont {J.~F.}\ \bibnamefont
  {Sherson}}, \bibinfo {author} {\bibfnamefont {C.}~\bibnamefont {Weitenberg}},
  \bibinfo {author} {\bibfnamefont {M.}~\bibnamefont {Endres}}, \bibinfo
  {author} {\bibfnamefont {M.}~\bibnamefont {Cheneau}}, \bibinfo {author}
  {\bibfnamefont {I.}~\bibnamefont {Bloch}}, \ and\ \bibinfo {author}
  {\bibfnamefont {S.}~\bibnamefont {Kuhr}},\ }\href {\doibase
  10.1038/nature09378} {\bibfield  {journal} {\bibinfo  {journal}
  {\href{https://www.nature.com/articles/nature09378}{Nature}}\ }\textbf
  {\bibinfo {volume} {467}},\ \bibinfo {pages} {68} (\bibinfo {year}
  {2010})}\BibitemShut {NoStop}%
\bibitem [{\citenamefont {Tai}\ \emph {et~al.}(2017)\citenamefont {Tai},
  \citenamefont {Lukin}, \citenamefont {Rispoli}, \citenamefont {Schittko},
  \citenamefont {Menke}, \citenamefont {Dan}, \citenamefont {Preiss},
  \citenamefont {Grusdt}, \citenamefont {Kaufman},\ and\ \citenamefont
  {Greiner}}]{r32}%
  \BibitemOpen
  \bibfield  {author} {\bibinfo {author} {\bibfnamefont {M.~E.}\ \bibnamefont
  {Tai}}, \bibinfo {author} {\bibfnamefont {A.}~\bibnamefont {Lukin}}, \bibinfo
  {author} {\bibfnamefont {M.}~\bibnamefont {Rispoli}}, \bibinfo {author}
  {\bibfnamefont {R.}~\bibnamefont {Schittko}}, \bibinfo {author}
  {\bibfnamefont {T.}~\bibnamefont {Menke}}, \bibinfo {author} {\bibfnamefont
  {B.}~\bibnamefont {Dan}}, \bibinfo {author} {\bibfnamefont {P.~M.}\
  \bibnamefont {Preiss}}, \bibinfo {author} {\bibfnamefont {F.}~\bibnamefont
  {Grusdt}}, \bibinfo {author} {\bibfnamefont {A.~M.}\ \bibnamefont {Kaufman}},
  \ and\ \bibinfo {author} {\bibfnamefont {M.}~\bibnamefont {Greiner}},\ }\href
  {\doibase 10.1038/nature22811} {\bibfield  {journal} {\bibinfo  {journal}
  {\href{https://www.nature.com/articles/nature22811/}{Nature}}\ }\textbf
  {\bibinfo {volume} {546}},\ \bibinfo {pages} {519} (\bibinfo {year}
  {2017})}\BibitemShut {NoStop}%
\bibitem [{\citenamefont {Aidelsburger}\ \emph {et~al.}(2013)\citenamefont
  {Aidelsburger}, \citenamefont {Atala}, \citenamefont {Lohse}, \citenamefont
  {Barreiro}, \citenamefont {Paredes},\ and\ \citenamefont {Bloch}}]{r33}%
  \BibitemOpen
  \bibfield  {author} {\bibinfo {author} {\bibfnamefont {M.}~\bibnamefont
  {Aidelsburger}}, \bibinfo {author} {\bibfnamefont {M.}~\bibnamefont {Atala}},
  \bibinfo {author} {\bibfnamefont {M.}~\bibnamefont {Lohse}}, \bibinfo
  {author} {\bibfnamefont {J.~T.}\ \bibnamefont {Barreiro}}, \bibinfo {author}
  {\bibfnamefont {B.}~\bibnamefont {Paredes}}, \ and\ \bibinfo {author}
  {\bibfnamefont {I.}~\bibnamefont {Bloch}},\ }\href {\doibase
  10.1103/PhysRevLett.111.185301} {\bibfield  {journal} {\bibinfo  {journal}
  {\href{https://journals.aps.org/prl/abstract/10.1103/PhysRevLett.111.185301}{Phys.
  Rev. Lett.}}\ }\textbf {\bibinfo {volume} {111}},\ \bibinfo {pages} {185301}
  (\bibinfo {year} {2013})}\BibitemShut {NoStop}%
\bibitem [{\citenamefont {Lohse}\ \emph {et~al.}(2018)\citenamefont {Lohse},
  \citenamefont {Schweizer}, \citenamefont {Price}, \citenamefont
  {Zilberberg},\ and\ \citenamefont {Bloch}}]{r35}%
  \BibitemOpen
  \bibfield  {author} {\bibinfo {author} {\bibfnamefont {M.}~\bibnamefont
  {Lohse}}, \bibinfo {author} {\bibfnamefont {C.}~\bibnamefont {Schweizer}},
  \bibinfo {author} {\bibfnamefont {H.~M.}\ \bibnamefont {Price}}, \bibinfo
  {author} {\bibfnamefont {O.}~\bibnamefont {Zilberberg}}, \ and\ \bibinfo
  {author} {\bibfnamefont {I.}~\bibnamefont {Bloch}},\ }\href {\doibase
  10.1038/nature25000} {\bibfield  {journal} {\bibinfo  {journal}
  {\href{https://www.nature.com/articles/nature25000}{Nature}}\ }\textbf
  {\bibinfo {volume} {553}},\ \bibinfo {pages} {55} (\bibinfo {year}
  {2018})}\BibitemShut {NoStop}%
\bibitem [{\citenamefont {Lohse}\ \emph {et~al.}(2015)\citenamefont {Lohse},
  \citenamefont {Schweizer}, \citenamefont {Zilberberg}, \citenamefont
  {Aidelsburger},\ and\ \citenamefont {Bloch}}]{r36}%
  \BibitemOpen
  \bibfield  {author} {\bibinfo {author} {\bibfnamefont {M.}~\bibnamefont
  {Lohse}}, \bibinfo {author} {\bibfnamefont {C.}~\bibnamefont {Schweizer}},
  \bibinfo {author} {\bibfnamefont {O.}~\bibnamefont {Zilberberg}}, \bibinfo
  {author} {\bibfnamefont {M.}~\bibnamefont {Aidelsburger}}, \ and\ \bibinfo
  {author} {\bibfnamefont {I.}~\bibnamefont {Bloch}},\ }\href {\doibase
  10.1038/nphys3584} {\bibfield  {journal} {\bibinfo  {journal}
  {\href{https://www.nature.com/articles/nphys3584}{Nature Physics}}\ }\textbf
  {\bibinfo {volume} {12}},\ \bibinfo {pages} {350} (\bibinfo {year}
  {2015})}\BibitemShut {NoStop}%
\bibitem [{\citenamefont {Mancini}\ \emph {et~al.}(2015)\citenamefont
  {Mancini}, \citenamefont {Pagano}, \citenamefont {Cappellini}, \citenamefont
  {Livi}, \citenamefont {Rider}, \citenamefont {Catani}, \citenamefont {Sias},
  \citenamefont {Zoller}, \citenamefont {Inguscio}, \citenamefont {Dalmonte},\
  and\ \citenamefont {Fallani}}]{r37}%
  \BibitemOpen
  \bibfield  {author} {\bibinfo {author} {\bibfnamefont {M.}~\bibnamefont
  {Mancini}}, \bibinfo {author} {\bibfnamefont {G.}~\bibnamefont {Pagano}},
  \bibinfo {author} {\bibfnamefont {G.}~\bibnamefont {Cappellini}}, \bibinfo
  {author} {\bibfnamefont {L.}~\bibnamefont {Livi}}, \bibinfo {author}
  {\bibfnamefont {M.}~\bibnamefont {Rider}}, \bibinfo {author} {\bibfnamefont
  {J.}~\bibnamefont {Catani}}, \bibinfo {author} {\bibfnamefont
  {C.}~\bibnamefont {Sias}}, \bibinfo {author} {\bibfnamefont {P.}~\bibnamefont
  {Zoller}}, \bibinfo {author} {\bibfnamefont {M.}~\bibnamefont {Inguscio}},
  \bibinfo {author} {\bibfnamefont {M.}~\bibnamefont {Dalmonte}}, \ and\
  \bibinfo {author} {\bibfnamefont {L.}~\bibnamefont {Fallani}},\ }\href
  {\doibase 10.1126/science.aaa8736} {\bibfield  {journal} {\bibinfo  {journal}
  {\href{https://science.sciencemag.org/content/349/6255/1510.full}{Science}}\
  }\textbf {\bibinfo {volume} {349}},\ \bibinfo {pages} {1510} (\bibinfo {year}
  {2015})}\BibitemShut {NoStop}%
\bibitem [{\citenamefont {Nakajima}\ \emph {et~al.}(2016)\citenamefont
  {Nakajima}, \citenamefont {Tomita}, \citenamefont {Taie}, \citenamefont
  {Ichinose}, \citenamefont {Ozawa}, \citenamefont {Wang}, \citenamefont
  {Troyer},\ and\ \citenamefont {Takahashi}}]{r38}%
  \BibitemOpen
  \bibfield  {author} {\bibinfo {author} {\bibfnamefont {S.}~\bibnamefont
  {Nakajima}}, \bibinfo {author} {\bibfnamefont {T.}~\bibnamefont {Tomita}},
  \bibinfo {author} {\bibfnamefont {S.}~\bibnamefont {Taie}}, \bibinfo {author}
  {\bibfnamefont {T.}~\bibnamefont {Ichinose}}, \bibinfo {author}
  {\bibfnamefont {H.}~\bibnamefont {Ozawa}}, \bibinfo {author} {\bibfnamefont
  {L.}~\bibnamefont {Wang}}, \bibinfo {author} {\bibfnamefont {M.}~\bibnamefont
  {Troyer}}, \ and\ \bibinfo {author} {\bibfnamefont {Y.}~\bibnamefont
  {Takahashi}},\ }\href {\doibase 10.1038/nphys3622} {\bibfield  {journal}
  {\bibinfo  {journal} {\href{https://www.nature.com/articles/nphys3622}{Nature
  Physics}}\ }\textbf {\bibinfo {volume} {12}},\ \bibinfo {pages} {296}
  (\bibinfo {year} {2016})}\BibitemShut {NoStop}%
\bibitem [{\citenamefont {Sun}\ \emph {et~al.}(2018)\citenamefont {Sun},
  \citenamefont {Wang}, \citenamefont {Xu}, \citenamefont {Yi}, \citenamefont
  {Zhang}, \citenamefont {Wu}, \citenamefont {Deng}, \citenamefont {Liu},
  \citenamefont {Chen},\ and\ \citenamefont {Pan}}]{r41}%
  \BibitemOpen
  \bibfield  {author} {\bibinfo {author} {\bibfnamefont {W.}~\bibnamefont
  {Sun}}, \bibinfo {author} {\bibfnamefont {B.~Z.}\ \bibnamefont {Wang}},
  \bibinfo {author} {\bibfnamefont {X.~T.}\ \bibnamefont {Xu}}, \bibinfo
  {author} {\bibfnamefont {C.~R.}\ \bibnamefont {Yi}}, \bibinfo {author}
  {\bibfnamefont {L.}~\bibnamefont {Zhang}}, \bibinfo {author} {\bibfnamefont
  {Z.}~\bibnamefont {Wu}}, \bibinfo {author} {\bibfnamefont {Y.}~\bibnamefont
  {Deng}}, \bibinfo {author} {\bibfnamefont {X.~J.}\ \bibnamefont {Liu}},
  \bibinfo {author} {\bibfnamefont {S.}~\bibnamefont {Chen}}, \ and\ \bibinfo
  {author} {\bibfnamefont {J.~W.}\ \bibnamefont {Pan}},\ }\href {\doibase
  10.1103/PhysRevLett.121.150401} {\bibfield  {journal} {\bibinfo  {journal}
  {\href{https://journals.aps.org/prl/abstract/10.1103/PhysRevLett.121.150401}{Phys.
  Rev. Lett.}}\ }\textbf {\bibinfo {volume} {121}},\ \bibinfo {pages} {150401}
  (\bibinfo {year} {2018})}\BibitemShut {NoStop}%
\bibitem [{\citenamefont {Wu}\ \emph {et~al.}(2016)\citenamefont {Wu},
  \citenamefont {Zhang}, \citenamefont {Sun}, \citenamefont {Xu}, \citenamefont
  {Wang}, \citenamefont {Ji}, \citenamefont {Deng}, \citenamefont {Chen},
  \citenamefont {Liu},\ and\ \citenamefont {Pan}}]{r42}%
  \BibitemOpen
  \bibfield  {author} {\bibinfo {author} {\bibfnamefont {Z.}~\bibnamefont
  {Wu}}, \bibinfo {author} {\bibfnamefont {L.}~\bibnamefont {Zhang}}, \bibinfo
  {author} {\bibfnamefont {W.}~\bibnamefont {Sun}}, \bibinfo {author}
  {\bibfnamefont {X.~T.}\ \bibnamefont {Xu}}, \bibinfo {author} {\bibfnamefont
  {B.~Z.}\ \bibnamefont {Wang}}, \bibinfo {author} {\bibfnamefont {S.~C.}\
  \bibnamefont {Ji}}, \bibinfo {author} {\bibfnamefont {Y.}~\bibnamefont
  {Deng}}, \bibinfo {author} {\bibfnamefont {S.}~\bibnamefont {Chen}}, \bibinfo
  {author} {\bibfnamefont {X.~J.}\ \bibnamefont {Liu}}, \ and\ \bibinfo
  {author} {\bibfnamefont {J.~W.}\ \bibnamefont {Pan}},\ }\href {\doibase
  10.1126/science.aaf6689} {\bibfield  {journal} {\bibinfo  {journal}
  {\href{https://science.sciencemag.org/content/354/6308/83.full}{Science}}\
  }\textbf {\bibinfo {volume} {354}},\ \bibinfo {pages} {83} (\bibinfo {year}
  {2016})}\BibitemShut {NoStop}%
\bibitem [{\citenamefont {Kim}\ \emph {et~al.}(2018)\citenamefont {Kim},
  \citenamefont {Zhu}, \citenamefont {Porto},\ and\ \citenamefont
  {Hafezi}}]{r51}%
  \BibitemOpen
  \bibfield  {author} {\bibinfo {author} {\bibfnamefont {H.}~\bibnamefont
  {Kim}}, \bibinfo {author} {\bibfnamefont {G.}~\bibnamefont {Zhu}}, \bibinfo
  {author} {\bibfnamefont {J.~V.}\ \bibnamefont {Porto}}, \ and\ \bibinfo
  {author} {\bibfnamefont {M.}~\bibnamefont {Hafezi}},\ }\href {\doibase
  10.1103/PhysRevLett.121.133002} {\bibfield  {journal} {\bibinfo  {journal}
  {\href{https://journals.aps.org/prl/abstract/10.1103/PhysRevLett.121.133002}{Phys.
  Rev. Lett.}}\ }\textbf {\bibinfo {volume} {121}},\ \bibinfo {pages} {133002}
  (\bibinfo {year} {2018})}\BibitemShut {NoStop}%
\bibitem [{\citenamefont {Chen}\ \emph {et~al.}(2018)\citenamefont {Chen},
  \citenamefont {Wang}, \citenamefont {Meng}, \citenamefont {Huang},
  \citenamefont {Cai}, \citenamefont {Wang}, \citenamefont {Zhu},\ and\
  \citenamefont {Zhang}}]{r52}%
  \BibitemOpen
  \bibfield  {author} {\bibinfo {author} {\bibfnamefont {L.}~\bibnamefont
  {Chen}}, \bibinfo {author} {\bibfnamefont {P.}~\bibnamefont {Wang}}, \bibinfo
  {author} {\bibfnamefont {Z.}~\bibnamefont {Meng}}, \bibinfo {author}
  {\bibfnamefont {L.}~\bibnamefont {Huang}}, \bibinfo {author} {\bibfnamefont
  {H.}~\bibnamefont {Cai}}, \bibinfo {author} {\bibfnamefont {D.~W.}\
  \bibnamefont {Wang}}, \bibinfo {author} {\bibfnamefont {S.~Y.}\ \bibnamefont
  {Zhu}}, \ and\ \bibinfo {author} {\bibfnamefont {J.}~\bibnamefont {Zhang}},\
  }\href {\doibase 10.1103/PhysRevLett.120.193601} {\bibfield  {journal}
  {\bibinfo  {journal}
  {\href{https://journals.aps.org/prl/abstract/10.1103/PhysRevLett.120.193601}{Phys.
  Rev. Lett.}}\ }\textbf {\bibinfo {volume} {120}},\ \bibinfo {pages} {193601}
  (\bibinfo {year} {2018})}\BibitemShut {NoStop}%
\bibitem [{\citenamefont {Liu}\ \emph {et~al.}(2019)\citenamefont {Liu},
  \citenamefont {Zhang}, \citenamefont {Ai}, \citenamefont {Gong},
  \citenamefont {Kawabata}, \citenamefont {Ueda},\ and\ \citenamefont
  {Nori}}]{r127}%
  \BibitemOpen
  \bibfield  {author} {\bibinfo {author} {\bibfnamefont {T.}~\bibnamefont
  {Liu}}, \bibinfo {author} {\bibfnamefont {Y.~R.}\ \bibnamefont {Zhang}},
  \bibinfo {author} {\bibfnamefont {Q.}~\bibnamefont {Ai}}, \bibinfo {author}
  {\bibfnamefont {Z.}~\bibnamefont {Gong}}, \bibinfo {author} {\bibfnamefont
  {K.}~\bibnamefont {Kawabata}}, \bibinfo {author} {\bibfnamefont
  {M.}~\bibnamefont {Ueda}}, \ and\ \bibinfo {author} {\bibfnamefont
  {F.}~\bibnamefont {Nori}},\ }\href {\doibase 10.1103/PhysRevLett.122.076801}
  {\bibfield  {journal} {\bibinfo  {journal} {Phys. Rev. Lett.}\ }\textbf
  {\bibinfo {volume} {122}},\ \bibinfo {pages} {076801} (\bibinfo {year}
  {2019})}\BibitemShut {NoStop}%
\bibitem [{\citenamefont {Sun}\ \emph {et~al.}(2011)\citenamefont {Sun},
  \citenamefont {Liu}, \citenamefont {Hemmerich},\ and\ \citenamefont
  {Das~Sarma}}]{r144}%
  \BibitemOpen
  \bibfield  {author} {\bibinfo {author} {\bibfnamefont {K.}~\bibnamefont
  {Sun}}, \bibinfo {author} {\bibfnamefont {W.~V.}\ \bibnamefont {Liu}},
  \bibinfo {author} {\bibfnamefont {A.}~\bibnamefont {Hemmerich}}, \ and\
  \bibinfo {author} {\bibfnamefont {S.}~\bibnamefont {Das~Sarma}},\ }\href
  {\doibase 10.1038/nphys2134} {\bibfield  {journal} {\bibinfo  {journal}
  {Nature Physics}\ }\textbf {\bibinfo {volume} {8}},\ \bibinfo {pages} {67}
  (\bibinfo {year} {2011})}\BibitemShut {NoStop}%
\bibitem [{\citenamefont {Li}\ \emph {et~al.}(2013)\citenamefont {Li},
  \citenamefont {Zhao},\ and\ \citenamefont {Liu}}]{r145}%
  \BibitemOpen
  \bibfield  {author} {\bibinfo {author} {\bibfnamefont {X.}~\bibnamefont
  {Li}}, \bibinfo {author} {\bibfnamefont {E.}~\bibnamefont {Zhao}}, \ and\
  \bibinfo {author} {\bibfnamefont {W.~V.}\ \bibnamefont {Liu}},\ }\href
  {\doibase 10.1038/ncomms2523} {\bibfield  {journal} {\bibinfo  {journal} {Nat
  Commun}\ }\textbf {\bibinfo {volume} {4}},\ \bibinfo {pages} {1523} (\bibinfo
  {year} {2013})}\BibitemShut {NoStop}%
\bibitem [{\citenamefont {Huang}\ \emph {et~al.}(2018)\citenamefont {Huang},
  \citenamefont {Wu},\ and\ \citenamefont {Liu}}]{r146}%
  \BibitemOpen
  \bibfield  {author} {\bibinfo {author} {\bibfnamefont {B.}~\bibnamefont
  {Huang}}, \bibinfo {author} {\bibfnamefont {Y.~H.}\ \bibnamefont {Wu}}, \
  and\ \bibinfo {author} {\bibfnamefont {W.~V.}\ \bibnamefont {Liu}},\ }\href
  {\doibase 10.1103/PhysRevLett.120.110603} {\bibfield  {journal} {\bibinfo
  {journal} {Phys Rev Lett}\ }\textbf {\bibinfo {volume} {120}},\ \bibinfo
  {pages} {110603} (\bibinfo {year} {2018})}\BibitemShut {NoStop}%
\bibitem [{\citenamefont {Huang}\ and\ \citenamefont {Liu}(2020)}]{r147}%
  \BibitemOpen
  \bibfield  {author} {\bibinfo {author} {\bibfnamefont {B.}~\bibnamefont
  {Huang}}\ and\ \bibinfo {author} {\bibfnamefont {W.~V.}\ \bibnamefont
  {Liu}},\ }\href {\doibase 10.1103/PhysRevLett.124.216601} {\bibfield
  {journal} {\bibinfo  {journal} {Phys Rev Lett}\ }\textbf {\bibinfo {volume}
  {124}},\ \bibinfo {pages} {216601} (\bibinfo {year} {2020})}\BibitemShut
  {NoStop}%
\bibitem [{\citenamefont {Hart}\ \emph {et~al.}(2015)\citenamefont {Hart},
  \citenamefont {Duarte}, \citenamefont {Yang}, \citenamefont {Liu},
  \citenamefont {Paiva}, \citenamefont {Khatami}, \citenamefont {Scalettar},
  \citenamefont {Trivedi}, \citenamefont {Huse},\ and\ \citenamefont
  {Hulet}}]{r73}%
  \BibitemOpen
  \bibfield  {author} {\bibinfo {author} {\bibfnamefont {R.~A.}\ \bibnamefont
  {Hart}}, \bibinfo {author} {\bibfnamefont {P.~M.}\ \bibnamefont {Duarte}},
  \bibinfo {author} {\bibfnamefont {T.~L.}\ \bibnamefont {Yang}}, \bibinfo
  {author} {\bibfnamefont {X.}~\bibnamefont {Liu}}, \bibinfo {author}
  {\bibfnamefont {T.}~\bibnamefont {Paiva}}, \bibinfo {author} {\bibfnamefont
  {E.}~\bibnamefont {Khatami}}, \bibinfo {author} {\bibfnamefont {R.~T.}\
  \bibnamefont {Scalettar}}, \bibinfo {author} {\bibfnamefont {N.}~\bibnamefont
  {Trivedi}}, \bibinfo {author} {\bibfnamefont {D.~A.}\ \bibnamefont {Huse}}, \
  and\ \bibinfo {author} {\bibfnamefont {R.~G.}\ \bibnamefont {Hulet}},\ }\href
  {\doibase 10.1038/nature14223} {\bibfield  {journal} {\bibinfo  {journal}
  {Nature}\ }\textbf {\bibinfo {volume} {519}},\ \bibinfo {pages} {211}
  (\bibinfo {year} {2015})}\BibitemShut {NoStop}%
\bibitem [{\citenamefont {Brown}\ \emph {et~al.}(2017)\citenamefont {Brown},
  \citenamefont {Mitra}, \citenamefont {Guardado-Sanchez}, \citenamefont
  {Schau{\ss}}, \citenamefont {Kondov}, \citenamefont {Khatami}, \citenamefont
  {Paiva}, \citenamefont {Trivedi}, \citenamefont {Huse},\ and\ \citenamefont
  {Bakr}}]{r74}%
  \BibitemOpen
  \bibfield  {author} {\bibinfo {author} {\bibfnamefont {P.~T.}\ \bibnamefont
  {Brown}}, \bibinfo {author} {\bibfnamefont {D.}~\bibnamefont {Mitra}},
  \bibinfo {author} {\bibfnamefont {E.}~\bibnamefont {Guardado-Sanchez}},
  \bibinfo {author} {\bibfnamefont {P.}~\bibnamefont {Schau{\ss}}}, \bibinfo
  {author} {\bibfnamefont {S.~S.}\ \bibnamefont {Kondov}}, \bibinfo {author}
  {\bibfnamefont {E.}~\bibnamefont {Khatami}}, \bibinfo {author} {\bibfnamefont
  {T.}~\bibnamefont {Paiva}}, \bibinfo {author} {\bibfnamefont
  {N.}~\bibnamefont {Trivedi}}, \bibinfo {author} {\bibfnamefont {D.~A.}\
  \bibnamefont {Huse}}, \ and\ \bibinfo {author} {\bibfnamefont {W.~S.}\
  \bibnamefont {Bakr}},\ }\href
  {https://science.sciencemag.org/content/357/6358/1385.full} {\bibfield
  {journal} {\bibinfo  {journal} {Science}\ }\textbf {\bibinfo {volume}
  {357}},\ \bibinfo {pages} {1385} (\bibinfo {year} {2017})}\BibitemShut
  {NoStop}%
\bibitem [{\citenamefont {Mazurenko}\ \emph {et~al.}(2017)\citenamefont
  {Mazurenko}, \citenamefont {Chiu}, \citenamefont {Ji}, \citenamefont
  {Parsons}, \citenamefont {Kanasz-Nagy}, \citenamefont {Schmidt},
  \citenamefont {Grusdt}, \citenamefont {Demler}, \citenamefont {Greif},\ and\
  \citenamefont {Greiner}}]{r75}%
  \BibitemOpen
  \bibfield  {author} {\bibinfo {author} {\bibfnamefont {A.}~\bibnamefont
  {Mazurenko}}, \bibinfo {author} {\bibfnamefont {C.~S.}\ \bibnamefont {Chiu}},
  \bibinfo {author} {\bibfnamefont {G.}~\bibnamefont {Ji}}, \bibinfo {author}
  {\bibfnamefont {M.~F.}\ \bibnamefont {Parsons}}, \bibinfo {author}
  {\bibfnamefont {M.}~\bibnamefont {Kanasz-Nagy}}, \bibinfo {author}
  {\bibfnamefont {R.}~\bibnamefont {Schmidt}}, \bibinfo {author} {\bibfnamefont
  {F.}~\bibnamefont {Grusdt}}, \bibinfo {author} {\bibfnamefont
  {E.}~\bibnamefont {Demler}}, \bibinfo {author} {\bibfnamefont
  {D.}~\bibnamefont {Greif}}, \ and\ \bibinfo {author} {\bibfnamefont
  {M.}~\bibnamefont {Greiner}},\ }\href {\doibase 10.1038/nature22362}
  {\bibfield  {journal} {\bibinfo  {journal} {Nature}\ }\textbf {\bibinfo
  {volume} {545}},\ \bibinfo {pages} {462} (\bibinfo {year}
  {2017})}\BibitemShut {NoStop}%
\bibitem [{\citenamefont {Mitra}\ \emph {et~al.}(2017)\citenamefont {Mitra},
  \citenamefont {Brown}, \citenamefont {Guardado-Sanchez}, \citenamefont
  {Kondov}, \citenamefont {Devakul}, \citenamefont {Huse}, \citenamefont
  {Schau{\ss}},\ and\ \citenamefont {Bakr}}]{r76}%
  \BibitemOpen
  \bibfield  {author} {\bibinfo {author} {\bibfnamefont {D.}~\bibnamefont
  {Mitra}}, \bibinfo {author} {\bibfnamefont {P.~T.}\ \bibnamefont {Brown}},
  \bibinfo {author} {\bibfnamefont {E.}~\bibnamefont {Guardado-Sanchez}},
  \bibinfo {author} {\bibfnamefont {S.~S.}\ \bibnamefont {Kondov}}, \bibinfo
  {author} {\bibfnamefont {T.}~\bibnamefont {Devakul}}, \bibinfo {author}
  {\bibfnamefont {D.~A.}\ \bibnamefont {Huse}}, \bibinfo {author}
  {\bibfnamefont {P.}~\bibnamefont {Schau{\ss}}}, \ and\ \bibinfo {author}
  {\bibfnamefont {W.~S.}\ \bibnamefont {Bakr}},\ }\href {\doibase
  10.1038/nphys4297} {\bibfield  {journal} {\bibinfo  {journal} {Nature
  Physics}\ }\textbf {\bibinfo {volume} {14}},\ \bibinfo {pages} {173}
  (\bibinfo {year} {2017})}\BibitemShut {NoStop}%
\bibitem [{\citenamefont {Diehl}\ \emph {et~al.}(2008)\citenamefont {Diehl},
  \citenamefont {Micheli}, \citenamefont {Kantian}, \citenamefont {Kraus},
  \citenamefont {B{\"u}chler},\ and\ \citenamefont {Zoller}}]{r22}%
  \BibitemOpen
  \bibfield  {author} {\bibinfo {author} {\bibfnamefont {S.}~\bibnamefont
  {Diehl}}, \bibinfo {author} {\bibfnamefont {A.}~\bibnamefont {Micheli}},
  \bibinfo {author} {\bibfnamefont {A.}~\bibnamefont {Kantian}}, \bibinfo
  {author} {\bibfnamefont {B.}~\bibnamefont {Kraus}}, \bibinfo {author}
  {\bibfnamefont {H.~P.}\ \bibnamefont {B{\"u}chler}}, \ and\ \bibinfo {author}
  {\bibfnamefont {P.}~\bibnamefont {Zoller}},\ }\href {\doibase
  10.1038/nphys1073} {\bibfield  {journal} {\bibinfo  {journal}
  {\href{https://www.nature.com/articles/nphys1073}{Nature Physics}}\ }\textbf
  {\bibinfo {volume} {4}},\ \bibinfo {pages} {878} (\bibinfo {year}
  {2008})}\BibitemShut {NoStop}%
\bibitem [{\citenamefont {Brazhnyi}\ \emph {et~al.}(2009)\citenamefont
  {Brazhnyi}, \citenamefont {Konotop}, \citenamefont {Perez-Garcia},\ and\
  \citenamefont {Ott}}]{r88}%
  \BibitemOpen
  \bibfield  {author} {\bibinfo {author} {\bibfnamefont {V.~A.}\ \bibnamefont
  {Brazhnyi}}, \bibinfo {author} {\bibfnamefont {V.~V.}\ \bibnamefont
  {Konotop}}, \bibinfo {author} {\bibfnamefont {V.~M.}\ \bibnamefont
  {Perez-Garcia}}, \ and\ \bibinfo {author} {\bibfnamefont {H.}~\bibnamefont
  {Ott}},\ }\href {\doibase 10.1103/PhysRevLett.102.144101} {\bibfield
  {journal} {\bibinfo  {journal} {Phys. Rev. Lett.}\ }\textbf {\bibinfo
  {volume} {102}},\ \bibinfo {pages} {144101} (\bibinfo {year}
  {2009})}\BibitemShut {NoStop}%
\bibitem [{\citenamefont {Diehl}\ \emph {et~al.}(2011)\citenamefont {Diehl},
  \citenamefont {Rico}, \citenamefont {Baranov},\ and\ \citenamefont
  {Zoller}}]{r23}%
  \BibitemOpen
  \bibfield  {author} {\bibinfo {author} {\bibfnamefont {S.}~\bibnamefont
  {Diehl}}, \bibinfo {author} {\bibfnamefont {E.}~\bibnamefont {Rico}},
  \bibinfo {author} {\bibfnamefont {M.~A.}\ \bibnamefont {Baranov}}, \ and\
  \bibinfo {author} {\bibfnamefont {P.}~\bibnamefont {Zoller}},\ }\href
  {\doibase 10.1038/nphys2106} {\bibfield  {journal} {\bibinfo  {journal}
  {\href{https://www.nature.com/articles/nphys2106}{Nature Physics}}\ }\textbf
  {\bibinfo {volume} {7}},\ \bibinfo {pages} {971} (\bibinfo {year}
  {2011})}\BibitemShut {NoStop}%
\bibitem [{\citenamefont {Zezyulin}\ \emph {et~al.}(2012)\citenamefont
  {Zezyulin}, \citenamefont {Konotop}, \citenamefont {Barontini},\ and\
  \citenamefont {Ott}}]{r89}%
  \BibitemOpen
  \bibfield  {author} {\bibinfo {author} {\bibfnamefont {D.~A.}\ \bibnamefont
  {Zezyulin}}, \bibinfo {author} {\bibfnamefont {V.~V.}\ \bibnamefont
  {Konotop}}, \bibinfo {author} {\bibfnamefont {G.}~\bibnamefont {Barontini}},
  \ and\ \bibinfo {author} {\bibfnamefont {H.}~\bibnamefont {Ott}},\ }\href
  {\doibase 10.1103/PhysRevLett.109.020405} {\bibfield  {journal} {\bibinfo
  {journal} {Phys. Rev. Lett.}\ }\textbf {\bibinfo {volume} {109}},\ \bibinfo
  {pages} {020405} (\bibinfo {year} {2012})}\BibitemShut {NoStop}%
\bibitem [{\citenamefont {Labouvie}\ \emph {et~al.}(2016)\citenamefont
  {Labouvie}, \citenamefont {Santra}, \citenamefont {Heun},\ and\ \citenamefont
  {Ott}}]{r24}%
  \BibitemOpen
  \bibfield  {author} {\bibinfo {author} {\bibfnamefont {R.}~\bibnamefont
  {Labouvie}}, \bibinfo {author} {\bibfnamefont {B.}~\bibnamefont {Santra}},
  \bibinfo {author} {\bibfnamefont {S.}~\bibnamefont {Heun}}, \ and\ \bibinfo
  {author} {\bibfnamefont {H.}~\bibnamefont {Ott}},\ }\href {\doibase
  10.1103/PhysRevLett.116.235302} {\bibfield  {journal} {\bibinfo  {journal}
  {\href{https://journals.aps.org/prl/abstract/10.1103/PhysRevLett.116.235302}{Phys.
  Rev. Lett.}}\ }\textbf {\bibinfo {volume} {116}},\ \bibinfo {pages} {235302}
  (\bibinfo {year} {2016})}\BibitemShut {NoStop}%
\bibitem [{\citenamefont {Lapp}\ \emph {et~al.}(2019)\citenamefont {Lapp},
  \citenamefont {Ang'ong'a}, \citenamefont {An},\ and\ \citenamefont
  {Gadway}}]{r25}%
  \BibitemOpen
  \bibfield  {author} {\bibinfo {author} {\bibfnamefont {S.}~\bibnamefont
  {Lapp}}, \bibinfo {author} {\bibfnamefont {J.}~\bibnamefont {Ang'ong'a}},
  \bibinfo {author} {\bibfnamefont {F.~A.}\ \bibnamefont {An}}, \ and\ \bibinfo
  {author} {\bibfnamefont {B.}~\bibnamefont {Gadway}},\ }\href {\doibase
  10.1088/1367-2630/ab1147} {\bibfield  {journal} {\bibinfo  {journal}
  {\href{https://www.nature.com/articles/s41467-019-08596-1}{New Journal of
  Physics}}\ }\textbf {\bibinfo {volume} {21}},\ \bibinfo {pages} {045006}
  (\bibinfo {year} {2019})}\BibitemShut {NoStop}%
\bibitem [{\citenamefont {Li}\ \emph {et~al.}(2019)\citenamefont {Li},
  \citenamefont {Harter}, \citenamefont {Liu}, \citenamefont {de~Melo},
  \citenamefont {Joglekar},\ and\ \citenamefont {Luo}}]{r26}%
  \BibitemOpen
  \bibfield  {author} {\bibinfo {author} {\bibfnamefont {J.}~\bibnamefont
  {Li}}, \bibinfo {author} {\bibfnamefont {A.~K.}\ \bibnamefont {Harter}},
  \bibinfo {author} {\bibfnamefont {J.}~\bibnamefont {Liu}}, \bibinfo {author}
  {\bibfnamefont {L.}~\bibnamefont {de~Melo}}, \bibinfo {author} {\bibfnamefont
  {Y.~N.}\ \bibnamefont {Joglekar}}, \ and\ \bibinfo {author} {\bibfnamefont
  {L.}~\bibnamefont {Luo}},\ }\href {\doibase 10.1038/s41467-019-08596-1}
  {\bibfield  {journal} {\bibinfo  {journal}
  {\href{https://www.nature.com/articles/s41467-019-08596-1}{Nat. Commun.}}\
  }\textbf {\bibinfo {volume} {10}},\ \bibinfo {pages} {855} (\bibinfo {year}
  {2019})}\BibitemShut {NoStop}%
\bibitem [{\citenamefont {Ren}\ \emph {et~al.}(2021)\citenamefont {Ren},
  \citenamefont {Liu}, \citenamefont {Zhao}, \citenamefont {He}, \citenamefont
  {Pak}, \citenamefont {Li},\ and\ \citenamefont {Jo}}]{r12}%
  \BibitemOpen
  \bibfield  {author} {\bibinfo {author} {\bibfnamefont {Z.}~\bibnamefont
  {Ren}}, \bibinfo {author} {\bibfnamefont {D.}~\bibnamefont {Liu}}, \bibinfo
  {author} {\bibfnamefont {E.~T.}\ \bibnamefont {Zhao}}, \bibinfo {author}
  {\bibfnamefont {C.~D.}\ \bibnamefont {He}}, \bibinfo {author} {\bibfnamefont
  {K.~k.}\ \bibnamefont {Pak}}, \bibinfo {author} {\bibfnamefont {J.~S.}\
  \bibnamefont {Li}}, \ and\ \bibinfo {author} {\bibfnamefont {G.~B.}\
  \bibnamefont {Jo}},\ }\href {https://arxiv.org/abs/2106.04874} {\  (\bibinfo
  {year} {2021})},\ \Eprint {http://arxiv.org/abs/arXiv:2106.04874}
  {arXiv:2106.04874} \BibitemShut {NoStop}%
\bibitem [{\citenamefont {Barontini}\ \emph {et~al.}(2013)\citenamefont
  {Barontini}, \citenamefont {Labouvie}, \citenamefont {Stubenrauch},
  \citenamefont {Vogler}, \citenamefont {Guarrera},\ and\ \citenamefont
  {Ott}}]{r21}%
  \BibitemOpen
  \bibfield  {author} {\bibinfo {author} {\bibfnamefont {G.}~\bibnamefont
  {Barontini}}, \bibinfo {author} {\bibfnamefont {R.}~\bibnamefont {Labouvie}},
  \bibinfo {author} {\bibfnamefont {F.}~\bibnamefont {Stubenrauch}}, \bibinfo
  {author} {\bibfnamefont {A.}~\bibnamefont {Vogler}}, \bibinfo {author}
  {\bibfnamefont {V.}~\bibnamefont {Guarrera}}, \ and\ \bibinfo {author}
  {\bibfnamefont {H.}~\bibnamefont {Ott}},\ }\href {\doibase
  10.1103/PhysRevLett.110.035302} {\bibfield  {journal} {\bibinfo  {journal}
  {\href{https://journals.aps.org/prl/abstract/10.1103/PhysRevLett.110.035302}{Phys.
  Rev. Lett.}}\ }\textbf {\bibinfo {volume} {110}},\ \bibinfo {pages} {035302}
  (\bibinfo {year} {2013})}\BibitemShut {NoStop}%
\bibitem [{\citenamefont {Takasu}\ \emph {et~al.}(2020)\citenamefont {Takasu},
  \citenamefont {Yagami}, \citenamefont {Ashida}, \citenamefont {Hamazaki},
  \citenamefont {Kuno},\ and\ \citenamefont {Takahashi}}]{r120}%
  \BibitemOpen
  \bibfield  {author} {\bibinfo {author} {\bibfnamefont {Y.}~\bibnamefont
  {Takasu}}, \bibinfo {author} {\bibfnamefont {T.}~\bibnamefont {Yagami}},
  \bibinfo {author} {\bibfnamefont {Y.}~\bibnamefont {Ashida}}, \bibinfo
  {author} {\bibfnamefont {R.}~\bibnamefont {Hamazaki}}, \bibinfo {author}
  {\bibfnamefont {Y.}~\bibnamefont {Kuno}}, \ and\ \bibinfo {author}
  {\bibfnamefont {Y.}~\bibnamefont {Takahashi}},\ }\href {\doibase
  10.1093/ptep/ptaa094} {\bibfield  {journal} {\bibinfo  {journal}
  {\href{https://academic.oup.com/ptep/article/2020/12/12A110/5905047}{Progress
  of Theoretical and Experimental Physics}}\ }\textbf {\bibinfo {volume}
  {2020}},\ \bibinfo {pages} {ptaa094} (\bibinfo {year} {2020})}\BibitemShut
  {NoStop}%
\bibitem [{\citenamefont {Dast}\ \emph {et~al.}(2014)\citenamefont {Dast},
  \citenamefont {Haag}, \citenamefont {Cartarius},\ and\ \citenamefont
  {Wunner}}]{r122}%
  \BibitemOpen
  \bibfield  {author} {\bibinfo {author} {\bibfnamefont {D.}~\bibnamefont
  {Dast}}, \bibinfo {author} {\bibfnamefont {D.}~\bibnamefont {Haag}}, \bibinfo
  {author} {\bibfnamefont {H.}~\bibnamefont {Cartarius}}, \ and\ \bibinfo
  {author} {\bibfnamefont {G.}~\bibnamefont {Wunner}},\ }\href {\doibase
  10.1103/PhysRevA.90.052120} {\bibfield  {journal} {\bibinfo  {journal}
  {\href{https://journals.aps.org/pra/abstract/10.1103/PhysRevA.90.052120}{Physical
  Review A}}\ }\textbf {\bibinfo {volume} {90}},\ \bibinfo {pages} {052120}
  (\bibinfo {year} {2014})}\BibitemShut {NoStop}%
\bibitem [{\citenamefont {Li}\ \emph {et~al.}(2021)\citenamefont {Li},
  \citenamefont {Mu}, \citenamefont {Lee},\ and\ \citenamefont {Gong}}]{r126}%
  \BibitemOpen
  \bibfield  {author} {\bibinfo {author} {\bibfnamefont {L.}~\bibnamefont
  {Li}}, \bibinfo {author} {\bibfnamefont {S.}~\bibnamefont {Mu}}, \bibinfo
  {author} {\bibfnamefont {C.~H.}\ \bibnamefont {Lee}}, \ and\ \bibinfo
  {author} {\bibfnamefont {J.}~\bibnamefont {Gong}},\ }\href {\doibase
  10.1038/s41467-021-25626-z} {\bibfield  {journal} {\bibinfo  {journal} {Nat.
  Commun.}\ }\textbf {\bibinfo {volume} {12}},\ \bibinfo {pages} {5294}
  (\bibinfo {year} {2021})}\BibitemShut {NoStop}%
\bibitem [{\citenamefont {Nakagawa}\ \emph {et~al.}(2018)\citenamefont
  {Nakagawa}, \citenamefont {Kawakami},\ and\ \citenamefont {Ueda}}]{r128}%
  \BibitemOpen
  \bibfield  {author} {\bibinfo {author} {\bibfnamefont {M.}~\bibnamefont
  {Nakagawa}}, \bibinfo {author} {\bibfnamefont {N.}~\bibnamefont {Kawakami}},
  \ and\ \bibinfo {author} {\bibfnamefont {M.}~\bibnamefont {Ueda}},\ }\href
  {\doibase 10.1103/PhysRevLett.121.203001} {\bibfield  {journal} {\bibinfo
  {journal} {Phys. Rev. Lett.}\ }\textbf {\bibinfo {volume} {121}},\ \bibinfo
  {pages} {203001} (\bibinfo {year} {2018})}\BibitemShut {NoStop}%
\bibitem [{\citenamefont {Nakagawa}\ \emph {et~al.}(2021)\citenamefont
  {Nakagawa}, \citenamefont {Kawakami},\ and\ \citenamefont {Ueda}}]{r129}%
  \BibitemOpen
  \bibfield  {author} {\bibinfo {author} {\bibfnamefont {M.}~\bibnamefont
  {Nakagawa}}, \bibinfo {author} {\bibfnamefont {N.}~\bibnamefont {Kawakami}},
  \ and\ \bibinfo {author} {\bibfnamefont {M.}~\bibnamefont {Ueda}},\ }\href
  {\doibase 10.1103/PhysRevLett.126.110404} {\bibfield  {journal} {\bibinfo
  {journal} {Phys Rev Lett}\ }\textbf {\bibinfo {volume} {126}},\ \bibinfo
  {pages} {110404} (\bibinfo {year} {2021})}\BibitemShut {NoStop}%
\bibitem [{\citenamefont {Pan}\ \emph {et~al.}(2020)\citenamefont {Pan},
  \citenamefont {Chen}, \citenamefont {Chen},\ and\ \citenamefont
  {Zhai}}]{r130}%
  \BibitemOpen
  \bibfield  {author} {\bibinfo {author} {\bibfnamefont {L.}~\bibnamefont
  {Pan}}, \bibinfo {author} {\bibfnamefont {X.}~\bibnamefont {Chen}}, \bibinfo
  {author} {\bibfnamefont {Y.}~\bibnamefont {Chen}}, \ and\ \bibinfo {author}
  {\bibfnamefont {H.}~\bibnamefont {Zhai}},\ }\href {\doibase
  10.1038/s41567-020-0889-6} {\bibfield  {journal} {\bibinfo  {journal} {Nature
  Physics}\ }\textbf {\bibinfo {volume} {16}},\ \bibinfo {pages} {767}
  (\bibinfo {year} {2020})}\BibitemShut {NoStop}%
\bibitem [{\citenamefont {Yamamoto}\ \emph {et~al.}(2019)\citenamefont
  {Yamamoto}, \citenamefont {Nakagawa}, \citenamefont {Adachi}, \citenamefont
  {Takasan}, \citenamefont {Ueda},\ and\ \citenamefont {Kawakami}}]{r132}%
  \BibitemOpen
  \bibfield  {author} {\bibinfo {author} {\bibfnamefont {K.}~\bibnamefont
  {Yamamoto}}, \bibinfo {author} {\bibfnamefont {M.}~\bibnamefont {Nakagawa}},
  \bibinfo {author} {\bibfnamefont {K.}~\bibnamefont {Adachi}}, \bibinfo
  {author} {\bibfnamefont {K.}~\bibnamefont {Takasan}}, \bibinfo {author}
  {\bibfnamefont {M.}~\bibnamefont {Ueda}}, \ and\ \bibinfo {author}
  {\bibfnamefont {N.}~\bibnamefont {Kawakami}},\ }\href {\doibase
  10.1103/PhysRevLett.123.123601} {\bibfield  {journal} {\bibinfo  {journal}
  {Phys. Rev. Lett.}\ }\textbf {\bibinfo {volume} {123}},\ \bibinfo {pages}
  {123601} (\bibinfo {year} {2019})}\BibitemShut {NoStop}%
\bibitem [{\citenamefont {Ashida}\ \emph {et~al.}(2020)\citenamefont {Ashida},
  \citenamefont {Gong},\ and\ \citenamefont {Ueda}}]{r133}%
  \BibitemOpen
  \bibfield  {author} {\bibinfo {author} {\bibfnamefont {Y.}~\bibnamefont
  {Ashida}}, \bibinfo {author} {\bibfnamefont {Z.}~\bibnamefont {Gong}}, \ and\
  \bibinfo {author} {\bibfnamefont {M.}~\bibnamefont {Ueda}},\ }\href {\doibase
  10.1080/00018732.2021.1876991} {\bibfield  {journal} {\bibinfo  {journal}
  {Advances in Physics}\ }\textbf {\bibinfo {volume} {69}},\ \bibinfo {pages}
  {249} (\bibinfo {year} {2020})}\BibitemShut {NoStop}%
\bibitem [{\citenamefont {Xu}\ \emph {et~al.}(2017)\citenamefont {Xu},
  \citenamefont {Wang},\ and\ \citenamefont {Duan}}]{r141}%
  \BibitemOpen
  \bibfield  {author} {\bibinfo {author} {\bibfnamefont {Y.}~\bibnamefont
  {Xu}}, \bibinfo {author} {\bibfnamefont {S.~T.}\ \bibnamefont {Wang}}, \ and\
  \bibinfo {author} {\bibfnamefont {L.~M.}\ \bibnamefont {Duan}},\ }\href
  {\doibase 10.1103/PhysRevLett.118.045701} {\bibfield  {journal} {\bibinfo
  {journal} {Phys Rev Lett}\ }\textbf {\bibinfo {volume} {118}},\ \bibinfo
  {pages} {045701} (\bibinfo {year} {2017})}\BibitemShut {NoStop}%
\bibitem [{\citenamefont {Bergholtz}\ \emph {et~al.}(2021)\citenamefont
  {Bergholtz}, \citenamefont {Budich},\ and\ \citenamefont {Kunst}}]{r136}%
  \BibitemOpen
  \bibfield  {author} {\bibinfo {author} {\bibfnamefont {E.~J.}\ \bibnamefont
  {Bergholtz}}, \bibinfo {author} {\bibfnamefont {J.~C.}\ \bibnamefont
  {Budich}}, \ and\ \bibinfo {author} {\bibfnamefont {F.~K.}\ \bibnamefont
  {Kunst}},\ }\href {\doibase 10.1103/RevModPhys.93.015005} {\bibfield
  {journal} {\bibinfo  {journal} {Reviews of Modern Physics}\ }\textbf
  {\bibinfo {volume} {93}},\ \bibinfo {pages} {015005} (\bibinfo {year}
  {2021})}\BibitemShut {NoStop}%
\bibitem [{\citenamefont {Lieu}\ \emph {et~al.}(2020)\citenamefont {Lieu},
  \citenamefont {McGinley},\ and\ \citenamefont {Cooper}}]{r143}%
  \BibitemOpen
  \bibfield  {author} {\bibinfo {author} {\bibfnamefont {S.}~\bibnamefont
  {Lieu}}, \bibinfo {author} {\bibfnamefont {M.}~\bibnamefont {McGinley}}, \
  and\ \bibinfo {author} {\bibfnamefont {N.~R.}\ \bibnamefont {Cooper}},\
  }\href {\doibase 10.1103/PhysRevLett.124.040401} {\bibfield  {journal}
  {\bibinfo  {journal} {Phys Rev Lett}\ }\textbf {\bibinfo {volume} {124}},\
  \bibinfo {pages} {040401} (\bibinfo {year} {2020})}\BibitemShut {NoStop}%
\bibitem [{\citenamefont {Yao}\ and\ \citenamefont
  {Wang}(2018)}]{yaoEdgeStatesTopological2018b}%
  \BibitemOpen
  \bibfield  {author} {\bibinfo {author} {\bibfnamefont {S.}~\bibnamefont
  {Yao}}\ and\ \bibinfo {author} {\bibfnamefont {Z.}~\bibnamefont {Wang}},\
  }\href {\doibase 10.1103/PhysRevLett.121.086803} {\bibfield  {journal}
  {\bibinfo  {journal} {Phys. Rev. Lett.}\ }\textbf {\bibinfo {volume} {121}},\
  \bibinfo {pages} {086803} (\bibinfo {year} {2018})}\BibitemShut {NoStop}%
\bibitem [{\citenamefont {Yao}\ \emph {et~al.}(2018)\citenamefont {Yao},
  \citenamefont {Song},\ and\ \citenamefont
  {Wang}}]{yaoNonHermitianChernBands2018b}%
  \BibitemOpen
  \bibfield  {author} {\bibinfo {author} {\bibfnamefont {S.}~\bibnamefont
  {Yao}}, \bibinfo {author} {\bibfnamefont {F.}~\bibnamefont {Song}}, \ and\
  \bibinfo {author} {\bibfnamefont {Z.}~\bibnamefont {Wang}},\ }\href {\doibase
  10.1103/PhysRevLett.121.136802} {\bibfield  {journal} {\bibinfo  {journal}
  {Phys. Rev. Lett.}\ }\textbf {\bibinfo {volume} {121}},\ \bibinfo {pages}
  {136802} (\bibinfo {year} {2018})}\BibitemShut {NoStop}%
\bibitem [{\citenamefont {Song}\ \emph
  {et~al.}(2019{\natexlab{a}})\citenamefont {Song}, \citenamefont {Yao},\ and\
  \citenamefont {Wang}}]{r123}%
  \BibitemOpen
  \bibfield  {author} {\bibinfo {author} {\bibfnamefont {F.}~\bibnamefont
  {Song}}, \bibinfo {author} {\bibfnamefont {S.}~\bibnamefont {Yao}}, \ and\
  \bibinfo {author} {\bibfnamefont {Z.}~\bibnamefont {Wang}},\ }\href {\doibase
  10.1103/PhysRevLett.123.246801} {\bibfield  {journal} {\bibinfo  {journal}
  {Phys. Rev. Lett.}\ }\textbf {\bibinfo {volume} {123}},\ \bibinfo {pages}
  {246801} (\bibinfo {year} {2019}{\natexlab{a}})}\BibitemShut {NoStop}%
\bibitem [{\citenamefont {Kunst}\ \emph {et~al.}(2018)\citenamefont {Kunst},
  \citenamefont {Edvardsson}, \citenamefont {Budich},\ and\ \citenamefont
  {Bergholtz}}]{PhysRevLett.121.026808}%
  \BibitemOpen
  \bibfield  {author} {\bibinfo {author} {\bibfnamefont {F.~K.}\ \bibnamefont
  {Kunst}}, \bibinfo {author} {\bibfnamefont {E.}~\bibnamefont {Edvardsson}},
  \bibinfo {author} {\bibfnamefont {J.~C.}\ \bibnamefont {Budich}}, \ and\
  \bibinfo {author} {\bibfnamefont {E.~J.}\ \bibnamefont {Bergholtz}},\ }\href
  {\doibase 10.1103/PhysRevLett.121.026808} {\bibfield  {journal} {\bibinfo
  {journal} {Phys. Rev. Lett.}\ }\textbf {\bibinfo {volume} {121}},\ \bibinfo
  {pages} {026808} (\bibinfo {year} {2018})}\BibitemShut {NoStop}%
\bibitem [{\citenamefont {Yokomizo}\ and\ \citenamefont
  {Murakami}(2019)}]{yokomizoNonBlochBandTheory2019a}%
  \BibitemOpen
  \bibfield  {author} {\bibinfo {author} {\bibfnamefont {K.}~\bibnamefont
  {Yokomizo}}\ and\ \bibinfo {author} {\bibfnamefont {S.}~\bibnamefont
  {Murakami}},\ }\href {\doibase 10.1103/PhysRevLett.123.066404} {\bibfield
  {journal} {\bibinfo  {journal} {Phys. Rev. Lett.}\ }\textbf {\bibinfo
  {volume} {123}},\ \bibinfo {pages} {066404} (\bibinfo {year}
  {2019})}\BibitemShut {NoStop}%
\bibitem [{\citenamefont {Zhang}\ \emph {et~al.}(2020)\citenamefont {Zhang},
  \citenamefont {Yang},\ and\ \citenamefont {Fang}}]{PhysRevLett.125.126402}%
  \BibitemOpen
  \bibfield  {author} {\bibinfo {author} {\bibfnamefont {K.}~\bibnamefont
  {Zhang}}, \bibinfo {author} {\bibfnamefont {Z.}~\bibnamefont {Yang}}, \ and\
  \bibinfo {author} {\bibfnamefont {C.}~\bibnamefont {Fang}},\ }\href {\doibase
  10.1103/PhysRevLett.125.126402} {\bibfield  {journal} {\bibinfo  {journal}
  {Phys. Rev. Lett.}\ }\textbf {\bibinfo {volume} {125}},\ \bibinfo {pages}
  {126402} (\bibinfo {year} {2020})}\BibitemShut {NoStop}%
\bibitem [{\citenamefont {Okuma}\ \emph {et~al.}(2020)\citenamefont {Okuma},
  \citenamefont {Kawabata}, \citenamefont {Shiozaki},\ and\ \citenamefont
  {Sato}}]{PhysRevLett.124.086801}%
  \BibitemOpen
  \bibfield  {author} {\bibinfo {author} {\bibfnamefont {N.}~\bibnamefont
  {Okuma}}, \bibinfo {author} {\bibfnamefont {K.}~\bibnamefont {Kawabata}},
  \bibinfo {author} {\bibfnamefont {K.}~\bibnamefont {Shiozaki}}, \ and\
  \bibinfo {author} {\bibfnamefont {M.}~\bibnamefont {Sato}},\ }\href {\doibase
  10.1103/PhysRevLett.124.086801} {\bibfield  {journal} {\bibinfo  {journal}
  {Phys. Rev. Lett.}\ }\textbf {\bibinfo {volume} {124}},\ \bibinfo {pages}
  {086801} (\bibinfo {year} {2020})}\BibitemShut {NoStop}%
\bibitem [{\citenamefont {Yang}\ \emph {et~al.}(2020)\citenamefont {Yang},
  \citenamefont {Zhang}, \citenamefont {Fang},\ and\ \citenamefont
  {Hu}}]{PhysRevLett.125.226402}%
  \BibitemOpen
  \bibfield  {author} {\bibinfo {author} {\bibfnamefont {Z.}~\bibnamefont
  {Yang}}, \bibinfo {author} {\bibfnamefont {K.}~\bibnamefont {Zhang}},
  \bibinfo {author} {\bibfnamefont {C.}~\bibnamefont {Fang}}, \ and\ \bibinfo
  {author} {\bibfnamefont {J.}~\bibnamefont {Hu}},\ }\href {\doibase
  10.1103/PhysRevLett.125.226402} {\bibfield  {journal} {\bibinfo  {journal}
  {Phys. Rev. Lett.}\ }\textbf {\bibinfo {volume} {125}},\ \bibinfo {pages}
  {226402} (\bibinfo {year} {2020})}\BibitemShut {NoStop}%
\bibitem [{\citenamefont {Martinez~Alvarez}\ \emph {et~al.}(2018)\citenamefont
  {Martinez~Alvarez}, \citenamefont {Barrios~Vargas},\ and\ \citenamefont
  {Foa~Torres}}]{PhysRevB.97.121401}%
  \BibitemOpen
  \bibfield  {author} {\bibinfo {author} {\bibfnamefont {V.~M.}\ \bibnamefont
  {Martinez~Alvarez}}, \bibinfo {author} {\bibfnamefont {J.~E.}\ \bibnamefont
  {Barrios~Vargas}}, \ and\ \bibinfo {author} {\bibfnamefont {L.~E.~F.}\
  \bibnamefont {Foa~Torres}},\ }\href {\doibase 10.1103/PhysRevB.97.121401}
  {\bibfield  {journal} {\bibinfo  {journal} {Phys. Rev. B}\ }\textbf {\bibinfo
  {volume} {97}},\ \bibinfo {pages} {121401} (\bibinfo {year}
  {2018})}\BibitemShut {NoStop}%
\bibitem [{\citenamefont {Lee}\ and\ \citenamefont
  {Thomale}(2019)}]{leeAnatomySkinModes2019b}%
  \BibitemOpen
  \bibfield  {author} {\bibinfo {author} {\bibfnamefont {C.~H.}\ \bibnamefont
  {Lee}}\ and\ \bibinfo {author} {\bibfnamefont {R.}~\bibnamefont {Thomale}},\
  }\href {\doibase 10.1103/PhysRevB.99.201103} {\bibfield  {journal} {\bibinfo
  {journal} {Phys. Rev. B}\ }\textbf {\bibinfo {volume} {99}},\ \bibinfo
  {pages} {201103} (\bibinfo {year} {2019})}\BibitemShut {NoStop}%
\bibitem [{\citenamefont {Song}\ \emph
  {et~al.}(2019{\natexlab{b}})\citenamefont {Song}, \citenamefont {Yao},\ and\
  \citenamefont {Wang}}]{PhysRevLett.123.170401}%
  \BibitemOpen
  \bibfield  {author} {\bibinfo {author} {\bibfnamefont {F.}~\bibnamefont
  {Song}}, \bibinfo {author} {\bibfnamefont {S.}~\bibnamefont {Yao}}, \ and\
  \bibinfo {author} {\bibfnamefont {Z.}~\bibnamefont {Wang}},\ }\href {\doibase
  10.1103/PhysRevLett.123.170401} {\bibfield  {journal} {\bibinfo  {journal}
  {Phys. Rev. Lett.}\ }\textbf {\bibinfo {volume} {123}},\ \bibinfo {pages}
  {170401} (\bibinfo {year} {2019}{\natexlab{b}})}\BibitemShut {NoStop}%
\bibitem [{\citenamefont {Lee}\ \emph {et~al.}(2019)\citenamefont {Lee},
  \citenamefont {Li},\ and\ \citenamefont {Gong}}]{PhysRevLett.123.016805}%
  \BibitemOpen
  \bibfield  {author} {\bibinfo {author} {\bibfnamefont {C.~H.}\ \bibnamefont
  {Lee}}, \bibinfo {author} {\bibfnamefont {L.}~\bibnamefont {Li}}, \ and\
  \bibinfo {author} {\bibfnamefont {J.}~\bibnamefont {Gong}},\ }\href {\doibase
  10.1103/PhysRevLett.123.016805} {\bibfield  {journal} {\bibinfo  {journal}
  {Phys. Rev. Lett.}\ }\textbf {\bibinfo {volume} {123}},\ \bibinfo {pages}
  {016805} (\bibinfo {year} {2019})}\BibitemShut {NoStop}%
\bibitem [{\citenamefont {Li}\ \emph {et~al.}(2020{\natexlab{a}})\citenamefont
  {Li}, \citenamefont {Lee},\ and\ \citenamefont
  {Gong}}]{PhysRevLett.124.250402}%
  \BibitemOpen
  \bibfield  {author} {\bibinfo {author} {\bibfnamefont {L.}~\bibnamefont
  {Li}}, \bibinfo {author} {\bibfnamefont {C.~H.}\ \bibnamefont {Lee}}, \ and\
  \bibinfo {author} {\bibfnamefont {J.}~\bibnamefont {Gong}},\ }\href {\doibase
  10.1103/PhysRevLett.124.250402} {\bibfield  {journal} {\bibinfo  {journal}
  {Phys. Rev. Lett.}\ }\textbf {\bibinfo {volume} {124}},\ \bibinfo {pages}
  {250402} (\bibinfo {year} {2020}{\natexlab{a}})}\BibitemShut {NoStop}%
\bibitem [{\citenamefont {Borgnia}\ \emph {et~al.}(2020)\citenamefont
  {Borgnia}, \citenamefont {Kruchkov},\ and\ \citenamefont
  {Slager}}]{PhysRevLett.124.056802}%
  \BibitemOpen
  \bibfield  {author} {\bibinfo {author} {\bibfnamefont {D.~S.}\ \bibnamefont
  {Borgnia}}, \bibinfo {author} {\bibfnamefont {A.~J.}\ \bibnamefont
  {Kruchkov}}, \ and\ \bibinfo {author} {\bibfnamefont {R.-J.}\ \bibnamefont
  {Slager}},\ }\href {\doibase 10.1103/PhysRevLett.124.056802} {\bibfield
  {journal} {\bibinfo  {journal} {Phys. Rev. Lett.}\ }\textbf {\bibinfo
  {volume} {124}},\ \bibinfo {pages} {056802} (\bibinfo {year}
  {2020})}\BibitemShut {NoStop}%
\bibitem [{\citenamefont {Longhi}(2019)}]{PhysRevResearch.1.023013}%
  \BibitemOpen
  \bibfield  {author} {\bibinfo {author} {\bibfnamefont {S.}~\bibnamefont
  {Longhi}},\ }\href {\doibase 10.1103/PhysRevResearch.1.023013} {\bibfield
  {journal} {\bibinfo  {journal} {Phys. Rev. Research}\ }\textbf {\bibinfo
  {volume} {1}},\ \bibinfo {pages} {023013} (\bibinfo {year}
  {2019})}\BibitemShut {NoStop}%
\bibitem [{\citenamefont {Longhi}(2020)}]{PhysRevLett.124.066602}%
  \BibitemOpen
  \bibfield  {author} {\bibinfo {author} {\bibfnamefont {S.}~\bibnamefont
  {Longhi}},\ }\href {\doibase 10.1103/PhysRevLett.124.066602} {\bibfield
  {journal} {\bibinfo  {journal} {Phys. Rev. Lett.}\ }\textbf {\bibinfo
  {volume} {124}},\ \bibinfo {pages} {066602} (\bibinfo {year}
  {2020})}\BibitemShut {NoStop}%
\bibitem [{\citenamefont {Yi}\ and\ \citenamefont
  {Yang}(2020)}]{PhysRevLett.125.186802}%
  \BibitemOpen
  \bibfield  {author} {\bibinfo {author} {\bibfnamefont {Y.}~\bibnamefont
  {Yi}}\ and\ \bibinfo {author} {\bibfnamefont {Z.}~\bibnamefont {Yang}},\
  }\href {\doibase 10.1103/PhysRevLett.125.186802} {\bibfield  {journal}
  {\bibinfo  {journal} {Phys. Rev. Lett.}\ }\textbf {\bibinfo {volume} {125}},\
  \bibinfo {pages} {186802} (\bibinfo {year} {2020})}\BibitemShut {NoStop}%
\bibitem [{\citenamefont {Li}\ \emph {et~al.}(2020{\natexlab{b}})\citenamefont
  {Li}, \citenamefont {Lee}, \citenamefont {Mu},\ and\ \citenamefont
  {Gong}}]{Li:2020aa}%
  \BibitemOpen
  \bibfield  {author} {\bibinfo {author} {\bibfnamefont {L.}~\bibnamefont
  {Li}}, \bibinfo {author} {\bibfnamefont {C.~H.}\ \bibnamefont {Lee}},
  \bibinfo {author} {\bibfnamefont {S.}~\bibnamefont {Mu}}, \ and\ \bibinfo
  {author} {\bibfnamefont {J.}~\bibnamefont {Gong}},\ }\href {\doibase
  10.1038/s41467-020-18917-4} {\bibfield  {journal} {\bibinfo  {journal}
  {Nature Communications}\ }\textbf {\bibinfo {volume} {11}},\ \bibinfo {pages}
  {5491} (\bibinfo {year} {2020}{\natexlab{b}})}\BibitemShut {NoStop}%
\bibitem [{\citenamefont {Zhang}\ \emph
  {et~al.}(2021{\natexlab{a}})\citenamefont {Zhang}, \citenamefont {Yang},\
  and\ \citenamefont {Fang}}]{r118}%
  \BibitemOpen
  \bibfield  {author} {\bibinfo {author} {\bibfnamefont {K.}~\bibnamefont
  {Zhang}}, \bibinfo {author} {\bibfnamefont {Z.}~\bibnamefont {Yang}}, \ and\
  \bibinfo {author} {\bibfnamefont {C.}~\bibnamefont {Fang}},\ }\href
  {https://arxiv.org/abs/2102.05059} {\  (\bibinfo {year}
  {2021}{\natexlab{a}})},\ \Eprint {http://arxiv.org/abs/arXiv:2102.05059}
  {arXiv:2102.05059} \BibitemShut {NoStop}%
\bibitem [{\citenamefont {Xiao}\ \emph {et~al.}(2020)\citenamefont {Xiao},
  \citenamefont {Deng}, \citenamefont {Wang}, \citenamefont {Zhu},
  \citenamefont {Wang}, \citenamefont {Yi},\ and\ \citenamefont
  {Xue}}]{Xiao:2020aa}%
  \BibitemOpen
  \bibfield  {author} {\bibinfo {author} {\bibfnamefont {L.}~\bibnamefont
  {Xiao}}, \bibinfo {author} {\bibfnamefont {T.}~\bibnamefont {Deng}}, \bibinfo
  {author} {\bibfnamefont {K.}~\bibnamefont {Wang}}, \bibinfo {author}
  {\bibfnamefont {G.}~\bibnamefont {Zhu}}, \bibinfo {author} {\bibfnamefont
  {Z.}~\bibnamefont {Wang}}, \bibinfo {author} {\bibfnamefont {W.}~\bibnamefont
  {Yi}}, \ and\ \bibinfo {author} {\bibfnamefont {P.}~\bibnamefont {Xue}},\
  }\href {\doibase 10.1038/s41567-020-0836-6} {\bibfield  {journal} {\bibinfo
  {journal} {Nature Physics}\ }\textbf {\bibinfo {volume} {16}},\ \bibinfo
  {pages} {761} (\bibinfo {year} {2020})}\BibitemShut {NoStop}%
\bibitem [{\citenamefont {Helbig}\ \emph {et~al.}(2020)\citenamefont {Helbig},
  \citenamefont {Hofmann}, \citenamefont {Imhof}, \citenamefont {Abdelghany},
  \citenamefont {Kiessling}, \citenamefont {Molenkamp}, \citenamefont {Lee},
  \citenamefont {Szameit}, \citenamefont {Greiter},\ and\ \citenamefont
  {Thomale}}]{Helbig:2020aa}%
  \BibitemOpen
  \bibfield  {author} {\bibinfo {author} {\bibfnamefont {T.}~\bibnamefont
  {Helbig}}, \bibinfo {author} {\bibfnamefont {T.}~\bibnamefont {Hofmann}},
  \bibinfo {author} {\bibfnamefont {S.}~\bibnamefont {Imhof}}, \bibinfo
  {author} {\bibfnamefont {M.}~\bibnamefont {Abdelghany}}, \bibinfo {author}
  {\bibfnamefont {T.}~\bibnamefont {Kiessling}}, \bibinfo {author}
  {\bibfnamefont {L.~W.}\ \bibnamefont {Molenkamp}}, \bibinfo {author}
  {\bibfnamefont {C.~H.}\ \bibnamefont {Lee}}, \bibinfo {author} {\bibfnamefont
  {A.}~\bibnamefont {Szameit}}, \bibinfo {author} {\bibfnamefont
  {M.}~\bibnamefont {Greiter}}, \ and\ \bibinfo {author} {\bibfnamefont
  {R.}~\bibnamefont {Thomale}},\ }\href {\doibase 10.1038/s41567-020-0922-9}
  {\bibfield  {journal} {\bibinfo  {journal} {Nature Physics}\ }\textbf
  {\bibinfo {volume} {16}},\ \bibinfo {pages} {747} (\bibinfo {year}
  {2020})}\BibitemShut {NoStop}%
\bibitem [{\citenamefont {Weidemann}\ \emph {et~al.}(2020)\citenamefont
  {Weidemann}, \citenamefont {Kremer}, \citenamefont {Helbig}, \citenamefont
  {Hofmann}, \citenamefont {Stegmaier}, \citenamefont {Greiter}, \citenamefont
  {Thomale},\ and\ \citenamefont {Szameit}}]{Weidemann311}%
  \BibitemOpen
  \bibfield  {author} {\bibinfo {author} {\bibfnamefont {S.}~\bibnamefont
  {Weidemann}}, \bibinfo {author} {\bibfnamefont {M.}~\bibnamefont {Kremer}},
  \bibinfo {author} {\bibfnamefont {T.}~\bibnamefont {Helbig}}, \bibinfo
  {author} {\bibfnamefont {T.}~\bibnamefont {Hofmann}}, \bibinfo {author}
  {\bibfnamefont {A.}~\bibnamefont {Stegmaier}}, \bibinfo {author}
  {\bibfnamefont {M.}~\bibnamefont {Greiter}}, \bibinfo {author} {\bibfnamefont
  {R.}~\bibnamefont {Thomale}}, \ and\ \bibinfo {author} {\bibfnamefont
  {A.}~\bibnamefont {Szameit}},\ }\href {\doibase 10.1126/science.aaz8727}
  {\bibfield  {journal} {\bibinfo  {journal} {Science}\ }\textbf {\bibinfo
  {volume} {368}},\ \bibinfo {pages} {311} (\bibinfo {year}
  {2020})}\BibitemShut {NoStop}%
\bibitem [{\citenamefont {Hofmann}\ \emph {et~al.}(2020)\citenamefont
  {Hofmann}, \citenamefont {Helbig}, \citenamefont {Schindler}, \citenamefont
  {Salgo}, \citenamefont {Brzezi\ifmmode~\acute{n}\else \'{n}\fi{}ska},
  \citenamefont {Greiter}, \citenamefont {Kiessling}, \citenamefont {Wolf},
  \citenamefont {Vollhardt}, \citenamefont {Kaba\ifmmode~\check{s}\else
  \v{s}\fi{}i}, \citenamefont {Lee}, \citenamefont {Bilu\ifmmode \check{s}\else
  \v{s}\fi{}i\ifmmode~\acute{c}\else \'{c}\fi{}}, \citenamefont {Thomale},\
  and\ \citenamefont {Neupert}}]{PhysRevResearch.2.023265}%
  \BibitemOpen
  \bibfield  {author} {\bibinfo {author} {\bibfnamefont {T.}~\bibnamefont
  {Hofmann}}, \bibinfo {author} {\bibfnamefont {T.}~\bibnamefont {Helbig}},
  \bibinfo {author} {\bibfnamefont {F.}~\bibnamefont {Schindler}}, \bibinfo
  {author} {\bibfnamefont {N.}~\bibnamefont {Salgo}}, \bibinfo {author}
  {\bibfnamefont {M.}~\bibnamefont {Brzezi\ifmmode~\acute{n}\else
  \'{n}\fi{}ska}}, \bibinfo {author} {\bibfnamefont {M.}~\bibnamefont
  {Greiter}}, \bibinfo {author} {\bibfnamefont {T.}~\bibnamefont {Kiessling}},
  \bibinfo {author} {\bibfnamefont {D.}~\bibnamefont {Wolf}}, \bibinfo {author}
  {\bibfnamefont {A.}~\bibnamefont {Vollhardt}}, \bibinfo {author}
  {\bibfnamefont {A.}~\bibnamefont {Kaba\ifmmode~\check{s}\else \v{s}\fi{}i}},
  \bibinfo {author} {\bibfnamefont {C.~H.}\ \bibnamefont {Lee}}, \bibinfo
  {author} {\bibfnamefont {A.}~\bibnamefont {Bilu\ifmmode \check{s}\else
  \v{s}\fi{}i\ifmmode~\acute{c}\else \'{c}\fi{}}}, \bibinfo {author}
  {\bibfnamefont {R.}~\bibnamefont {Thomale}}, \ and\ \bibinfo {author}
  {\bibfnamefont {T.}~\bibnamefont {Neupert}},\ }\href {\doibase
  10.1103/PhysRevResearch.2.023265} {\bibfield  {journal} {\bibinfo  {journal}
  {Phys. Rev. Research}\ }\textbf {\bibinfo {volume} {2}},\ \bibinfo {pages}
  {023265} (\bibinfo {year} {2020})}\BibitemShut {NoStop}%
\bibitem [{\citenamefont {Haga}\ \emph {et~al.}(2021)\citenamefont {Haga},
  \citenamefont {Nakagawa}, \citenamefont {Hamazaki},\ and\ \citenamefont
  {Ueda}}]{r124}%
  \BibitemOpen
  \bibfield  {author} {\bibinfo {author} {\bibfnamefont {T.}~\bibnamefont
  {Haga}}, \bibinfo {author} {\bibfnamefont {M.}~\bibnamefont {Nakagawa}},
  \bibinfo {author} {\bibfnamefont {R.}~\bibnamefont {Hamazaki}}, \ and\
  \bibinfo {author} {\bibfnamefont {M.}~\bibnamefont {Ueda}},\ }\href {\doibase
  10.1103/PhysRevLett.127.070402} {\bibfield  {journal} {\bibinfo  {journal}
  {Phys. Rev. Lett.}\ }\textbf {\bibinfo {volume} {127}},\ \bibinfo {pages}
  {070402} (\bibinfo {year} {2021})}\BibitemShut {NoStop}%
\bibitem [{\citenamefont {Claes}\ and\ \citenamefont {Hughes}(2021)}]{r125}%
  \BibitemOpen
  \bibfield  {author} {\bibinfo {author} {\bibfnamefont {J.}~\bibnamefont
  {Claes}}\ and\ \bibinfo {author} {\bibfnamefont {T.~L.}\ \bibnamefont
  {Hughes}},\ }\href
  {https://journals.aps.org/prb/abstract/10.1103/PhysRevB.103.L140201}
  {\bibfield  {journal} {\bibinfo  {journal} {Physical Review B}\ }\textbf
  {\bibinfo {volume} {103}},\ \bibinfo {pages} {L140201} (\bibinfo {year}
  {2021})}\BibitemShut {NoStop}%
\bibitem [{\citenamefont {Sun}\ \emph {et~al.}(2021)\citenamefont {Sun},
  \citenamefont {Zhu},\ and\ \citenamefont {Hughes}}]{r131}%
  \BibitemOpen
  \bibfield  {author} {\bibinfo {author} {\bibfnamefont {X.~Q.}\ \bibnamefont
  {Sun}}, \bibinfo {author} {\bibfnamefont {P.}~\bibnamefont {Zhu}}, \ and\
  \bibinfo {author} {\bibfnamefont {T.~L.}\ \bibnamefont {Hughes}},\ }\href
  {\doibase 10.1103/PhysRevLett.127.066401} {\bibfield  {journal} {\bibinfo
  {journal} {Phys. Rev. Lett.}\ }\textbf {\bibinfo {volume} {127}},\ \bibinfo
  {pages} {066401} (\bibinfo {year} {2021})}\BibitemShut {NoStop}%
\bibitem [{\citenamefont {Schindler}\ and\ \citenamefont {Prem}(2021)}]{r112}%
  \BibitemOpen
  \bibfield  {author} {\bibinfo {author} {\bibfnamefont {F.}~\bibnamefont
  {Schindler}}\ and\ \bibinfo {author} {\bibfnamefont {A.}~\bibnamefont
  {Prem}},\ }\href {\doibase 10.1103/PhysRevB.104.L161106} {\bibfield
  {journal} {\bibinfo  {journal} {Physical Review B}\ }\textbf {\bibinfo
  {volume} {104}},\ \bibinfo {pages} {L161106} (\bibinfo {year}
  {2021})}\BibitemShut {NoStop}%
\bibitem [{\citenamefont {Vecsei}\ \emph {et~al.}(2021)\citenamefont {Vecsei},
  \citenamefont {Denner}, \citenamefont {Neupert},\ and\ \citenamefont
  {Schindler}}]{r140}%
  \BibitemOpen
  \bibfield  {author} {\bibinfo {author} {\bibfnamefont {P.~M.}\ \bibnamefont
  {Vecsei}}, \bibinfo {author} {\bibfnamefont {M.~M.}\ \bibnamefont {Denner}},
  \bibinfo {author} {\bibfnamefont {T.}~\bibnamefont {Neupert}}, \ and\
  \bibinfo {author} {\bibfnamefont {F.}~\bibnamefont {Schindler}},\ }\href
  {\doibase 10.1103/PhysRevB.103.L201114} {\bibfield  {journal} {\bibinfo
  {journal} {Physical Review B}\ }\textbf {\bibinfo {volume} {103}},\ \bibinfo
  {pages} {L201114} (\bibinfo {year} {2021})}\BibitemShut {NoStop}%
\bibitem [{\citenamefont {Hu}\ and\ \citenamefont {Zhao}(2021)}]{r142}%
  \BibitemOpen
  \bibfield  {author} {\bibinfo {author} {\bibfnamefont {H.}~\bibnamefont
  {Hu}}\ and\ \bibinfo {author} {\bibfnamefont {E.}~\bibnamefont {Zhao}},\
  }\href {\doibase 10.1103/PhysRevLett.126.010401} {\bibfield  {journal}
  {\bibinfo  {journal} {Phys Rev Lett}\ }\textbf {\bibinfo {volume} {126}},\
  \bibinfo {pages} {010401} (\bibinfo {year} {2021})}\BibitemShut {NoStop}%
\bibitem [{\citenamefont {Yang}(2020)}]{r64}%
  \BibitemOpen
  \bibfield  {author} {\bibinfo {author} {\bibfnamefont {Z.}~\bibnamefont
  {Yang}},\ }\href {https://arxiv.org/abs/2012.03333} {\  (\bibinfo {year}
  {2020})},\ \Eprint {http://arxiv.org/abs/arXiv:2012.03333} {arXiv:2012.03333}
  \BibitemShut {NoStop}%
\bibitem [{\citenamefont {Ghatak}\ \emph {et~al.}(2020)\citenamefont {Ghatak},
  \citenamefont {Brandenbourger}, \citenamefont {Wezel},\ and\ \citenamefont
  {Coulais}}]{r57}%
  \BibitemOpen
  \bibfield  {author} {\bibinfo {author} {\bibfnamefont {A.}~\bibnamefont
  {Ghatak}}, \bibinfo {author} {\bibfnamefont {M.}~\bibnamefont
  {Brandenbourger}}, \bibinfo {author} {\bibfnamefont {J.}~\bibnamefont
  {Wezel}}, \ and\ \bibinfo {author} {\bibfnamefont {C.}~\bibnamefont
  {Coulais}},\ }\href {\doibase 10.1073/pnas.2010580117} {\bibfield  {journal}
  {\bibinfo  {journal}
  {\href{https://www.pnas.org/content/117/47/29561}{Proceedings of the National
  Academy of Sciences}}\ }\textbf {\bibinfo {volume} {117}},\ \bibinfo {pages}
  {29561} (\bibinfo {year} {2020})}\BibitemShut {NoStop}%
\bibitem [{\citenamefont {Xiao}\ \emph {et~al.}(2021)\citenamefont {Xiao},
  \citenamefont {Deng}, \citenamefont {Wang}, \citenamefont {Wang},
  \citenamefont {Yi},\ and\ \citenamefont {Xue}}]{r54}%
  \BibitemOpen
  \bibfield  {author} {\bibinfo {author} {\bibfnamefont {L.}~\bibnamefont
  {Xiao}}, \bibinfo {author} {\bibfnamefont {T.}~\bibnamefont {Deng}}, \bibinfo
  {author} {\bibfnamefont {K.}~\bibnamefont {Wang}}, \bibinfo {author}
  {\bibfnamefont {Z.}~\bibnamefont {Wang}}, \bibinfo {author} {\bibfnamefont
  {W.}~\bibnamefont {Yi}}, \ and\ \bibinfo {author} {\bibfnamefont
  {P.}~\bibnamefont {Xue}},\ }\href {\doibase 10.1103/PhysRevLett.126.230402}
  {\bibfield  {journal} {\bibinfo  {journal}
  {\href{https://journals.aps.org/prl/abstract/10.1103/PhysRevLett.126.230402}{Phys.
  Rev. Lett}}\ }\textbf {\bibinfo {volume} {126}},\ \bibinfo {pages} {230402}
  (\bibinfo {year} {2021})}\BibitemShut {NoStop}%
\bibitem [{\citenamefont {Palacios}\ \emph {et~al.}(2021)\citenamefont
  {Palacios}, \citenamefont {Tchoumakov}, \citenamefont {Guix}, \citenamefont
  {Pagonabarraga}, \citenamefont {S{\'a}nchez},\ and\ \citenamefont
  {Grushin}}]{r94}%
  \BibitemOpen
  \bibfield  {author} {\bibinfo {author} {\bibfnamefont {L.~S.}\ \bibnamefont
  {Palacios}}, \bibinfo {author} {\bibfnamefont {S.}~\bibnamefont
  {Tchoumakov}}, \bibinfo {author} {\bibfnamefont {M.}~\bibnamefont {Guix}},
  \bibinfo {author} {\bibfnamefont {I.}~\bibnamefont {Pagonabarraga}}, \bibinfo
  {author} {\bibfnamefont {S.}~\bibnamefont {S{\'a}nchez}}, \ and\ \bibinfo
  {author} {\bibfnamefont {A.~G.}\ \bibnamefont {Grushin}},\ }\href {\doibase
  10.1038/s41467-021-24948-2} {\bibfield  {journal} {\bibinfo  {journal} {Nat.
  Commun.}\ }\textbf {\bibinfo {volume} {12}},\ \bibinfo {pages} {4691}
  (\bibinfo {year} {2021})}\BibitemShut {NoStop}%
\bibitem [{\citenamefont {Zhang}\ \emph
  {et~al.}(2021{\natexlab{b}})\citenamefont {Zhang}, \citenamefont {Tian},
  \citenamefont {Jiang}, \citenamefont {Lu},\ and\ \citenamefont {Chen}}]{r95}%
  \BibitemOpen
  \bibfield  {author} {\bibinfo {author} {\bibfnamefont {X.}~\bibnamefont
  {Zhang}}, \bibinfo {author} {\bibfnamefont {Y.}~\bibnamefont {Tian}},
  \bibinfo {author} {\bibfnamefont {J.-H.}\ \bibnamefont {Jiang}}, \bibinfo
  {author} {\bibfnamefont {M.-H.}\ \bibnamefont {Lu}}, \ and\ \bibinfo {author}
  {\bibfnamefont {Y.-F.}\ \bibnamefont {Chen}},\ }\href
  {https://arxiv.org/abs/2102.09825} {\  (\bibinfo {year}
  {2021}{\natexlab{b}})},\ \Eprint {http://arxiv.org/abs/arXiv:2102.09825}
  {arXiv:2102.09825} \BibitemShut {NoStop}%
\bibitem [{\citenamefont {Zou}\ \emph {et~al.}(2021)\citenamefont {Zou},
  \citenamefont {Chen}, \citenamefont {He}, \citenamefont {Bao}, \citenamefont
  {Lee}, \citenamefont {Sun},\ and\ \citenamefont {Zhang}}]{r96}%
  \BibitemOpen
  \bibfield  {author} {\bibinfo {author} {\bibfnamefont {D.}~\bibnamefont
  {Zou}}, \bibinfo {author} {\bibfnamefont {T.}~\bibnamefont {Chen}}, \bibinfo
  {author} {\bibfnamefont {W.}~\bibnamefont {He}}, \bibinfo {author}
  {\bibfnamefont {J.}~\bibnamefont {Bao}}, \bibinfo {author} {\bibfnamefont
  {C.~H.}\ \bibnamefont {Lee}}, \bibinfo {author} {\bibfnamefont
  {H.}~\bibnamefont {Sun}}, \ and\ \bibinfo {author} {\bibfnamefont
  {X.}~\bibnamefont {Zhang}},\ }\href {https://arxiv.org/abs/2104.11260} {\
  (\bibinfo {year} {2021})},\ \Eprint {http://arxiv.org/abs/arXiv:2104.11260}
  {arXiv:2104.11260} \BibitemShut {NoStop}%
\bibitem [{\citenamefont {Zhang}\ \emph
  {et~al.}(2021{\natexlab{c}})\citenamefont {Zhang}, \citenamefont {Yang},
  \citenamefont {Ge}, \citenamefont {Guan}, \citenamefont {Chen}, \citenamefont
  {Yan}, \citenamefont {Chen}, \citenamefont {Xi}, \citenamefont {Li},
  \citenamefont {Jia}, \citenamefont {Yuan}, \citenamefont {Sun}, \citenamefont
  {Chen},\ and\ \citenamefont {Zhang}}]{r98}%
  \BibitemOpen
  \bibfield  {author} {\bibinfo {author} {\bibfnamefont {L.}~\bibnamefont
  {Zhang}}, \bibinfo {author} {\bibfnamefont {Y.}~\bibnamefont {Yang}},
  \bibinfo {author} {\bibfnamefont {Y.}~\bibnamefont {Ge}}, \bibinfo {author}
  {\bibfnamefont {Y.-j.}\ \bibnamefont {Guan}}, \bibinfo {author}
  {\bibfnamefont {Q.}~\bibnamefont {Chen}}, \bibinfo {author} {\bibfnamefont
  {Q.}~\bibnamefont {Yan}}, \bibinfo {author} {\bibfnamefont {F.}~\bibnamefont
  {Chen}}, \bibinfo {author} {\bibfnamefont {R.~.}\ \bibnamefont {Xi}},
  \bibinfo {author} {\bibfnamefont {Y.}~\bibnamefont {Li}}, \bibinfo {author}
  {\bibfnamefont {D.}~\bibnamefont {Jia}}, \bibinfo {author} {\bibfnamefont
  {S.-q.}\ \bibnamefont {Yuan}}, \bibinfo {author} {\bibfnamefont {H.-x.}\
  \bibnamefont {Sun}}, \bibinfo {author} {\bibfnamefont {H.}~\bibnamefont
  {Chen}}, \ and\ \bibinfo {author} {\bibfnamefont {B.}~\bibnamefont {Zhang}},\
  }\href {https://arxiv.org/abs/2104.08844} {\  (\bibinfo {year}
  {2021}{\natexlab{c}})},\ \Eprint {http://arxiv.org/abs/arXiv:2104.08844}
  {arXiv:2104.08844} \BibitemShut {NoStop}%
\bibitem [{\citenamefont {Zhai}(2012)}]{r14}%
  \BibitemOpen
  \bibfield  {author} {\bibinfo {author} {\bibfnamefont {H.}~\bibnamefont
  {Zhai}},\ }\href {\doibase 10.1142/s0217979212300010} {\bibfield  {journal}
  {\bibinfo  {journal}
  {\href{https://www.worldscientific.com/doi/abs/10.1142/S0217979212300010}{International
  Journal of Modern Physics B}}\ }\textbf {\bibinfo {volume} {26}},\ \bibinfo
  {pages} {1230001} (\bibinfo {year} {2012})}\BibitemShut {NoStop}%
\bibitem [{SM3()}]{SM3}%
  \BibitemOpen
  \href@noop {} {\bibinfo  {journal} {Here, we define the recoil momentum
  $\hbar q_{r}$ and recoil energy $E_{r}=\hbar^{2}q_{r}^{2}/{2m}$ as the units
  of momentum and energy, respectively. The corresponding derivation of the
  dimensionless Hamiltonian can be found in the supplemental material}\
  }\BibitemShut {NoStop}%
\bibitem [{SM1({\natexlab{a}})}]{SM12}%
  \BibitemOpen
\bibfield  {journal} {  }\href@noop {} {\bibfield  {journal} {\bibinfo
  {journal} {The degeneracy splitting proposed here is for the whole spectrum.
  There may be few intersection points in the spectrum at which no splitting
  occurs, so that the state formed by these corresponding eigenstates may not
  be localized at the boundary, that is, an extended state.}\ }
  ({\natexlab{a}})}\BibitemShut {NoStop}%
\bibitem [{SM5()}]{SM5}%
  \BibitemOpen
  \href@noop {} {\bibinfo  {journal} {More precisely, $\langle
  s_z\rangle=\langle u^R_\pm(k)|s_z|u^R_\pm(k)\rangle$, where
  $|u^R_\pm(k)\rangle$ is the right-eigenstate~\cite{Brody_2013} of
  Eq.~\ref{Eq1} with eigenvalue $E_\pm(k)$}\ }\BibitemShut {NoStop}%
\bibitem [{SM1({\natexlab{b}})}]{SM1}%
  \BibitemOpen
\bibfield  {journal} {  }\href@noop {} {\bibfield  {journal} {\bibinfo
  {journal} {Here we note that the wave functions we plotted are
  $|\psi_{E_i,\uparrow}(x)|^2+|\psi_{E_i, \downarrow}(x)|^2$, where
  $\psi_{E_i,\uparrow/\downarrow}(x)$ are the spin up/down components of the
  eigenfunction whose eigenvalue is $E_i$}\ } ({\natexlab{b}})}\BibitemShut
  {NoStop}%
\bibitem [{SM2()}]{SM2}%
  \BibitemOpen
  \href@noop {} {\bibinfo  {journal} {See Supplemental Material for details}\
  }\BibitemShut {NoStop}%
\bibitem [{SM7()}]{SM7}%
  \BibitemOpen
\bibfield  {journal} {  }\href@noop {} {\bibinfo  {journal} {For example, here
  the points $A/C$ represent the position at which $\Im [k_+]=1/100$; and $D/J$
  $\Im[k_-]=1/100$}\ }\BibitemShut {NoStop}%
\bibitem [{\citenamefont {Kawabata}\ \emph {et~al.}(2019)\citenamefont
  {Kawabata}, \citenamefont {Shiozaki}, \citenamefont {Ueda},\ and\
  \citenamefont {Sato}}]{r67}%
  \BibitemOpen
\bibfield  {journal} {  }\bibfield  {author} {\bibinfo {author} {\bibfnamefont
  {K.}~\bibnamefont {Kawabata}}, \bibinfo {author} {\bibfnamefont
  {K.}~\bibnamefont {Shiozaki}}, \bibinfo {author} {\bibfnamefont
  {M.}~\bibnamefont {Ueda}}, \ and\ \bibinfo {author} {\bibfnamefont
  {M.}~\bibnamefont {Sato}},\ }\href {\doibase 10.1103/PhysRevX.9.041015}
  {\bibfield  {journal} {\bibinfo  {journal} {Physical Review X}\ }\textbf
  {\bibinfo {volume} {9}},\ \bibinfo {pages} {041015} (\bibinfo {year}
  {2019})}\BibitemShut {NoStop}%
\bibitem [{\citenamefont {Gong}\ \emph {et~al.}(2018)\citenamefont {Gong},
  \citenamefont {Ashida}, \citenamefont {Kawabata}, \citenamefont {Takasan},
  \citenamefont {Higashikawa},\ and\ \citenamefont {Ueda}}]{PhysRevX.8.031079}%
  \BibitemOpen
  \bibfield  {author} {\bibinfo {author} {\bibfnamefont {Z.}~\bibnamefont
  {Gong}}, \bibinfo {author} {\bibfnamefont {Y.}~\bibnamefont {Ashida}},
  \bibinfo {author} {\bibfnamefont {K.}~\bibnamefont {Kawabata}}, \bibinfo
  {author} {\bibfnamefont {K.}~\bibnamefont {Takasan}}, \bibinfo {author}
  {\bibfnamefont {S.}~\bibnamefont {Higashikawa}}, \ and\ \bibinfo {author}
  {\bibfnamefont {M.}~\bibnamefont {Ueda}},\ }\href {\doibase
  10.1103/PhysRevX.8.031079} {\bibfield  {journal} {\bibinfo  {journal} {Phys.
  Rev. X}\ }\textbf {\bibinfo {volume} {8}},\ \bibinfo {pages} {031079}
  (\bibinfo {year} {2018})}\BibitemShut {NoStop}%
\bibitem [{SM9()}]{SM9}%
  \BibitemOpen
  \href@noop {} {\bibinfo  {journal} {In the SM, we use an exact solvable model
  to demonstrate this point}\ }\BibitemShut {NoStop}%
\bibitem [{SM1({\natexlab{c}})}]{SM11}%
  \BibitemOpen
\bibfield  {journal} {  }\href@noop {} {\bibfield  {journal} {\bibinfo
  {journal} {It should be noted that our theory is consistent with the current
  experimental measurements of Ref.~\cite{r12}. Although the momentum resolved
  Rabi oscillation detected in Ref.~\cite{r12} agrees with the free particle
  spectral, we cannot say the system does not have NHSE. The reason is that the
  NHSE cannot be detected by the short-time correlations far away from the
  boundary. In the supplementary material~\cite{SM2}, we provide a detailed
  discussion and simulations to demonstrate this point}\ }
  ({\natexlab{c}})}\BibitemShut {NoStop}%
\bibitem [{SM1({\natexlab{d}})}]{SM10}%
  \BibitemOpen
  \href@noop {} {\bibfield  {journal} {\bibinfo  {journal} {Experimentally,
  since the box potential has been realized in Ref.~\cite{r148s,r149s}, we
  believe that the measurement of the distribution of fermion density over time
  which is often used in ultracold atom technology~\cite{r31,r121,r38}, can be
  used to observe the DSE}\ } ({\natexlab{d}})}\BibitemShut {NoStop}%
\bibitem [{\citenamefont {Brody}(2013)}]{Brody_2013}%
  \BibitemOpen
  \bibfield  {author} {\bibinfo {author} {\bibfnamefont {D.~C.}\ \bibnamefont
  {Brody}},\ }\href {\doibase 10.1088/1751-8113/47/3/035305} {\bibfield
  {journal} {\bibinfo  {journal} {Journal of Physics A: Mathematical and
  Theoretical}\ }\textbf {\bibinfo {volume} {47}},\ \bibinfo {pages} {035305}
  (\bibinfo {year} {2013})}\BibitemShut {NoStop}%
\bibitem [{\citenamefont {Gaunt}\ \emph {et~al.}(2013)\citenamefont {Gaunt},
  \citenamefont {Schmidutz}, \citenamefont {Gotlibovych}, \citenamefont
  {Smith},\ and\ \citenamefont {Hadzibabic}}]{r148s}%
  \BibitemOpen
  \bibfield  {author} {\bibinfo {author} {\bibfnamefont {A.~L.}\ \bibnamefont
  {Gaunt}}, \bibinfo {author} {\bibfnamefont {T.~F.}\ \bibnamefont
  {Schmidutz}}, \bibinfo {author} {\bibfnamefont {I.}~\bibnamefont
  {Gotlibovych}}, \bibinfo {author} {\bibfnamefont {R.~P.}\ \bibnamefont
  {Smith}}, \ and\ \bibinfo {author} {\bibfnamefont {Z.}~\bibnamefont
  {Hadzibabic}},\ }\href {\doibase 10.1103/PhysRevLett.110.200406} {\bibfield
  {journal} {\bibinfo  {journal} {Phys. Rev. Lett.}\ }\textbf {\bibinfo
  {volume} {110}},\ \bibinfo {pages} {200406} (\bibinfo {year}
  {2013})}\BibitemShut {NoStop}%
\bibitem [{\citenamefont {Mukherjee}\ \emph {et~al.}(2017)\citenamefont
  {Mukherjee}, \citenamefont {Yan}, \citenamefont {Patel}, \citenamefont
  {Hadzibabic}, \citenamefont {Yefsah}, \citenamefont {Struck},\ and\
  \citenamefont {Zwierlein}}]{r149s}%
  \BibitemOpen
  \bibfield  {author} {\bibinfo {author} {\bibfnamefont {B.}~\bibnamefont
  {Mukherjee}}, \bibinfo {author} {\bibfnamefont {Z.}~\bibnamefont {Yan}},
  \bibinfo {author} {\bibfnamefont {P.~B.}\ \bibnamefont {Patel}}, \bibinfo
  {author} {\bibfnamefont {Z.}~\bibnamefont {Hadzibabic}}, \bibinfo {author}
  {\bibfnamefont {T.}~\bibnamefont {Yefsah}}, \bibinfo {author} {\bibfnamefont
  {J.}~\bibnamefont {Struck}}, \ and\ \bibinfo {author} {\bibfnamefont {M.~W.}\
  \bibnamefont {Zwierlein}},\ }\href {\doibase 10.1103/PhysRevLett.118.123401}
  {\bibfield  {journal} {\bibinfo  {journal} {Phys Rev Lett}\ }\textbf
  {\bibinfo {volume} {118}},\ \bibinfo {pages} {123401} (\bibinfo {year}
  {2017})}\BibitemShut {NoStop}%
\bibitem [{\citenamefont {Weitenberg}\ \emph {et~al.}(2011)\citenamefont
  {Weitenberg}, \citenamefont {Endres}, \citenamefont {Sherson}, \citenamefont
  {Cheneau}, \citenamefont {Schauss}, \citenamefont {Fukuhara}, \citenamefont
  {Bloch},\ and\ \citenamefont {Kuhr}}]{r121}%
  \BibitemOpen
  \bibfield  {author} {\bibinfo {author} {\bibfnamefont {C.}~\bibnamefont
  {Weitenberg}}, \bibinfo {author} {\bibfnamefont {M.}~\bibnamefont {Endres}},
  \bibinfo {author} {\bibfnamefont {J.~F.}\ \bibnamefont {Sherson}}, \bibinfo
  {author} {\bibfnamefont {M.}~\bibnamefont {Cheneau}}, \bibinfo {author}
  {\bibfnamefont {P.}~\bibnamefont {Schauss}}, \bibinfo {author} {\bibfnamefont
  {T.}~\bibnamefont {Fukuhara}}, \bibinfo {author} {\bibfnamefont
  {I.}~\bibnamefont {Bloch}}, \ and\ \bibinfo {author} {\bibfnamefont
  {S.}~\bibnamefont {Kuhr}},\ }\href {\doibase 10.1038/nature09827} {\bibfield
  {journal} {\bibinfo  {journal}
  {\href{https://www.nature.com/articles/nature09827}{Nature}}\ }\textbf
  {\bibinfo {volume} {471}},\ \bibinfo {pages} {319} (\bibinfo {year}
  {2011})}\BibitemShut {NoStop}%
\end{thebibliography}%
\bibliographystyle{apsrev4-1}

\onecolumngrid
\newpage


\section*{Supplementary material}

# Supplemental Material for " Theoretical Prediction of Non-Hermitian Skin-effect in Ultracold Atom Systems "

Sibo Guo[1,2,∗] Chenxiao Dong[1,2,∗] Fuchun Zhang[3,†] Jiangping Hu[1,2,3,‡] and Zhesen Yang[3§]

[1]*Beijing National Laboratory for Condensed Matter Physics,*
*and Institute of Physics, Chinese Academy of Sciences, Beijing 100190, China*
[2]*School of Physical Sciences, University of Chinese Academy of Sciences, Beijing 100049, China and*
[3]*Kavli Institute for Theoretical Sciences, University of Chinese Academy of Sciences, Beijing 100190, China*
(Dated: October 1, 2022)

## SI. THE NUMERICAL RESULTS UNDER SOME REALISTIC POTENTIALS

In this section, we discuss the robustness of non-Hermitian skin effects (NHSE) to different trapping potentials and make a simple comparison with the corresponding results measured in Ref. [1]. This section is organized as follows. In Sec.I-A, we give the numerical results for other potential situations that are more practical in the experiment [2–4]. Then we calculate the Rabi oscillation in different potentials with the Gaussian wave packet as the initial state to show that our theoretical prediction is consistent with the measurement in Ref. [1] in Sec.I-B.

### SI-A. The robustness of NHSE

Here we discuss the existence of skin modes in three different trapping potentials, which are relatively common in experiments [2–4]. For the convenience of the following discussion, we write down our static Schrödinger equation

$$[H(\hat{k}) + V(x)]\psi_E(x) = E\psi_E(x), \tag{1}$$

and make the parameters dimensionless in energy units $E_r = \frac{\hbar^2 q_r^2}{2m}$ and momentum units $\hbar q_r$, so the corresponding dimensionless Hamiltonian is

$$H(\hat{k}') + V(x) = \begin{pmatrix} \left(\hat{k}' - 1\right)^2 + \delta/2 - i\Gamma_\uparrow/2 + V(x) & \Omega_R/2 \\ \Omega_R/2 & \left(\hat{k}' + 1\right)^2 - \delta/2 - i\Gamma_\downarrow/2 + V(x) \end{pmatrix}, \tag{2}$$

where $\hat{k}' = -i\partial_x/q_r$.

As shown in FIG. S1 (a)-(c), here we have plotted the eigenvalues and several eigenfunctions of Eq. 1 under the following three types of trapping potentials, i.e.

$$V_a(x) = -V_0 e^{-\frac{x^2}{2\sigma}} \ , \quad V_b(x) = \begin{cases} -V_0 + ax^2, & -20 < x < 20 \\ 0, & \text{otherwise} \end{cases} \ , \quad V_c(x) = \begin{cases} -V_0, & -20 < x < 20 \\ 0, & \text{otherwise} \end{cases} \ , \tag{3}$$

where $V_0 = 1.5$, $\sigma = 50$, $a = 1.5/400$, and $V_a$, $V_b$ and $V_c$ represent the Gaussian potential, the harmonic potential and finite deep square well, respectively. The forms of the trapping potential are represented by the gray regions in FIG. S1 (a2/a4), (b2/b4), (c2/c4), respectively. Here $V_a(x)$ can be regarded as the realistic trapping potential in experiments. In each row of FIG. S1, the left two sub-figures represent the case with NHSE (that is $(\delta, \Omega_R, \Gamma_\uparrow, \Gamma_\downarrow) = (4, 9/2, 6, 6/13)$ which is the same as the one calculated in the main text), while the right two sub-figures represent the case without NHSE (that is $(\delta, \Omega_R, \Gamma_\uparrow, \Gamma_\downarrow) = (0, 1 + 2i, 0, 0)$, where only the imaginary part of $\Omega_R$ contributes to the non-Hermitian term). From these numerical results, one can find that the NHSE is robust to the external trapping potentials.

### SI-B. Comparison with Ref. [1]

It can be seen from the Ref. [1] that the results measured by the author in the experiment are consistent with the free space theory, which may mean that the system does not have NHSE, or NHSE does not seem to be robust to external potential, that is, it is inconsistent with our theoretical prediction. In fact, the experimental results in Ref. [1] do not contradict our theory. Next, we will demonstrate that our theory is consistent with Ref. [1].

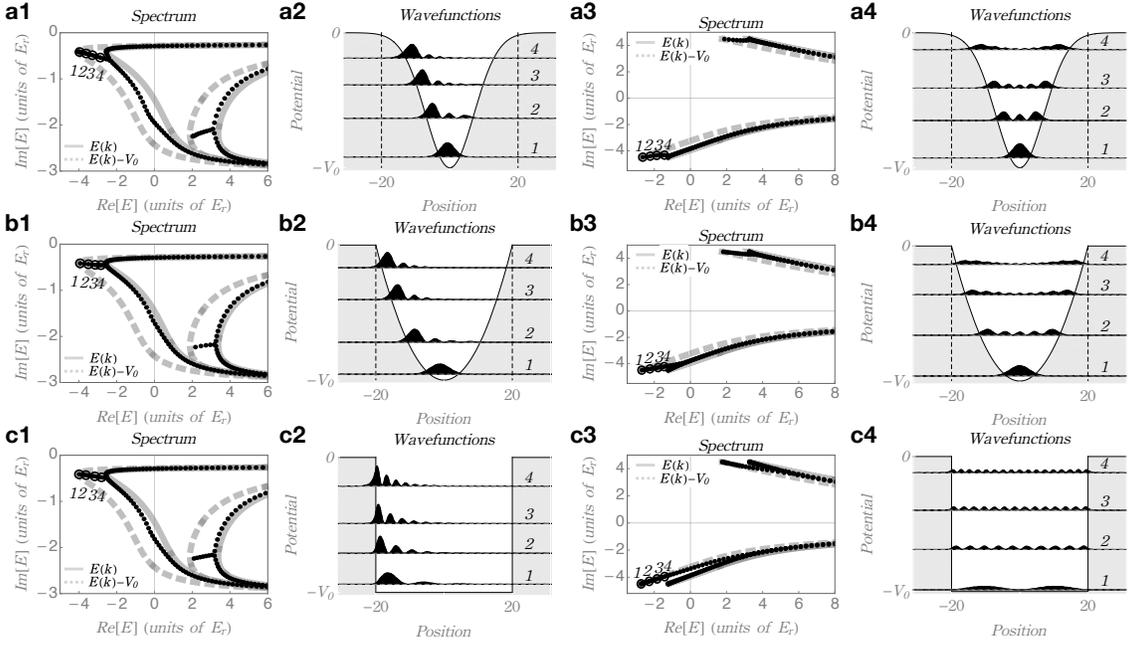

FIG. S1. **The robustness of NHSE to different trapping potentials.** The first column shows the energy spectrum of the system, where the gray solid/dotted lines represent the energy spectrum under the free case with the potential constant 0 and -1.5, respectively, and the black dots represents the open boundary spectrum under the potential $V_a$, $V_b$ and $V_c$, respectively. The second column shows that the distribution of the eigenstates $(|\psi_\uparrow(x)|^2 + |\psi_\downarrow(x)|^2)$ of the system in real space, where $1, 2, 3$ and $4$ correspond to the four cycles in the first column, respectively. Similarly, the third and fourth column also describes the spectrum and wave function. The difference between them is that the non-Hermitian terms in the first two columns and the last two columns come from the Raman coupling $\Omega_R \in \mathbb{C}$ and the loss $\Gamma_{\uparrow,\downarrow} \neq 0$, respectively. The result is that the spectra under the open boundary and periodic boundary in the third column are almost identical, which also means that the eigenstates of the system are extended as shown in the figure in the fourth column.

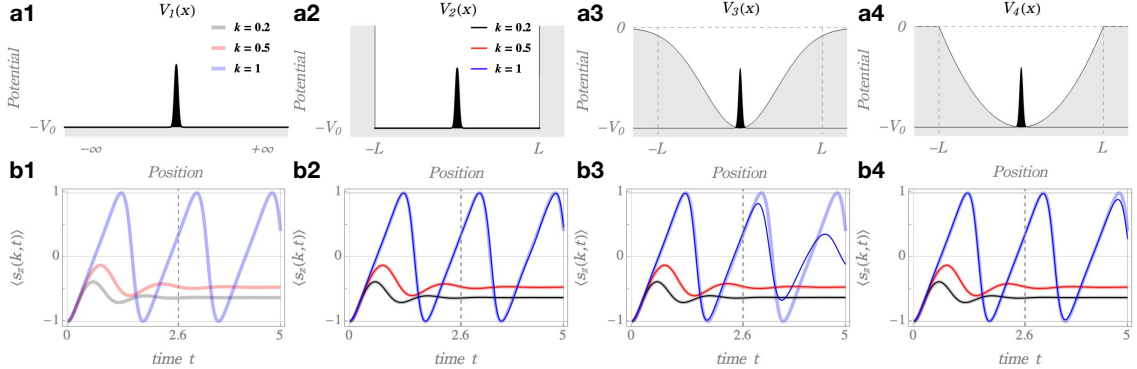

FIG. S2. **Rabi oscillations under different external potentials.** In (a1)-(a4), the black and gray parts represent the initial wave function and the specific shape of the external potential corresponding to Eq. 4, respectively. In (b1)-(b4), blue, red and black represent the Rabi oscillation curves with initial momentum of $-0.2$, $-0.4$ and $-1$ respectively, and the dashed black lines represent the maximum time that the experiment has taken place. In (b2)-(b4), the results in free space corresponding to (b1) are also drawn, so it can be seen that the oscillations in the four potentials are almost the same in the time range from 0 to 2.6.

To demonstrate this point, we calculate the momentum resolved Rabi oscillation with and without different trapping potentials. As shown in FIG. S2 (a), the trapping potential $V(x)$ is chosen to be

$$V_1(x) = -V_0 \ , \quad V_2(x) = \begin{cases} -V_0, & -L < x < L \\ \infty, & \text{otherwise} \end{cases} \ , \quad V_3(x) = -V_0 e^{-\frac{x^2}{2\sigma}} \ , \quad V_4(x) = \begin{cases} -V_0 + ax^2, & -L < x < L \\ 0, & \text{otherwise} \end{cases} \ . \quad (4)$$

Based on Ref. [5, 6], the length scale of the trapping potential is about $2 \times 8\mu m$, which is about $2 \times 55.56$ in the units of $1/q_r$. Without loss of generality, we choose $L = 50$ here, and other parameters are chosen to be $V_0 = 1.5$, $\sigma = 500$,

$a = 0.0006$. Another important quantity in the experiment is the longest evolution time, which is about $300\mu s$, that is, about $2.655$ in the units of $\hbar/E_r$. This value has been represented by the dashed lines in FIG. S2 (b). To simulate the experimental setups, we start with a Gaussian wave packet as an initial state, whose form is

$$\psi(x, t=0) = Ne^{-x^2/2} \otimes | \downarrow \rangle. \tag{5}$$

By solving the differential equation

$$i\partial_t \psi(x, t) = \Big[ H(\hat{k}) + V(x) \Big] \psi(x, t), \tag{6}$$

one can obtain $\psi(x, t) = \langle x|\psi(t)\rangle$. Based on this, the momentum resolved Rabi oscillation detected in Ref. [36] can be expressed as

$$s_z(k, t) = \frac{|\phi_{k\uparrow}(t)|^2 - |\phi_{k\downarrow}(t)|^2}{|\phi_{k\uparrow}(t)|^2 + |\phi_{k\downarrow}(t)|^2}, \tag{7}$$

where $\phi_{k\uparrow}(t) = \langle k \uparrow |\psi(t)\rangle$ and $\phi_{k\downarrow}(t) = \langle k \downarrow |\psi(t)\rangle$. As shown in FIG. S2 (b), the Rabbi oscillation with different resolved momentum are plotted. One can find that the result is not sensitive to the trapping potentials.

As discussed in the main text, the reason for the above phenomenon is that the NHSE cannot be detected by the *short time correlations* far away from the boundary. This can be understood from the fact that when the wave packet is excited far away from the boundary, its short-time evolution cannot feel the presence or absence of boundaries. To

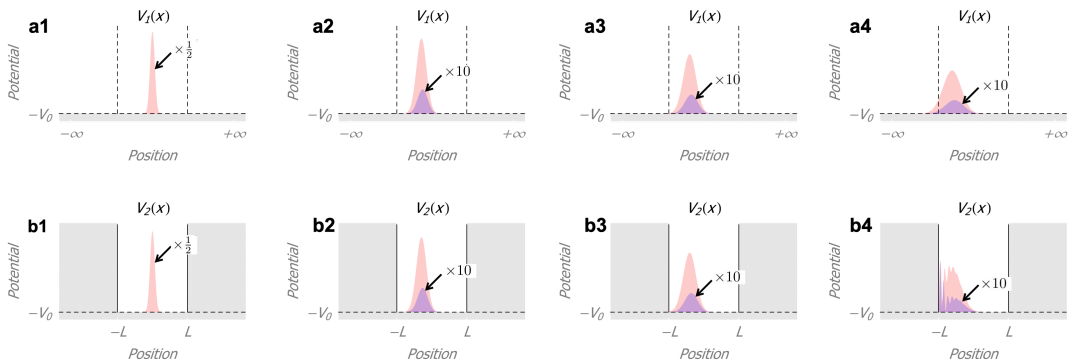

FIG. S3. **Evolution of wave packet in the free case and the infinite deep potential well.** (a) and (b) represents the evolution of the wave packet in the free case and the infinite deep potential well, where (a1)-(a4)/(b1)-(b4) correspond to time $t = 0, 2, 2.7, 4$, and the pink and purple components represent the spin-down and -up, respectively. The numbers $1/2$ and $10$ in (a) and (b) represent the proportion by which the respective spin components are reduced or amplified.

be more precise, considering the time evolution of the wave packet $\psi(x, t=0) = N\exp[-(x-x_0)^2 + ip_0(x-x_0)] \otimes | \downarrow \rangle$ $(x_0 = 0$ and $p_0 = -2)$ under the potential $V_1(x)$ and $V_2(x)$ in Eq. 4, FIG. S3 shows the comparison between the corresponding dynamics. One can find that, once of the wave packet stays far away from the boundary, the corresponding short time dynamics is similar to the free particle case, which is in line with physical intuitions. This inspires us to define a boundary-reaching time $t_s$ (that is FIG. S3 b3), before which $(t \leq t_s)$, we have

$$\begin{aligned} |\psi_1(t)\rangle &:= e^{-i[H(k)+V_1(x)]t/\hbar}|\psi(t=0)\rangle \\ &\simeq |\psi_2(t)\rangle := e^{-i[H(k)+V_2(x)]t/\hbar}|\psi(t=0)\rangle, \end{aligned} \tag{8}$$

where $V_1(x)$ and $V_2(x)$ are shown in Eq. 4 or FIG. S3. Since $|\psi_1(t)\rangle \simeq |\psi_2(t)\rangle$, the corresponding momentum resolved Rabi oscillation must be the similar, i.e.

$$\begin{aligned} s_{1z}(k, t) &= \frac{|\langle k \uparrow |\psi_1(t)\rangle|^2 - |\langle k \downarrow |\psi_1(t)\rangle|^2}{|\langle k \uparrow |\psi_1(t)\rangle|^2 + |\langle k \downarrow |\psi_1(t)\rangle|^2} \\ &\simeq s_{2z}(k, t) = \frac{|\langle k \uparrow |\psi_2(t)\rangle|^2 - |\langle k \downarrow |\psi_2(t)\rangle|^2}{|\langle k \uparrow |\psi_2(t)\rangle|^2 + |\langle k \downarrow |\psi_2(t)\rangle|^2}. \end{aligned} \tag{9}$$

Therefore, the above discussion proves that our theory is in agreement with the experimental results in Ref. [1].

## SII. GENERALIZED WAVE VECTOR THEORY

In this section, we give a detailed elaboration of generalized wave vector (GWV) theory. This section is organized as follows. In Sec.II-A, we first review the generalized Brillouin zone (GBZ) theory in the lattice model and generalize it to the continuum model, which we call the GWV theory. Next in Sec.II-B we give the more generalized and more strict GWV theory for multi-band cases. Then in Sec.II-C we apply our two-band model as an example to give the GWV condition used to determine the continuum spectrum and the detailed proof of the GWV theory. Finally, we demonstrate briefly that the coordinate components of eigenstates have the same exponential decay term in Sec.II-D.

### SII-A. Single-band GWV theory

The one-dimensional GBZ theory describes the asymptotic behavior of the lattice model with open boundary condition (OBC) as the size of the chain becomes larger. We consider the following one-dimensional single-band lattice model

$$H = \sum_{n=1}^{L} \sum_{j=-p}^{q} a_j \hat{c}_j^\dagger \hat{c}_{n+j}. \tag{10}$$

We define the characteristic polynomial as

$$f(\beta, E) = \sum_{j=-p}^{q} a_j \beta^j - E. \tag{11}$$

In the case of no special symmetry, the GBZ condition $|\beta_p| = |\beta_{p+1}|$ can be obtained under the OBC for large $L$, where $\beta_j$ is the $j$th solution of $f(\beta, E) = 0$ with the order $|\beta_1| \leq |\beta_2| \leq \cdots \leq |\beta_{p+q}|$. Then one can get the continuum bands combining GBZ condition and $f(\beta, E) = 0$.

Here, we propose a similar theory named GWV theory which describes the asymptotic behavior of one-dimensional infinitely deep square potential well as the length of the well becomes larger. Let's take the single-band model as an example to illustrate this GWV theory. Considering the well-posed problem [7] described by following stationary equation under one-dimensional infinitely deep square potential well

$$E\psi(x) = \sum_{j=0}^{n} a_j \left(-i\frac{\partial}{\partial x}\right)^j \psi(x), \tag{12}$$

and the boundary conditions are

$$\begin{cases} \psi(x)|_{x=0} = 0 \\ \frac{d}{dx}\psi(x)|_{x=0} = 0 \\ \vdots \\ \frac{d^{p-1}}{d^{p-1}x}\psi(x)|_{x=0} = 0 \end{cases}, \quad \begin{cases} \psi(x)|_{x=L} = 0 \\ \frac{d}{dx}\psi(x)|_{x=L} = 0 \\ \vdots \\ \frac{d^{q-1}}{d^{q-1}x}\psi(x)|_{x=L} = 0 \end{cases}, \tag{13}$$

where $a_j \in \mathbb{C}$, $E$ is the eigenvalue, $p + q = n$ (the requirement that the number of boundary conditions is given by the well-posed problem [7]). Particularly when $n = 2$, Eq.12 becomes stationary Schrödinger equation. Then the corresponding characteristic polynomial is

$$f(k, E) = \sum_{j=0}^{n} a_j k^j - E. \tag{14}$$

The GWV theory points out that for large $L$, the asymptotic solution $(E_n, \psi_n(x))$ of the stationary problem gives the GWV condition $\text{Im}(k_p) = \text{Im}(k_{p+1})$, where $k_j$ is the $j$th solution of $f(k, E) = 0$ with the order $\text{Im}(k_1) \geqslant \text{Im}(k_2) \geqslant \cdots \geqslant \text{Im}(k_n)$. And the continuum energy spectrum can be obtained by using GWV condition and Eq.14.

## SII-B. Multi-band GWV theory

As single-band GBZ theory is generalized to the multi-band GBZ theory, the single-band GWV theory can be generalized to the multi-band GWV theory. Here we give a more mathematical form of the multi-band GWV theory.

We consider a well-posed system of high-order linear differential equations:

$$\begin{cases} \frac{\partial}{\partial t}\psi_1(x,t) = \sum_{l=1}^m \sum_{j=0}^n a_{l,1,j}\left(-i\frac{\partial}{\partial x}\right)^j \psi_l(x,t) \\ \vdots \\ \frac{\partial}{\partial t}\psi_m(x,t) = \sum_{l=1}^m \sum_{j=0}^n a_{l,m,j}\left(-i\frac{\partial}{\partial x}\right)^j \psi_l(x,t) \end{cases}, \tag{15}$$

where $a_{l,l,j} \in \mathbb{C}$

Then we define its corresponding stationary equation

$$E\psi(x) = \begin{pmatrix} \sum_{j=0}^n a_{1,1,j}\left(-i\frac{\partial}{\partial x}\right)^j & \cdots & \sum_{j=0}^n a_{1,m,j}\left(-i\frac{\partial}{\partial x}\right)^j \\ \vdots & \vdots & \vdots \\ \sum_{j=0}^n a_{m,1,j}\left(-i\frac{\partial}{\partial x}\right)^j & \cdots & \sum_{j=0}^n a_{m,m,j}\left(-i\frac{\partial}{\partial x}\right)^j \end{pmatrix}\psi(x), \tag{16}$$

Here, $E \in \mathbb{C}$, called the eigenvalue,

$$\psi(x) = \begin{pmatrix} \psi_1(x) \\ \psi_2(x) \\ \vdots \\ \psi_m(x) \end{pmatrix}, \tag{17}$$

called the eigenstate.

The general solution of Eq.15 is composed of the solution of Eq.16:

$$\psi(x,t) = \sum_k c_k e^{iE_k t}\psi_{E_k}(x) \tag{18}$$

Here $c_k$ is the complex number determined by the boundary conditions, and $\psi_{E_k}(x)$ is a eigenstate solution of Eq.16 with the eigenvalue $E = E_k$.

The GWV theory gives the asymptotic solution of Eq.16 with the Dirichlet boundary conditions:

$$\begin{cases} \psi(x)|_{x=0} = 0 \\ \frac{d}{dx}\psi(x)|_{x=0} = 0 \\ \vdots \\ \frac{d^{p-1}}{d^{p-1}x}\psi(x)|_{x=0} = 0 \end{cases}, \quad \begin{cases} \psi(x)|_{x=L} = 0 \\ \frac{d}{dx}\psi(x)|_{x=L} = 0 \\ \vdots \\ \frac{d^{q-1}}{d^{q-1}x}\psi(x)|_{x=L} = 0 \end{cases} \tag{19}$$

where $p + q = n$ (the requirement that the number of boundary conditions is given by the well-posed problem [7]).

The corresponding characteristic polynomial follows

$$f(k,E) = \det\left[\begin{pmatrix} \sum_{j=0}^n a_{1,1,j}k^j & \cdots & \sum_{j=0}^n a_{1,m,j}k^j \\ \vdots & \vdots & \vdots \\ \sum_{j=0}^n a_{m,1,j}k^j & \cdots & \sum_{j=0}^n a_{m,m,j}k^j \end{pmatrix} - EI_{m,m}\right]. \tag{20}$$

Here $I_{m,m}$ is an identity matrix of order $m$.

The GWV theory points out that if $L$ is large and $f(k,E)$ is the $nm$ order polynomial about the variant $k$, the asymptotic solution $(E_l, \psi_l(x))$ of the stationary problem gives the GWV condition $\text{Im}(k_{mp}) = \text{Im}(k_{mp+1})$, where $k_j$ is the $j$th solution of $f(k,E) = 0$ with the order $\text{Im}(k_1) \geqslant \text{Im}(k_2) \geqslant \cdots \geqslant \text{Im}(k_{nm})$. And the continuum energy spectrum can be obtained by using GWV condition and Eq.14.

In the next section, we will give a concise proof method for the two-band model in the main text, which is also applicable to the other model but will not be repeated here.

**SII-C. Proof of the GWV condition for two-band model**

In this subsection, we will demonstrate GWV condition in the single-band model is also true for Eq.2 in the main text by using a similar method proposed in Ref. [8]. Plugging a particular solution $e^{ikx}(\phi_\uparrow, \phi_\downarrow)^T$ into eigenequation, we have

$$
\begin{cases}
[\frac{\hbar^2}{2m}(k - q_r)^2 + \frac{\delta}{2} - i\frac{\Gamma_\uparrow}{2}]\phi_\uparrow + \frac{\Omega_R}{2}\phi_\downarrow = E\phi_\uparrow \\
\frac{\Omega_R}{2}\phi_\uparrow + [\frac{\hbar^2}{2m}(k + q_r)^2 - \frac{\delta}{2} - i\frac{\Gamma_\downarrow}{2}]]\phi_\downarrow = E\phi_\downarrow
\end{cases}. \tag{21}
$$

From Eq. 21, we get the corresponding characteristic polynomial and the relation between $\phi_\uparrow$ and $\phi_\downarrow$

$$
f(k, E) \equiv \begin{vmatrix} \frac{\hbar^2}{2m}(k - q_r)^2 + \frac{\delta}{2} - i\frac{\Gamma_\uparrow}{2} - E & \frac{\Omega_R}{2} \\ \frac{\Omega_R}{2} & \frac{\hbar^2}{2m}(k + q_r)^2 - \frac{\delta}{2} - i\frac{\Gamma_\downarrow}{2} - E \end{vmatrix}, \tag{22}
$$

$$
\phi_\downarrow = -\frac{\Omega_R}{2} \frac{1}{\frac{\hbar^2}{2m}(k + q_r)^2 - \frac{\delta}{2} - i\frac{\Gamma_\downarrow}{2} - E} \phi_\uparrow \equiv g(k, E)\phi_\uparrow. \tag{23}
$$

We know that there are four $k$ values for fixed energy $E$ in Eq. 22, so we superimpose the four solutions to construct a general solution, that is,

$$
\psi(x) = e^{ik_1 x} \begin{pmatrix} \phi_\uparrow^{(1)} \\ \phi_\downarrow^{(1)} \end{pmatrix} + e^{ik_2 x} \begin{pmatrix} \phi_\uparrow^{(2)} \\ \phi_\downarrow^{(2)} \end{pmatrix} + e^{ik_3 x} \begin{pmatrix} \phi_\uparrow^{(3)} \\ \phi_\downarrow^{(3)} \end{pmatrix} + e^{ik_4 x} \begin{pmatrix} \phi_\uparrow^{(4)} \\ \phi_\downarrow^{(4)} \end{pmatrix}. \tag{24}
$$

Here, we suppose that the four $k$ values are not equal to each other. Plugging Eq. 24 into the OBC $\psi(x)|_{x=0} = \psi(x)|_{x=L} = 0$, we have

$$
\begin{cases}
\phi_\uparrow^{(1)} + \phi_\uparrow^{(2)} + \phi_\uparrow^{(3)} + \phi_\uparrow^{(4)} = 0, \\
\phi_\downarrow^{(1)} + \phi_\downarrow^{(2)} + \phi_\downarrow^{(3)} + \phi_\downarrow^{(4)} = 0, \\
e^{ik_1 L}\phi_\uparrow^{(1)} + e^{ik_2 L}\phi_\uparrow^{(2)} + e^{ik_3 L}\phi_\uparrow^{(3)} + e^{ik_4 L}\phi_\uparrow^{(4)} = 0, \\
e^{ik_1 L}\phi_\downarrow^{(1)} + e^{ik_2 L}\phi_\downarrow^{(2)} + e^{ik_3 L}\phi_\downarrow^{(3)} + e^{ik_4 L}\phi_\downarrow^{(4)} = 0.
\end{cases} \tag{25}
$$

The requirement of non-trivial solution gives the following condition

$$
\begin{vmatrix} 1 & 1 & 1 & 1 \\ g(k_1, E) & g(k_2, E) & g(k_3, E) & g(k_4, E) \\ e^{ik_1 L} & e^{ik_2 L} & e^{ik_3 L} & e^{ik_4 L} \\ e^{ik_1 L} g(k_1, E) & e^{ik_2 L} g(k_2, E) & e^{ik_3 L} g(k_3, E) & e^{ik_4 L} g(k_4, E) \end{vmatrix} = \sum_{\{j_1 j_2 j_3 j_4\} \in S_4} G(\{j_1 j_2 j_3 j_4\}, E) e^{ik_{j_3} L} e^{ik_{j_4} L} = 0, \tag{26}
$$

where $S_4$ represents all the permutations of $(1234)$, $G(\{j_1 j_2 j_3 j_4\}, E) = (-1)^{P(j_1 j_2 j_3 j_4)}(j_1)^0 g(k_{j_2}, E) g(k_{j_4}, E)$, $P$ represents permutation operation. If $\text{Im}[k_2] > \text{Im}[k_3]$, the leading term in Eq.26 is

$$
[G(\{1234\}, E) + G(\{2134\}, E) + G(\{1243\}, E) + G(\{2143\}, E)]e^{ik_3 L} e^{ik_4 L} = 0. \tag{27}
$$

Futher simplifing Eq.27 to

$$[g(k_2, E) - g(k_1, E)][g(k_4, E) - g(k_3, E)] = 0. \tag{28}$$

Next we demonstrate that $g(k_j, E) \neq g(k_m, E)$ for any $k_j \neq k_m$ by using contradiction. Supposing that there are $k_j$, $k_m$ ($k_j \neq k_m$) such that $g(k_j, E) = g(k_m, E)$, we have

$$k_m = -k_j - 2q_r. \tag{29}$$

In addition, $k_j$ and $k_m$ also correspond to the same energy E, so from Eq. 22 we have

$$\begin{cases} [\frac{\hbar^2}{2m}(k_j - q_r)^2 + \frac{\delta}{2} - i\frac{\Gamma_\uparrow}{2} - E][\frac{\hbar^2}{2m}(k_j + q_r)^2 + \frac{\delta}{2} - i\frac{\Gamma_\downarrow}{2} - E] = (\frac{\Omega_R}{2})^2 \\ [\frac{\hbar^2}{2m}(-k_j - 3q_r)^2 + \frac{\delta}{2} - i\frac{\Gamma_\uparrow}{2} - E][\frac{\hbar^2}{2m}(k_j + q_r)^2 + \frac{\delta}{2} - i\frac{\Gamma_\downarrow}{2} - E] = (\frac{\Omega_R}{2})^2. \end{cases} \tag{30}$$

The left-hand side of the Eq. 30 gives

$$k_j = -q_r. \tag{31}$$

Combining Eq. 29 and Eq. 31 gives $k_j = k_m$. It's obvious that this result isn't consistent with supposing $k_j \neq k_m$. Thus Eq. 27 is not true. Then we need $\text{Im}[k_2] = \text{Im}[k_3]$ and there are two leading terms in Eq. 26

$$[G(\{1234\}, E) + G(\{2134\}, E) + G(\{1243\}, E) + G(\{2143\}, E)]e^{ik_3 L}e^{ik_4 L}$$

$$+ [G(\{1324\}, E) + G(\{3124\}, E) + G(\{1342\}, E) + G(\{3142\}, E)]e^{ik_2 L}e^{ik_4 L} = 0. \tag{32}$$

Simplifying the Eq. 26 to

$$e^{i(k_2 - k_3)L} = -\frac{[g(k_2, E) - g(k_1, E)][g(k_4, E) - g(k_3, E)]}{[g(k_3, E) - g(k_1, E)][g(k_2, E) - g(k_4, E)]}. \tag{33}$$

Eq.33 gives the continuum energy spectrum for large $L$. The above discussion proves that GWV condition

$$\text{Im}[k_p] = \text{Im}[k_{p+1}] \quad (\text{here} \;\; p = 2 \;\; \text{in our model}) \tag{34}$$

still holds in our two-band model.

In addition, if we consider the interval of the infinitely deep square potential well changing from $[0, L]$ to $[-L, L]$, then the OBC becomes

$$\psi(x)|_{x=-L} = \psi(x)|_{x=L} = 0. \tag{35}$$

Plugging Eq. 24 into Eq.35, we have

$$\begin{cases} e^{-ik_1 L}\phi_\uparrow^{(1)} + e^{-ik_2 L}\phi_\uparrow^{(2)} + e^{-ik_3 L}\phi_\uparrow^{(3)} + e^{-ik_4 L}\phi_\uparrow^{(4)} = 0, \\ e^{-ik_1 L}\phi_\downarrow^{(1)} + e^{-ik_2 L}\phi_\downarrow^{(2)} + e^{-ik_3 L}\phi_\downarrow^{(3)} + e^{-ik_4 L}\phi_\downarrow^{(4)} = 0, \\ e^{ik_1 L}\phi_\uparrow^{(1)} + e^{ik_2 L}\phi_\uparrow^{(2)} + e^{ik_3 L}\phi_\uparrow^{(3)} + e^{ik_4 L}\phi_\uparrow^{(4)} = 0, \\ e^{ik_1 L}\phi_\downarrow^{(1)} + e^{ik_2 L}\phi_\downarrow^{(2)} + e^{ik_3 L}\phi_\downarrow^{(3)} + e^{ik_4 L}\phi_\downarrow^{(4)} = 0. \end{cases} \tag{36}$$

The determinant of the corresponding coefficient is

$$\begin{vmatrix} e^{-ik_1 L} & e^{-ik_2 L} & e^{-ik_3 L} & e^{-ik_4 L} \\ e^{-ik_1 L}g(k_1, E) & e^{-ik_2 L}g(k_2, E) & e^{-ik_3 L}g(k_3, E) & e^{-ik_4 L}g(k_4, E) \\ e^{ik_1 L} & e^{ik_2 L} & e^{ik_3 L} & e^{ik_4 L} \\ e^{ik_1 L}g(k_1, E) & e^{ik_2 L}g(k_2, E) & e^{ik_3 L}g(k_3, E) & e^{ik_4 L}g(k_4, E) \end{vmatrix} = \sum_{\{j_1 j_2 j_3 j_4\} \in S_4} G(\{j_1 j_2 j_3 j_4\}, E)e^{-ik_{j_1} L}e^{-ik_{j_2} L}e^{ik_{j_3} L}e^{ik_{j_4} L} = 0. \tag{37}$$

If $\text{Im}[k_2] > \text{Im}[k_3]$, the leading term in Eq.37 is

$$[G(\{1234\}, E) + G(\{2134\}, E) + G(\{1243\}, E) + G(\{2143\}, E)]e^{-ik_1 L}e^{-ik_2 L}e^{ik_3 L}e^{ik_4 L} = 0. \tag{38}$$

By futher simplifing Eq.38, we can get the same equation as Eq.28, so we can konw from the above discussion that the case of $\text{Im}[k_2] > \text{Im}[k_3]$ is not true. Then we also need $\text{Im}[k_2] = \text{Im}[k_3]$ and the two leading terms in Eq.37 are

$$[G(\{1234\}, E) + G(\{2134\}, E) + G(\{1243\}, E) + G(\{2143\}, E)]e^{-ik_1 L}e^{-ik_2 L}e^{ik_3 L}e^{ik_4 L}$$

$$+[G(\{1324\}, E) + G(\{3124\}, E) + G(\{1342\}, E) + G(\{3142\}, E)]e^{-ik_1 L}e^{-ik_3 L}e^{ik_2 L}e^{ik_4 L} = 0. \tag{39}$$

Similarly, we can obtain Eq.33 by simplifying Eq.39. Therefore, we can get the same GWV condition for the OBC in Eq. 35.

### SII-D. Further discussion on eigenstate

Based on the discussion in the previous subsections, we can conclude that the solution $(E_n, \psi_n(x))$ of the stationary problem Eq. 21 with large $L$ satisfy $f(k, E) = 0$ in Eq. 22 and $\text{Im}(k_2) = \text{Im}(k_3)$. Now, we rewrite Eq. 24 as

$$\psi(x) = c_1 e^{ik_1 x} \begin{pmatrix} \phi_\uparrow^{(1)} \\ \phi_\downarrow^{(1)} \end{pmatrix} + c_2 e^{ik_2 x} \begin{pmatrix} \phi_\uparrow^{(2)} \\ \phi_\downarrow^{(2)} \end{pmatrix} + c_3 e^{ik_3 x} \begin{pmatrix} \phi_\uparrow^{(3)} \\ \phi_\downarrow^{(3)} \end{pmatrix} + c_4 e^{ik_4 x} \begin{pmatrix} \phi_\uparrow^{(4)} \\ \phi_\downarrow^{(4)} \end{pmatrix}. \tag{40}$$

Here, we discuss the specific values of these coefficient $c_j, j = 1, 2, 3, 4$. We take the OBC in Eq. 35 and then plug Eq. 40 into Eq. 35 to get

$$\begin{cases} e^{-ik_1 L}c_1\phi_\uparrow^{(1)} + e^{-ik_2 L}c_2\phi_\uparrow^{(2)} + e^{-ik_3 L}c_3\phi_\uparrow^{(3)} + e^{-ik_4 L}c_4\phi_\uparrow^{(4)} = 0, \\ e^{-ik_1 L}g(k_1, E)c_1\phi_\uparrow^{(1)} + e^{-ik_2 L}g(k_2, E)c_2\phi_\uparrow^{(2)} + e^{-ik_3 L}g(k_3, E)c_3\phi_\uparrow^{(3)} + e^{-ik_4 L}g(k_4, E)c_4\phi_\uparrow^{(4)} = 0, \\ e^{ik_1 L}c_1\phi_\uparrow^{(1)} + e^{ik_2 L}c_2\phi_\uparrow^{(2)} + e^{ik_3 L}c_3\phi_\uparrow^{(3)} + e^{ik_4 L}c_4\phi_\uparrow^{(4)} = 0, \\ e^{ik_1 L}g(k_1, E)c_1\phi_\uparrow^{(1)} + e^{ik_2 L}g(k_2, E)c_2\phi_\uparrow^{(2)} + e^{ik_3 L}g(k_3, E)c_3\phi_\uparrow^{(3)} + e^{ik_4 L}g(k_4, E)c_4\phi_\uparrow^{(4)} = 0. \end{cases} \tag{41}$$

Adding up the first two and the last two lines in Eq. 41, respectively, we get

$$\sum_{j=1}^4 [1 + g(k_j, E)]e^{-ik_j L}c_j\phi_\uparrow^{(j)} = 0 \tag{42}$$

and

$$\sum_{j=1}^4 [1 + g(k_j, E)]e^{ik_j L}c_j\phi_\uparrow^{(j)} = 0. \tag{43}$$

For large $L$, the remaining leading term in Eq. 42 and Eq. 43 are $[1 + g(k_1, E)]e^{-ik_1 L}c_1\phi_\uparrow^{(1)} = 0$ and $[1 + g(k_4, E)]e^{ik_4 L}c_4\phi_\uparrow^{(4)} = 0$, respectively. In general, $g(k, E) \neq -1$, then we have $c_1 = c_4 = 0$. Therefore, the coordinate components of the eigenstates have the same exponential decay term, indicating that spin-up and -down components have the same localized length.

### SIII. APPLICATION OF GWV THEORY TO OUR MODEL

In this section, we discuss in detail the results of applying GWV theory to our model, some of which have been shown in Fig.3 in the main text. Here we will make a further explanation for the GWV curves and the localization of eigenstates. This section is organized as follows. In Sec.SIII-A, we present two methods for calculating GWVs, in

which the absolute value of its imaginary part, $|\text{Im}[k]|$, is used to illustrate which state is localized or delocalized. After obtaining GWVs, in Sec.SIII-B, we briefly discuss the difference between the number of peaks in the GWVs $k_+$ and $k_-$ in terms of the different complex spectrum regions covered by the bands $\text{Re}[E_\pm(k_\pm)]$. And then to see GWV more fully as the parameters change, in Sec.SIII-C, we show a picture of GWV with $\Gamma_\downarrow/\Gamma_\uparrow$ as an independent variable, and we define a boundary of transition from localized state to delocalized state, which is similar to the "mobility edge" in Hermitian disordered systems. Finally, we further show the changes in the strength of NHSE under different parameters shown in Fig. S8.

### SIII-A. Calculation of $|\text{Im}[k]|$

In this subsection, we give two methods to obtain $|\text{Im}[k]|$ characterizing the localization length of eigenstates in the main text.

(1) Let's start with the first method. This method requires the numerical eigenvalues of the Schrödinger equation. As shown in FIG.3 (d) in the main text, the color dots represent the corresponding numerical eigenvalues. Here, for each eigenvalue $E_i$, we can calculate the corresponding $|\text{Im}\, k(E_i)|$ (see the next paragraph), and represent them with dot size. One can find that, with the increase of $\text{Re}\, E_i$, the dot size tends to zero, indicating that the corresponding eigenstate becomes extended.

Now we demonstrate the calculation procedure. (i) calculate the eigenvalues of the Schrödinger equation of our model numerically; (ii) substitute each eigenvalue $E_i$ into the characteristic polynomial $f(E, k) = \det[E - H(k)] = 0$, and calculate the corresponding solutions, which can be labeled by $k_j(E_i)$ for the $j$-th root; (iii) sort $k_j(E_i)$ by the their imaginary parts, say, $\text{Im}\, k_1(E_i) \leq \text{Im}\, k_2(E_i) \leq \text{Im}\, k_3(E_i) \leq \text{Im}\, k_4(E_i)$; (iv) according to the GWV theory, we must have $\text{Im}\, k_2(E_i) = \text{Im}\, k_3(E_i)$ (please refer to section II.B for details of the derivation), and one can define $\text{Im}\, k(E_i) := \text{Im}\, k_2(E_i) = \text{Im}\, k_3(E_i)$ to characterize the localization properties of the eigenstate. FIG.3 (d) in the main text shows the corresponding results of the above procedure.

(2) Next, we discuss the other method. This method is the auxiliary generalized wave vector (aGWV) theory, which is an analytical result that does not require the numerical eigenvalues of the Schrödinger equation. As shown in FIG.3 (a) and (b) in the main text, the thin solid gray lines represent the auxiliary generalized wave vectors, and the thick solid colorful lines represent the generalized wave vector, i.e. $k_+$ and $k_-$ corresponding to the red and blue curves in Fig. S4, respectively. Mapping them onto the complex energy plane, one can obtain FIG.3 (c) in the main text, in which the curve $A - B - C$ represents the energy flow for $E_+(k_+)$ and $D - E - F - G - H - I - J$ for $E_-(k_-)$. Here $E, B, I$ represent a common point, and $F, H$ represent another common point in the complex energy plane.

Then we give the calculation process of aGWV theory by generalizing the aGBZ method [9] to our continuum model. We know the GWV condition is $\text{Im}[k_2] = \text{Im}[k_3]$ from the Eq. 34, so the corresponding set of $\{k_2, k_3\}$ is a subset of solutions to the following equations

$$\begin{cases} f(k_1 = k_R + ik_I, E) = 0 \\ f(k_2 = k_R' + ik_I, E) = 0 \end{cases},\tag{44}$$

where $f(k, E)$ is represented by Eq. 22, $k_1$ and $k_2$ have the same imaginary part but different real part. Eq. 22 should have common roots with respect to energy $E$, so we eliminate $E$ first using the resultant

$$R_E[f(k_1 = k_R + ik_I, E), f(k_2 = k_R' + ik_I, E)] \equiv F(k_R, k_R', k_I) = 0.\tag{45}$$

The fact that the real and imaginary parts of $F(k_R, k_R', k_I)$ are equal to zero tells us that the resultant with respect to $k_R'$ should also be zero, that is,

$$R_{k_R'}[\text{Re}[F], \text{Im}[F]] \equiv G(k_R, k_I) = 0.\tag{46}$$

The solution set $\{(k_R, k_I)\}$ for Eq. 46 is called the auxiliary generalized wave vector (aGWV) corresponding to a bunch of analytic arcs. We need to mark each arc between the two branch points and the specific calculation method is as follows: (i) plug $k_s = k_{R,s} + ik_{I,s}$ solved from Eq. 46 into Eq. 22, then obtain two energies marked as

$E_1$ and $E_2$; (ii) plug two energies from (i) into Eq. 22 again, and get two set of solutions $(k_1^{(j)}, k_2^{(j)}, k_3^{(j)}, k_4^{(j)})_{E_j}$ corresponding energy $E_j$, where $j = 1, 2$, $\mathrm{Im}[k_1^{(j)}] \geqslant \mathrm{Im}[k_2^{(j)}] \geqslant \mathrm{Im}[k_3^{(j)}] \geqslant \mathrm{Im}[k_4^{(j)}]$; (iii) check which set of solutions have at least two $k_m^{(j)}$ and $k_{m+1}^{(j)}$ such that $\mathrm{Im}[k_m^{(j)}] = \mathrm{Im}[k_{m+1}^{(j)}] = k_{I,s}$, then we label this analytic arc with $(m, m+1; E)$ until reaching the branch point. After the procedure above, we obtain the aGWV in Fig. S4, which includes generalized wave vectors (GWVs) formed by arcs with $(2, 3; E_{1/2})$ giving the continuum spectrum. Thus functional relationship between $k_R$ and $k_I$ given by the discussion above is piecewise analytic. This idea of resultant method is clear but really tedious, so we can use Mathematica to calculate it directly.

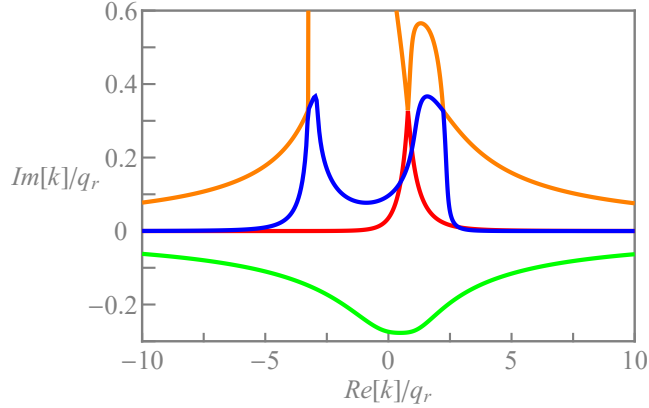

FIG. S4. **aGWV of continuum two-band model discussed in the main text.** Orange and green represent the aGWV labeld (1,2) and (3,4). Red and blue correspond to the GWVs that determine the continuum spectrum $E_+$ and $E_-$ as shown in Fig. 3 (c) in the main text, respectively. The parameters are $(\delta, \Omega_R, \Gamma_\uparrow, \Gamma_\downarrow) = (4, 9/2, 6, 6/13)E_r$.

Therefore using $\mathrm{Im}[k]$ we can define the localized and delocalized eigenstates. Generally, this definition is artificial and dependent on the system size. In our treatment, when $|\mathrm{Im}\, k_\pm| < 1/L$, the corresponding state is defined as a delocalized state, while when $|\mathrm{Im}\, k_\pm| > 1/L$, the corresponding state is defined as a localized state. Here, L is the system length and we assume L=100. In FIG.3 (a) and (b) in the main text, $A, C, D, J$ represent the corresponding points whose $|\mathrm{Im}\, k_\pm| = 0.01$. From FIG.3 (c) in the main text, one can find that the left sides of $A/J$ and $C/D$ belong to the localized states and the right sides belong to the delocalized states.

### SIII-B. Further discussion of curves $k_+$ and $k_-$

In this subsection, we discuss further Fig.3(a) and (b) in the main text, where we mainly emphasize the difference between the number of peaks in the GWVs $k_+$ and $k_-$ in terms of the different complex spectrum regions covered by the bands $\mathrm{Re}[E_\pm(k_\pm)]$. From Fig.3(a) and (b), we can see that the GWV curves $k_+$ and $k_-$ have one peak and two peaks, respectively. In general, this is not a universal phenomenon, and the difference between the number of peaks can be understood from the fact that $\mathrm{Re}[E_+(k_+)]$ and $\mathrm{Re}[E_-(k_-)]$ cover different regions. And then from the point of view of band splitting, we can roughly explain the reason why $E_\pm(k_\pm)$ covers different energy spectra.

First, we explain the difference in the number of peaks in this paragraph. As can be seen from Fig.3 in the main text, the arcs $A - B$ in (a) and $J - I$ in (b) correspond to the same energy spectrum in (c); $B - C$ in (a) and $E - D$ in (b) also correspond to the same energy spectrum in (c). If $E$ and $I$ are the same points in (b), the number of peaks in $k_+$ and $k_-$ will be the same. However, here $E$ and $I$ are different points in $k_-$ and they are connected by the arc $E - F - G - H - I$. Therefore, it is possible to have two peaks in $k_-$, namely in (b). To be more precise, when $\mathrm{Im}\, k_-(G) < \mathrm{Im}\, k_-(E) = \mathrm{Im}\, k_-(I)$, as the case in our model, the $k_-$ must have at least two peaks.

And then in this paragraph, we'll briefly explain why $E_\pm(k_\pm)$ covers different spectra. This phenomenon can be understood in terms of the presence of band splitting from the presence of $\delta\sigma_z$ term. As shown in FIG. S5 (b), here we have plotted the dispersion relation of the two GWVs, i.e. $\mathrm{Re}\, E_\pm(k_\pm)$ and $\mathrm{Re}\, E_{+/-}(k_{+/-}) \pm |\mathrm{Im}\, E_{+/-}(k_{+/-})|/2$ represented by the thick solid lines and thin solid lines, respectively. One can find that due to the presence of $\delta\sigma_z$ term, the real parts of the two bands $E_+$ and $E_-$ splits, where points $E, B$ and $I$ have the same real part of energy, and the contribution to the complex spectrum below this real part of energy, namely the curve $E - F - G - H - I$, comes entirely from the $E_-$ band.

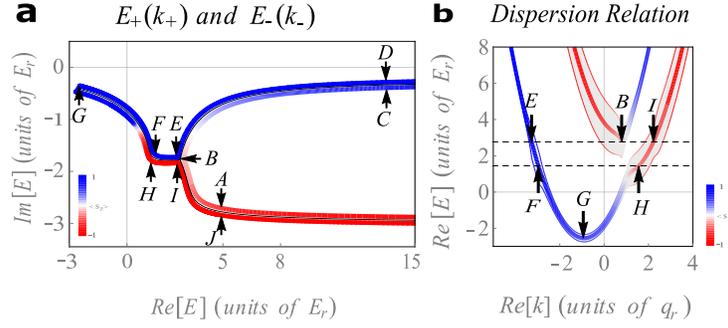

**a** $E_+(k_+)$ and $E_-(k_-)$

**b** *Dispersion Relation*

FIG. S5. (a) is Fig. 3 (c) in the main text. (b) shows the dispersion with the real parts of $k$ and $E$ as the horizontal and vertical coordinates, which shows the splitting between $\mathrm{Re}[E_+]$ and $\mathrm{Re}[E_-]$.

By contrast, when $\delta = 0$, as shown in FIG. S6 (c), the real parts of $E_+$ and $E_-$, represented by red and blue points, cover the same region, which may need to be satisfied by having the same number of peaks in $k_+$ and $k_-$. From FIG. S6 (a) and (b), we can see that there are indeed two peaks in the two generalized wave vector curves, where point $C$ is the common point between them and is mapped to points $C_+$ and $C_-$ in FIG. S6 (c) respectively under the action of $E_+$ and $E_-$. Further, we show FIG. S6 (d) with $\mathrm{Re}[k]$ and $\mathrm{Re}[E]$ as the horizontal and vertical coordinate, from which we can see that there is no splitting between $\mathrm{Re}[E_+]$ and $\mathrm{Re}[E_-]$, which means that both $E_+$ and $E_-$ may contribute to the degeneracy of the complex spectrum. And it can be seen from FIG. S6 (c) that $E_+$ and $E_-$ maps $A - B - C/C - D - E$ in $k_+/k_-$ is mapped to the curve $A - B - C_-/C_+ - D - E$ in the complex spectrum, that is, the blue and red branches are covered twice respectively.

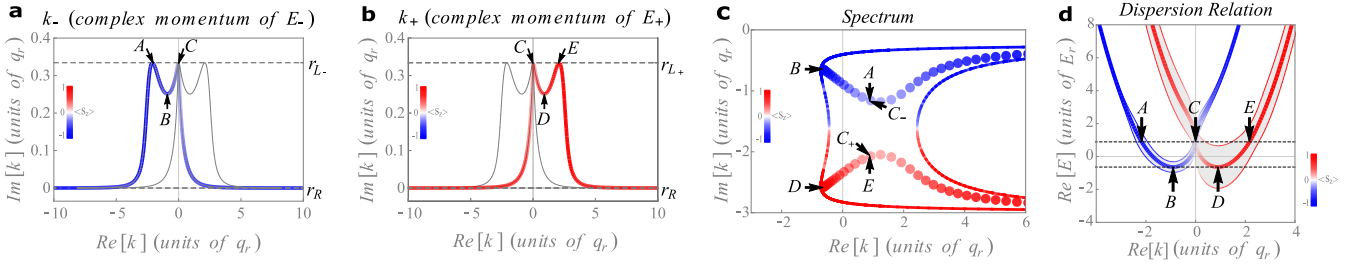

**a** $k_-$ *(complex momentum of $E_+$)*

**b** $k_+$ *(complex momentum of $E_+$)*

**c** *Spectrum*

**d** *Dispersion Relation*

FIG. S6. **The result in the case of parameter** $(\delta, \Omega_R, \Gamma_\uparrow, \Gamma_\downarrow) = (0, 4, 6, 6/13)E_r$. (a) and (b) are GWV curves $k_+$ and $k_-$, respectively. The discrete points and continuous solid lines in (c) are the energy spectra under open and periodic boundary conditions respectively, where points $A, B, C_-/C_+, D, E$ correspond to points $A, B, C/C, D, E$ in (a)/(b). In (d) $\mathrm{Re}\,E_\pm(k_\pm)$ and $\mathrm{Re}\,E_{+/-}(k_{+/-}) \pm |\mathrm{Im}\,E_{+/-}(k_{+/-})|/2$ represented by the thick solid lines and thin solid lines, respectively, and there is no splitting between $\mathrm{Re}[E_+]$ and $\mathrm{Re}[E_-]$.

### SIII-C. Non-reciprocal edge

In this subsection, we clearly show the results of GWV under different parameters, where we take $\Gamma_\downarrow/\Gamma_\uparrow$ as an independent variable. And then we find a boundary of transition from localized state to delocalized state, which can be used to separate the region between localized and delocalized eigenstates.

As shown in FIG. S7, we fixed $(\delta, \Omega_R) = (4, 9/2)$ and took $\Gamma_\downarrow/\Gamma_\uparrow$ as the variable to show the change of $\mathrm{Im}[k]$, where FIG. S7 (a) and (b) correspond to $E_+$ and $E_-$, respectively. In FIG. S7 (a) and (b), we can indeed see that for fixed $\Gamma_\downarrow/\Gamma_\uparrow$ values, $\mathrm{Im}\,k$ tends to zero when $\mathrm{Re}\,k$ reaches a certain value. This transition is similar to the transition from localization to de-localization in Hermitian case. We call the boundary of this transition as *non-reciprocal edge* here.

Here, we explain why we call this transition boundary a "non-reciprocal edge". In general, the phrase "mobility edge" in the Hermitian system is used to refer to the localize-delocalize transition in a disordered system. The occurrence of localization in a disordered system means that electrons cannot move freely compared with delocalized cases. In this sense, we call this transition a "mobility edge". However, in the non-Hermitian systems with NHSE, the emergence of NHSE, which is characterized by the appearance of localized eigenstates at the boundary, does not imply that the particles cannot move freely. As discussed in Ref. [10, 11], when the NHSE appears, the quantum

tunneling from one side to the other side can even be enhanced, while the opposite is suppressed. This is a type of "non-reciprocal" transport phenomenon, which motivates us to define this transition as "non-reciprocal edge".

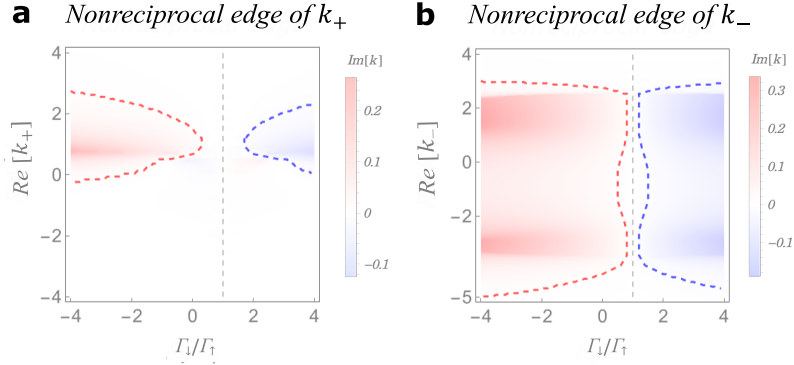

FIG. S7. **The non-reciprocal edge of the model.** (a) and (b) show the relation between $\Gamma_\downarrow/\Gamma_\uparrow$, $\mathrm{Re}\,k$, and $\mathrm{Im}\,k$ correspond to $k_+$ and $k_-$, respectively. $\mathrm{Im}[k]$ reaches zero when $\mathrm{Re}[k]$ and $\Gamma_\downarrow/\Gamma_\uparrow$ reach certain critical points, which form the non-reciprocal edge that distinguishes the localized and delocalized states.

### SIII-D. The strength of NHSE as functions of parameters

In this subsection, we show the strength of NHSE under different parameters, which shows the robustness of the NHSE. In order to characterize the strength of the NHSE, we can introduce the maximum and minimum values of the imaginary part of $k_\pm$, which are denoted as $r_L$ and $r_R$ respectively, i.e. $r_L := \max\{\mathrm{Im}\,k_+, \mathrm{Im}\,k_-\}$ and $r_R := \min\{\mathrm{Im}\,k_+, \mathrm{Im}\,k_-\}$.

Firstly, we fixed the parameter $\Gamma_\uparrow = 6$ and took $(\Gamma_\downarrow/\Gamma_\uparrow, \delta)$ as an independent variable to clearly show the strength of skin effect represented by $r_L$ and $r_R$. As can be seen from Fig. S8(a) and (b), when $\Gamma_\downarrow/\Gamma_\uparrow < 1\ (>1)$, $r_L > 0\ (=0)$ and $r_R = 0\ (>0)$, which implies that all the eigenstates are localized at the left (right) boundary. However, when $\Gamma_\downarrow = \Gamma_\uparrow$, we have $r_L = r_R = 0$, that is, the system has no skin effect and all the eigenstates are extended.

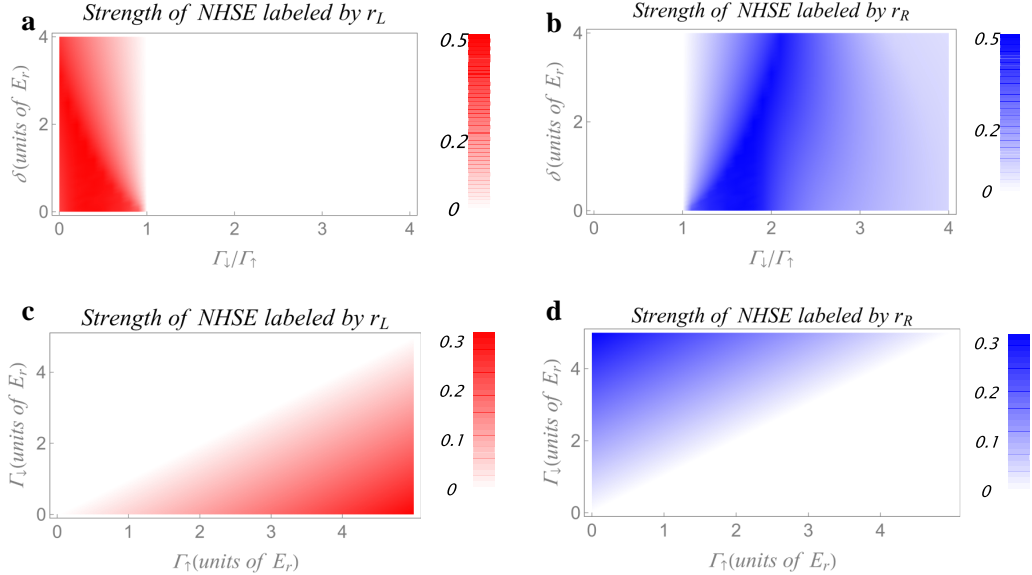

FIG. S8. (a) and (b) represent the strength of NHSE for different values of $\delta$ and $\Gamma_\downarrow$ under fixed $(\Omega_R, \Gamma_\downarrow) = (9/2, 6)$, whose color strength is proportional to the value of $r_{L/R}$. (c) and (d) show the calculated $r_R$ and $r_L$ as functions of $(\Gamma_\uparrow, \Gamma_\downarrow)$ for the parameter $(\delta, \Omega_R) = (4, 9/2)$.

Further, we fixed the parameter $\delta = 4$ and took $(\Gamma_\downarrow, \Gamma_\uparrow)$ as an independent variable to clearly show the strength of

skin effect represented by $r_L$ and $r_R$. Then we can find that the results in Fig. S8 (c)-(d) are consistent with Fig. S8 (a)-(b).

## SIV. THEORY OF DYNAMIC STICKY EFFECT

In this section, we give a detailed theoretical explanation of the dynamic behavior of Gaussian wave packets, which are used in the main text as an experimental signal to detect the existence of non-Hermitian skin effects in the system. This section is organized as follows. In Sec.SIV-A, we give the detailed calculation of group velocity (gv) of Gaussian wave packet in the Hermitian and non-Hermitian cases, in which there is no dynamic sticky effect (DSE). To explain the DSE of the system in the non-Hermitian case, in Sec.SIV-B, we take an exact solvable spinless model as an example to show that the DSE is the result of spectrum degeneracy splitting, which is mentioned in Fig.1 in the main text. And then in Sec.SIV-C, we give a numerical demonstration of the DSE on the two-band model discussed in the main text. Finally, in Sec.SIV-D, we further discuss the influence of different strengths of dissipation on the dynamic behavior of wave packets.

### SIV-A. The group velocity of wave packet and the normal boundary scattering

In this subsection, we discuss the calculation of the gv for Gaussian wave packet in detail as shown in Fig. 4 (a), (b) and (d) in the main text, which indicates the normal scattering dynamics near the boundary in spinful cases. We first review the definition of gv. The wave functions of the two monochromatic waves propagating in the x-direction with different wave numbers and frequencies are

$$\psi_1 = A cos(k_1 x - \omega_1 t), \qquad \psi_2 = A cos(k_2 x - \omega_2 t), \qquad (47)$$

where $k_j$ and $\omega_j$ are the wave numbers and angular frequencies of $\psi_j$, $j = 1, 2$. The wave packet formed by the superposition of $\psi_1$ and $\psi_2$ is

$$\psi = \psi_1 + \psi_2 = \bar{A} cos(\bar{k} x - \bar{\omega} t), \qquad (48)$$

where $\bar{A} = 2A cos(k_m x - \omega_m t)$, $\bar{k} = \frac{k_1 + k_2}{2}$, $\bar{\omega} = \frac{\omega_1 + \omega_2}{2}$, $k_m = \frac{k_1 - k_2}{2}$, $\omega_m = \frac{\omega_1 - \omega_2}{2}$. The phase velocity and gv are then defined as $v_p = \frac{\bar{\omega}}{\bar{k}}, v_g = \frac{\Delta \omega}{\Delta k}$. And when the difference between the angular frequencies of two waves is very small, i.e. $\Delta \omega \approx 0$, then the gv can be written as

$$v_g = \frac{\partial \omega}{\partial k}. \qquad (49)$$

Thus, we can use Eq. 49 to calculate the gv before and after hitting the boundary for the initial state

$$(\psi_\uparrow, \psi_\downarrow)|_{t=0} = (0, A exp[-(x - x_0)^2/20 + ik_0 x])^T. \qquad (50)$$

For the Hermitian case without skin effect, Fig. 4 (a) and (b) in the main text show that the initial states with $(x_0, k_0) = (40, -3)$ and $(40, -5)$ have similar reflection behavior in the range of $20t_R$.

For $k_0 = -3$, there are two energies $E_+(k_2 = -3)$ and $E_-(k_5 = -3)$ as shown in Fig. S9 (a1), where $k_2$ and $k_5$ are marked by discrete points including other momenta in Fig. S9 (a1), so the group velocities of the two trajectories before reaching the boundary are $\frac{\partial E_+}{\partial k}|_{k=k_2}$ and $\frac{\partial E_-}{\partial k}|_{k=k_5}$ corresponding to the reciprocal of the slope of the dashed line and solid line labeled by $k_2$ and $k_5$ in Fig. S9 (b1), respectively. After reflected by the left boundary, $k_2$ and $k_5$ are scattered into $k_3$, $k_4$ and $k_6$, and the group velocities of the corresponding reflected waves become $\frac{\partial E_+}{\partial k}|_{k=k_3}$, $\frac{\partial E_-}{\partial k}|_{k=k_4}$ and $\frac{\partial E_-}{\partial k}|_{k=k_6}$ represented by two dashed lines and one solid line labeled by $k_3$, $k_4$ and $k_6$ in Fig. S9 (b), respectively. Specically, the gv of the two dashed lines marked by $k_7$ and $k_8$ are almost the same so that those wave packets don't split in real space, but we can distinguish them in the momentum space, shown in the Fig S9 (d1). Approximately, within $20t_R$ the two dashed lines in Fig. S9 (b1) are again reflected by the right boundary causing both $k_3$, $k_4$ to return to $k_2$, that is, the gv goes back to $\frac{\partial E_+}{\partial k}|_{k=k_2}$. It is noteworthy that reflection not only changes momentum but also changes the mean value of spin z-component $\langle \hat{s}_z \rangle$, indicating that spin-orbit coupling is revealed by reflection at the boundary.

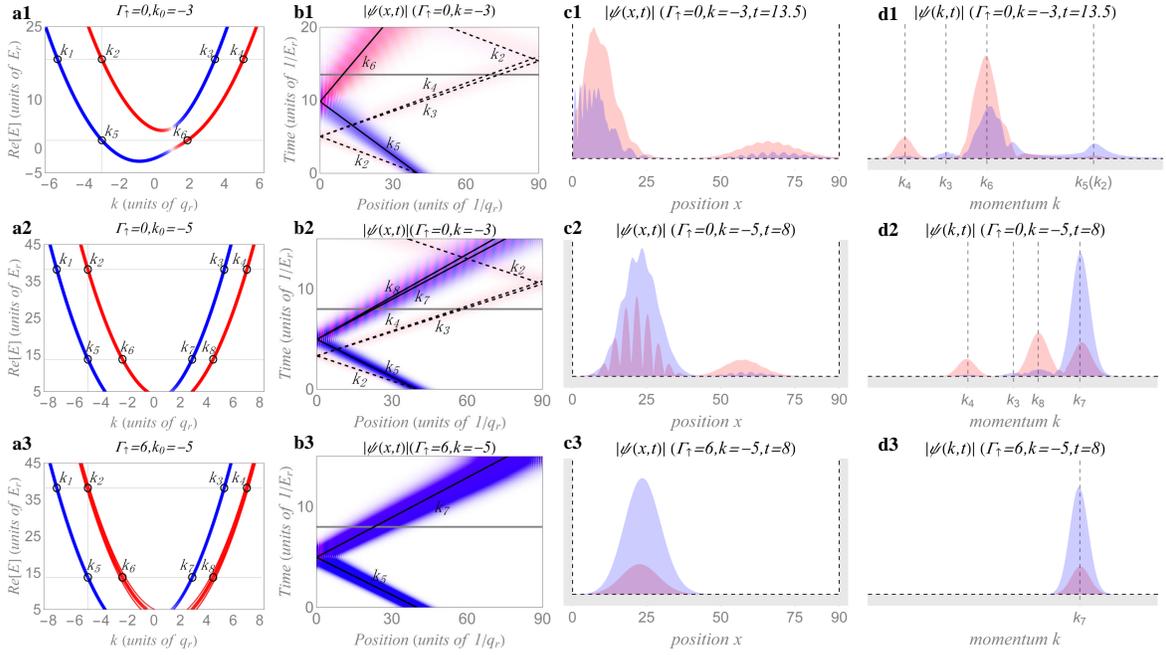

FIG. S9. **The dispersion relation and the change of gv in the normal boundary scattering** (a1-d1), (a2-d2) and (a3-d3) show the dynamic behavior of the system corresponding to the parameters $(\delta, \Omega_R, \Gamma_\uparrow, \Gamma_\downarrow) = (4, 9/2, 0, 0)E_r$ and $(4, 9/2, 6, 6/13)E_r$, respectively. (a1)-(a3) represent the free particle spectrum under the periodic boundary condition, in which the momentum points marked on the same horizontal lines correspond to the same real part of the energy. And (b1)-(b3) show the scattering behavior of Gaussian wave packet with a different momentum, in which the solid and dashed lines represent two separated trajectories. The momentum marked on each trajectory in the second column corresponds to points marked with the same value in the first column, respectively, and the behavior of reflected waves in (b1-b3) can be accurately predicted by the gv. (c) and (d) show the wave function in the real space and the momentum space at a certain time marked by a horizontal solid line in (b), where red and blue represent the component of spin-up and -down, respectively, and it can be seen from (d) that the reflected wave at this time is indeed composed of multiple monochromatic waves marked by different momenta. Here the color in (a) and (b) represents the mean values of the $z$-component spin, $\langle s_z \rangle$.

Similarly within $15t_R$, the evolution of the initial state with $k_0 = -5$ also shows two trajectories as shown in the Fig. 4 (b) in the main text, and Fig. S9 (a2) shows that the presence of momentum $k_1, k_2, k_3, k_4$ and $k_5, k_6, k_7, k_8$ corresponding to energy $E_+(k_2 = -5)$ and $E_-(k_5 = -5)$, respectively. So, for the dashed line in Fig. S9 (b2), left boundary reflection changes the momentum from $k_2$ to $k_3$ and $k_4$, and then back to $k_2$ after reflection from the right boundary. And for the solid line in Fig. S9 (b2), a reflection by the left boundary changes momentum from $k_5$ to $k_7$ and $k_8$. Accordingly, the group velocities of the two trajectories also change according to the corresponding change of momentum, and Fig. S9 (b2) shows that the change process of gv of the initial state with $k_0 = -5$. Similarly, the gv of the two solid lines marked by $k_7$ and $k_8$ are also almost the same, which causes the two monochromatic waves to interfere with each other as the main components of the reflected waves to form the clear coherent stripes in Fig. S9 (b2).

Therefore, from the above results, for the Hermitian case without dissipation, the evolution of each spin component of the wave packet will have two paths. And through a reflection on the boundary, the momentum will be converted to other values with the same energy, and the corresponding gv can also be calculated in the traditional way.

For the non-Hermitian case with skin effect but without the DSE, since the energy changes from real to complex, the definition of gv also changes into $\frac{\partial \text{Re}\,[E]}{\partial k}$, where $\text{Re}\,[E]$ is the real part of the energy. In this case, the evolution of the initial state with $k_0 = -5$ shows only one trajectory as shown in Fig. 4 (d) in the main text (the reason for this phenomenon will be explained below). Further, Fig. S9 (a3) shows that the presence of momentum $k_1, k_2, k_3, k_4$ and $k_5, k_6, k_7, k_8$ corresponding to the same real part of the energy $E_+(k_2)$ and $E_-(k_5)$, respectively. So for the solid line in Fig. S9 (b3), a reflection by the left boundary changes momentum from $k_5$ to $k_7$ and the gv is a good description of the reflection behavior.

## SIV-B. DSE in a spinless model

In this subsection, we provide a physical picture for the emergence of the DSE, or equivalent say, why the eigenstates of the above Hamiltonian are not the superposition of plane waves ($\text{Im } k = 0$), but non-plane waves ($\text{Im } k \neq 0$). We can use the following spinless model to illustrate this point,

$$[-\hbar^2 \partial_x / 2m + \lambda \hbar \partial_x + V(x)] \psi_E(x) = E \psi_E(x), \tag{51}$$

where

$$V(x) = \begin{cases} \infty, & |x| > L \\ 0. & |x| \leq L \end{cases} \tag{52}$$

(1) We first exactly solve the eigenstates, GWV, and spectrum of this spinless model. Two particular solutions of Eq. 51 can be solved simply as

$$\phi_1(x) = e^{ikx}, \quad \phi_2(x) = e^{-ikx} e^{-\frac{2\lambda m}{\hbar} x} \tag{53}$$

with the corresponding eigenvalue

$$E = \frac{\hbar^2 k^2}{2m} - i\lambda \hbar k, \tag{54}$$

and the dashed line in Fig. S10 (a) is the spectrum in free space when $k$ is a real number in this formula. We superimpose the two particular solutions linearly to construct a general solution

$$\psi(x) = c_1 \phi_1(x) + c_2 \phi_2(x), \tag{55}$$

and then according to the boundary condition $\psi(x)|_{x=L} = \psi(x)|_{x=-L} = 0$, we can get

$$\begin{cases} c_1 e^{-ikL} + c_2 e^{ikL} e^{\frac{2\lambda m}{\hbar} L} = 0 \\ c_1 e^{ikL} + c_2 e^{-ikL} e^{-\frac{2\lambda m}{\hbar} L} = 0 \end{cases} \tag{56}$$

Here we set $k = k_R + i k_I$, and given that the determinant of coefficient Eq. 56 is 0 we obtain

$$e^{2L(2ik_R - 2k_I + \frac{2\lambda m}{\hbar})} = 1, \tag{57}$$

that is,

$$\begin{cases} k_R = \frac{n\pi}{2L}, & n \in \mathbb{Z} \\ k_I = \frac{\lambda m}{\hbar}. \end{cases} \tag{58}$$

Eq. 58 gives the GWV of this spinless model, where it is a series of discrete evenly spaced points with fixed imaginary parts as shown in Fig. S10 (c). By substituting Eq. 58 into Eq. 54, we can get the spectrum under the open boundary condition

$$E = \frac{\hbar^2}{2m} \left(\frac{n\pi}{2L}\right)^2 + \frac{m\lambda^2}{2}, \tag{59}$$

which are correspondingly also some discrete points as shown in Fig. S10 (a). By further substituting Eq. 58 into Eq. 53, we can obtain

$$\begin{cases} \phi_1(x) = e^{-\frac{\lambda m}{\hbar} x} e^{i \frac{n\pi}{L} x}, \\ \phi_2(x) = e^{-\frac{\lambda m}{\hbar} x} e^{-i \frac{n\pi}{L} x}, \end{cases} \tag{60}$$

which shows that all the eigenstates of the system have the same localization length under the constraints of these two boundary conditions.

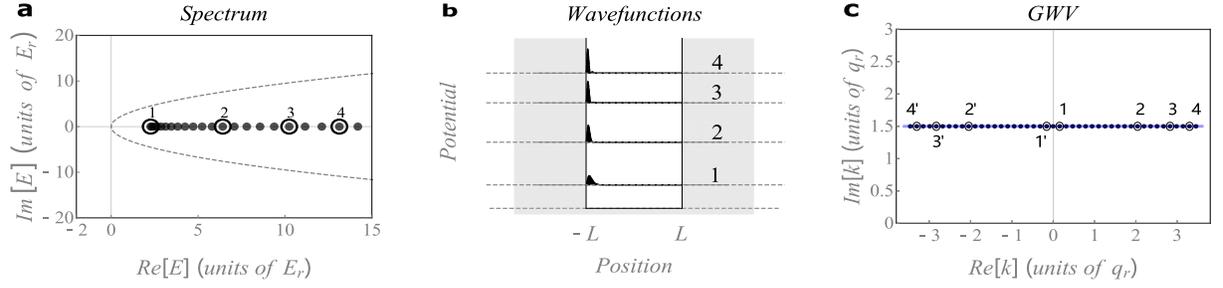

FIG. S10. **Exact solution of the spinless model.** (a) shows the energy spectrum in free space and infinitely deep square well, which correspond to gray dashed line and black dots, respectively. The wave function in (b) corresponds to the eigenstates belonging to the four eigenvalues in (a), respectively, and the gray part represents the infinite deep square potential well. The solid blue line and discrete points in (c) represent the GWV given by Eq. 58 as $L$ equals 10 and tends to infinity, respectively, where the points numbered $1(1')$, $2(2')$, $3(3')$ and $4(4')$ correspond to the eigenvalues numbered 1, 2, 3 and 4 in (a) respectively, both of which are double degeneracy.

(2) We then demonstrate that the model has a DSE from degeneracy splitting. Here we consider the system to be in a semi-infinite deep square well

$$V(x) = \begin{cases} \infty, & x < 0 \\ 0, & x \geq 0 \end{cases} \tag{61}$$

where there is only one effective boundary condition $\psi(x)|_{x=0} = 0$. This will not result in $k$ taking only complex numbers as in Eq. 53, that is, $k$ can take real numbers. For any given $k_0 \in \mathbb{R}$, one can obtain another $k' = -k_0 + \frac{i2\lambda m}{\hbar}$, whose energies are the same, i.e. $E(k_0) = E(k')$. Since $\lambda > 0$, all the non-plane waves $e^{ik'x}$ are localized at the left boundary, which satisfies the boundary condition at $+\infty$, i.e. $|\psi(x \to \infty)| < \infty$. In this case, the corresponding general solution is the same as Eq. 55, but by substituting it into the boundary condition at $x = 0$, we only get $c_1 = -c_2$, that is,

$$\psi(x,t) = Ne^{-iE(k_0)t}(e^{ik_0x} - e^{-ik_0x}e^{-\frac{2\lambda m}{\hbar}x}) \equiv \psi_I(x,t) - \psi_R(x,t). \tag{62}$$

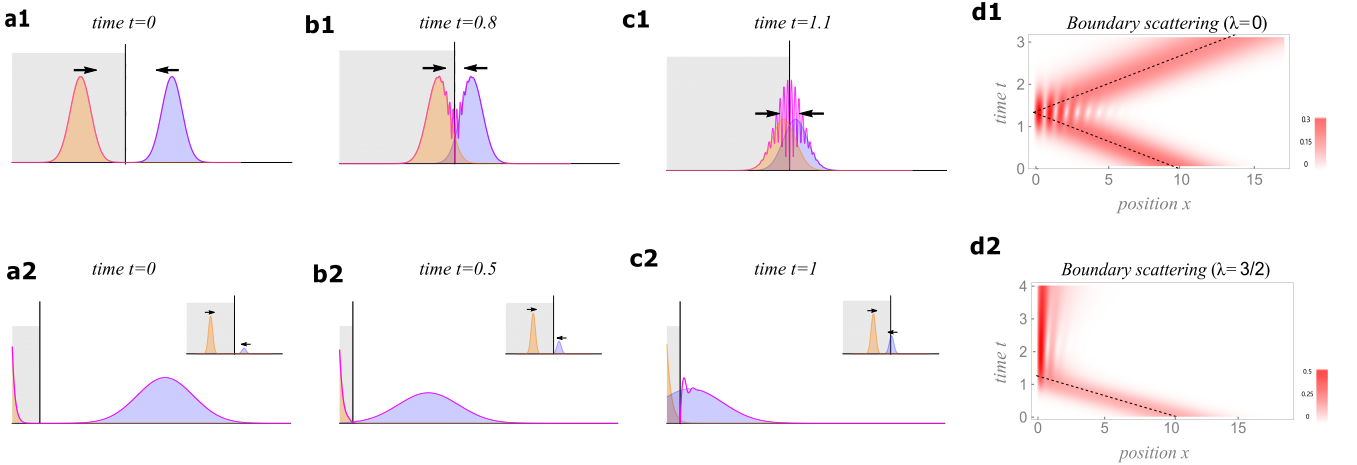

FIG. S11. **The DSE of an exact solvable model.** The blue, yellow and pink in (a1-c1) represent $\psi_I(x,t)$, $\psi_R(x,t)$ and their superposition at different times in the Hermitian case. (d1) represents the evolution of the Gaussian wave packet at $\lambda = 0$, which is reflected back directly after reaching the boundary. And its spatial distribution at time $t = 0, 0.8$, and 1.1 correspond to the pink curve in (a1), (b1) and (c1) where $x \geq 0$. Similarly, (a2-d2) represent the results of the non-Hermitian case where $\lambda = 3/2$. In (a2-c2), we give $\psi_I(x,t)$ (blue), $\psi_R(x,t)$ (yellow) and their superposition (pink) at different times, and it should be noted that wave functions $\psi_I(x,t)$ $(\psi_R(x,t))$ in the illustrations at time $t = 0, 0.5$ and 1 is scaled to 3.6 $(\frac{1}{2.5 \times 10^{21}})$, $\frac{1}{62.5}$ $(\frac{1}{8.1 \times 10^{18}})$ and $\frac{1}{3.6 \times 10^4}$ $(\frac{1}{3.9 \times 10^{16}})$ of the original values, respectively, which clearly shows the relative magnitude and position of $\psi_I(x,t)$ and $\psi_R(x,t)$. The DSE of this model is shown in (d2), where the spatial distribution at $t = 0, 0.5$ and 1.0 corresponds to the pink curve in (a2), (b2) and (c2), respectively.

Here $\psi_{I/R}(x, t)$ represent the incident and reflection waves, respectively. More generally. the initial states can be expanded into the superposition of a series of plane waves, and it can be known from Eq. 53 that when any plane wave with momentum $k$ and energy $E(k)$ meets the left boundary, the reflection wave expression is non-plane wave $\phi_2$ in Eq. 53. Then it can be concluded that the reflection wave $\psi_R(x, t)$ relative to the incident state $\psi_I(x, t)$ is

$$\psi_R(x, t) = \psi_I(-x, t)e^{-2\lambda m x}. \tag{63}$$

From the above result, one can find that when $\lambda = 0$ (Hermitian case), the reflection wave is the inversion partner of the incident wave as shown in Fig. S11 (a1-c1). The initial state here is a Gaussian wave packet, $\psi(x, t = 0) \propto \exp[-(x - x_0)^2/10 + ip_0 x]$ whose center is at $x_0 = 10$ and mean value of momentum $p_0 = -4$, and Fig. S11 (d1) shows that it is reflected directly by the left boundary. However, for the non-Hermitian case with $\lambda = 3/2$, the reflection wave becomes the inversion partner of the incident wave times a localization factor. This implies all the reflection waves at any given time will be localized at the left boundary, which is nothing but the DSE, and this is indeed the case from the results shown in Fig. S11 (d2). From the above analysis, one can find that the DSE origins from the degeneracy splitting, which implies that the solution Eq. 51 cannot be written as the superposition of plane waves.

### SIV-C. DSE in the two-band model

For our two-band model in the main text, there is degeneracy splitting in the free particle spectrum so that the reflected wave can no longer be a superposition of plane waves but a series of non-plane waves, which means that the DSE will occur. Therefore, here we combine the degeneracy splitting of the free particle spectrum to numerically give the dynamic behavior of the wave packet when the initial momentum changes from $-3$ to $-5$, which will show a transition from DSE to normal reflection.

It can be seen from FIG. S12 (a1-a4) that the splitting of energy corresponding to the momentum of the reflection wave gradually decreases, which means that DSE is gradually weakened. Specifically, for the momentum $k = -3$, the system has a significant DSE, and when the momentum becomes $-4$, the incident wave is directly reflected after reaching the boundary, which means that there is a transition between $-3$ and $-4$. In addition, we notice that in FIG. S12 (a), when momentum changes from $-3$ to $-4$, the horizontal solid line intersecting with band $E_-$ at two points $k_1$ and $k_2$ in FIG. S12 (a1) rises gradually until it intersects with two bands at four points $k_1$, $k_2$, $k_3$ and $k_4$ in FIG. S12 (a3), which also implies that there will be an intermediate momentum that makes this horizontal solid line tangent to band $E_+$. Combining these two points, we show the dynamics of the system at momentum $k = -3.3758$ in the second column. It can be seen that in FIG. S12 (a2) the horizontal line is tangent to the band $E_+$ at point $k_2$, and FIG. S12 (b2) and (c2) also show the dynamics of the reflected wave, which include both the components that are stuck and those that are directly reflected by the boundary.

In particular, we note that when the initial momentum $k_0 = -5$, the solid line trajectory marked by $k_3$ given by gv is in good agreement with the evolution of reflected waves, which means that the energy splitting at $k_3$ is so small that the condition of equal energy is approximately satisfied. Furthermore, the energy spectrum in FIG. S12 is plotted on the complex energy plane, which clearly shows the degeneracy splitting at each momentum point. It can be seen from FIG. S12 (d1-d4) that, compared with other momenta, the energy of the incident and reflected wave is indeed approximately equal when the initial momentum $k_0 = -5$.

### SIV-D. The effect of dissipative strength on dynamics in the two-band model

From FIG.4 in the main text and FIG. S12, we can see that non-Hermitian not only brings DSE to the dynamics of the system but also change the evolutionary trajectory of the system from two in the Hermitian case to one in the non-Hermitian case. In addition, this phenomenon and DSE should only appear when the strength of dissipation reaches a certain value. Based on the above consideration, in this subsection, we start from the magnitude of the imaginary part of the energy after the degeneracy splitting of bands to explain qualitatively how the increase of dissipation term affects the disappearance of the trajectory marked by the dashed line and the appearance of DSE.

For the parameters selected by us, the imaginary part of the system eigenvalue is negative. Therefore, for a non-plane wave eigenstate, the smaller the imaginary part of its eigenvalue is, the faster its amplitude will decay over time. So we can consider that for a wave packet with a given initial momentum, it will have two incident trajectories, and

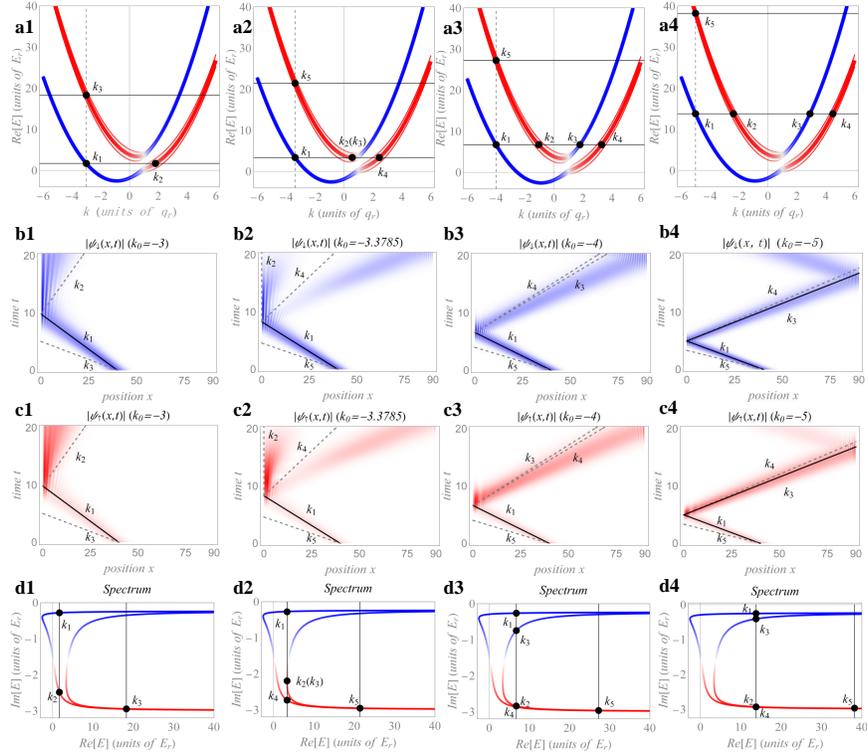

FIG. S12. **Dynamic behavior of the system with different initial momenta.** (a1-a4) show the free particle spectrum of the system, in which points marked by different momenta represent the energies corresponding to different incident momenta and corresponding reflected momenta. The figures in the middle two-row show the dynamical behavior of wave packets with different initial momenta, where (b1-b4) and (c1-c4) correspond to the spin-up and -down components, respectively. (d1-d4) show the free particle spectrum on the complex energy plane, which clearly shows the magnitude of the imaginary of the energy at different momentum points after degeneracy splitting. The parameters in Hamiltonian are $(\delta, \Omega_R, \Gamma_\uparrow, \Gamma_\downarrow) = (4, 9/2, 6, 6/13)E_r$.

then we can imagine that one with the smaller imaginary part of the energy will decay faster, and its amplitude will decay faster and faster as the dissipation increases. As for the reflection behavior, with the increase of the strength of dissipation, the splitting of the free particle spectrum should gradually become larger, so that the reflected behavior with small initial momentum gradually changes from normal reflection behavior to DSE.

In the following, we fix $\Gamma_\downarrow = \Gamma_\uparrow/13$ and change $\Gamma_\uparrow$ to prove the above statement numerically, where the initial momentum of the wave packet is $k = -3$. Firstly, we investigate the change of incident wave trajectory with the increase of $\Gamma_\uparrow$. FIG. S13(a) and (b) show that the free particle spectrum of the system and the evolution of the trajectory of the spin-up component as $\Gamma_\uparrow$ increases from 0 to 6. For the Hermitian case of $\Gamma_\uparrow = 0$, the spectrum is real and neither trajectory is decayed. With the increase of $\Gamma_\uparrow$, compared with point $k_1$, the imaginary part of the energy at $k_2$ in FIG. S13(a) gradually decreases, which makes the trajectory corresponding to the dashed line in FIG. S13 (b2-b5) decay faster and faster. Fig. S13 (c) further shows the distribution of wave packet at time $t = 3$ corresponding to the dashed line in Fig. S13 (b), from which the vanishing behavior of the dashed line can be clearly seen.

Then we describe the behavior of the reflected wave as $\Gamma_\uparrow$ changes. As shown in FIG. S14 (a1-a5), we show the free particle spectrum corresponding to $\Gamma_\uparrow$ from 0 to 6, respectively. It can be seen that the free particle spectrum at $k_2$ splits gradually as $\Gamma_\uparrow$ increases, which means the DSE will become more and more obvious through the discussion in Section SIV-B. And it can also be clearly seen from FIG. S14(b-c) that, as $\Gamma_\uparrow$ increases from 0 to 6, the DSE gradually appears. To further explain the dynamics of the system, we display the free particle spectrum on the complex plane in FIG. S14(d), as can be seen, with the increase of $\Gamma_\uparrow$, the imaginary part of $E_-(k_2)$ and $E_+(k_3)$ change greatly compared with $E_-(k_1)$, which means that the dashed line branch of the incident wave will decay quickly and the behavior of reflected wave will deviate from the description of gv. These two points can also be clearly seen in FIG. S14(b) and (c), which once again proves the above conclusion about the disappearance of the trajectory corresponding to the dashed line.

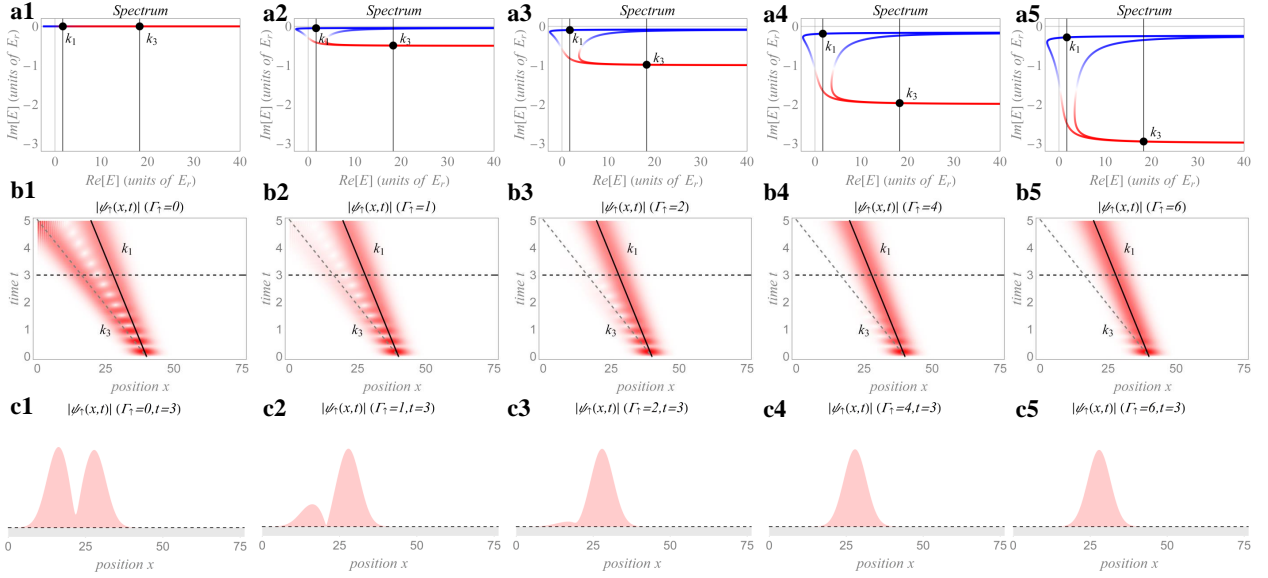

FIG. S13. **Dynamic behavior of incident waves under different strengths of dissipation.** (a1-a5) show the free particle spectrum with different dissipations on the complex energy plane, which show the magnitude of the imaginary of the energy at each momentum as the dissipation increases. In (b1-b5), the dashed line marked by $k_3$ in the incident wave gradually disappears, which is due to the fact that the imaginary part of $E_+(k_3)$ becomes smaller and smaller relative to $E_-(k_1)$ with the increase of the strength of dissipation. In particular, in (c1-c5) we show the spatial distribution of the wave function at time $t = 3$ corresponding to the horizontal dashed line in (b), which can more clearly see the disappearing behavior of the trajectory marked by $k_3$. Here in (b) and (c) we only show the evolution of the spin-up component of the wave packet. The initial momentum of the wave packet is $k_0 = -3$ and other parameters in Hamiltonian are $(\delta, \Omega_R, \Gamma_\downarrow) = (4, 9/2, \Gamma_\uparrow/13)E_r$.

---

[*] These two authors contributed equally.
[†] Corresponding author: fuchun@ucas.ac.cn
[‡] Corresponding author: jphu@iphy.ac.cn
[§] Corresponding author: yangzs@ucas.ac.cn

[1] Z. Ren, D. Liu, E. T. Zhao, C. D. He, K. k. Pak, J. S. Li, and G. B. Jo, (2021), arXiv:2106.04874.
[2] V. Bretin, S. Stock, Y. Seurin, and J. Dalibard, Phys. Rev. Lett. **92**, 050403 (2004).
[3] J.-X. Hou, Journal of Low Temperature Physics **177**, 305 (2014).
[4] D.-W. Zhang, Y.-Q. Zhu, Y. X. Zhao, H. Yan, and S.-L. Zhu, Advances in Physics **67**, 253 (2019).
[5] S. Nakajima, T. Tomita, S. Taie, T. Ichinose, H. Ozawa, L. Wang, M. Troyer, and Y. Takahashi, Nature Physics **12**, 296 (2016).
[6] Y. Takasu, K. Maki, K. Komori, T. Takano, K. Honda, M. Kumakura, T. Yabuzaki, and Y. Takahashi, Phys Rev Lett **91**, 040404 (2003).
[7] L. C. Evans, Partial differential equations (2010).
[8] K. Yokomizo and S. Murakami, Phys. Rev. Lett. **123**, 066404 (2019).
[9] Z. Yang, K. Zhang, C. Fang, and J. Hu, Phys. Rev. Lett. **125**, 226402 (2020).
[10] Y. Yi and Z. Yang, Phys. Rev. Lett. **125**, 186802 (2020).
[11] W.-T. Xue, M.-R. Li, Y.-M. Hu, F. Song, and Z. Wang, Physical Review B **103** (2021), 10.1103/PhysRevB.103.L241408.

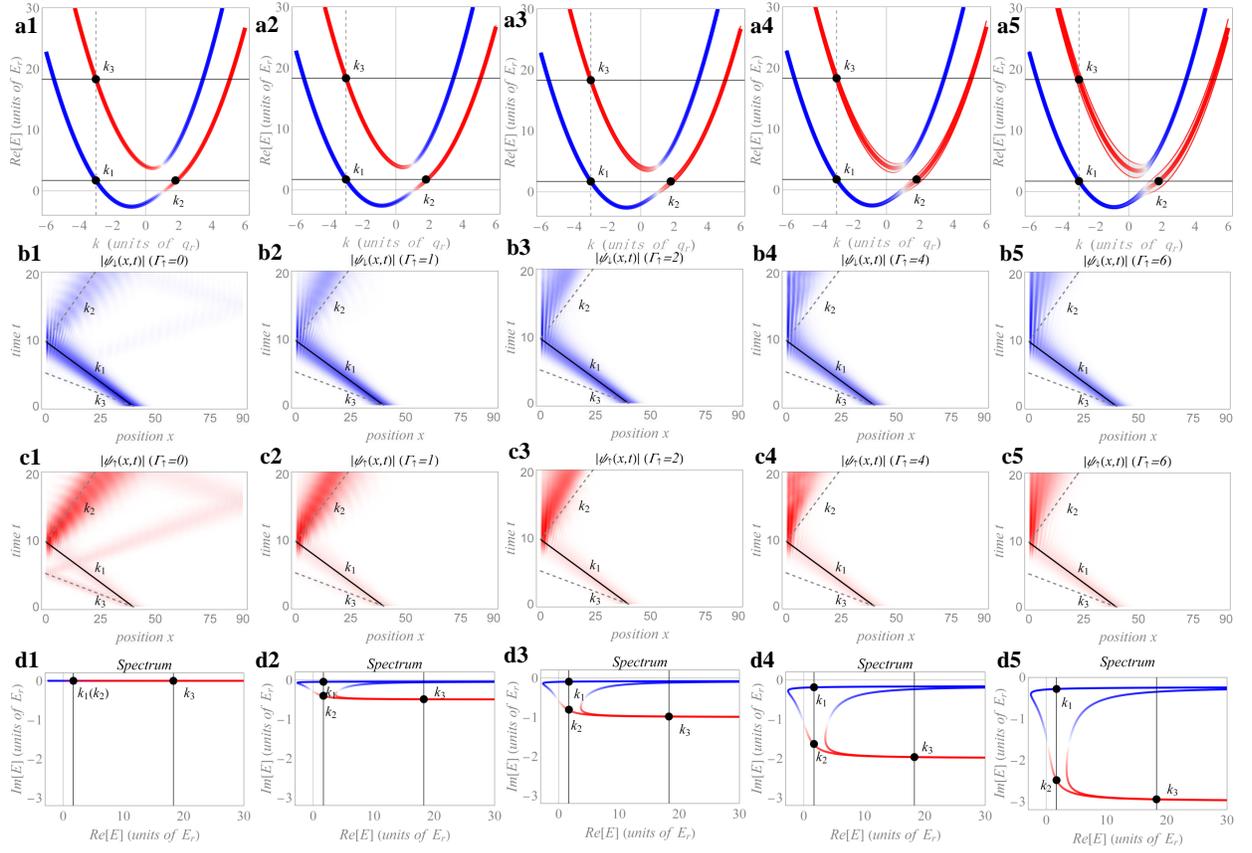

FIG. S14. **The DSE of reflection waves under different strengths of dissipation.** (a1-a5) and (d1-d5) show the free particle spectrum of the system under different strengths of dissipation, from which it can be seen that the splitting of $E_-(k_2)$ and $E_+(k_3)$ gradually increases with the increase of dissipation. The figures in the middle two rows represent the changes in the DSE of the system, where (b1-b5) and (c1-c5) correspond to the spin-up and -down components of the wave packet, respectively. The initial momentum of the wave packet is $k_0 = -3$ and other parameters of the Hamiltonian are $(\delta, \Omega_R, \Gamma_\downarrow) = (4, 9/2, \Gamma_\uparrow/13)E_r$.



\end{document}